\documentclass[a4paper,fleqn]{cas-sc}

\usepackage{float}
\usepackage[T1]{fontenc}
\usepackage[utf8]{inputenc}
\usepackage{lmodern}
\usepackage{threeparttable}
\usepackage{microtype}
\usepackage{amsmath,amssymb}
\usepackage{booktabs,longtable,array,multirow,tabularx}
\usepackage[most]{tcolorbox}
\usepackage{colortbl}
\usepackage{orcidlink}
\usepackage{orcidlink}
\usepackage{graphicx}
\usepackage{adjustbox}
\usepackage{lastpage}
\usepackage{caption}
\usepackage{xcolor}
\usepackage{enumitem}
\usepackage{tikz}
\usetikzlibrary{positioning,arrows.meta,shapes.misc,calc,trees,shapes.geometric}

\usepackage{pgfplots}
\usepackage{fontawesome5}
\pgfplotsset{compat=1.17}
\usepackage[numbers,sort&compress]{natbib}
\usetikzlibrary{shapes.multipart}
\usepackage{xurl}
\usepackage{hyperref}
\usepackage{xspace}
\usepackage[acronym,nomain,nonumberlist,toc]{glossaries}
\makeglossaries
\newacronym{ai}{AI}{Artificial Intelligence}
\newacronym{ml}{ML}{Machine Learning}
\newacronym{ids}{IDS}{Intrusion Detection System}
\newacronym{ips}{IPS}{Intrusion Prevention System}
\newacronym{ndr}{NDR}{Network Detection and Response}
\newacronym{c2}{C2}{Command and Control}
\newacronym{ddos}{DDoS}{Distributed Denial of Service}
\newacronym{iot}{IoT}{Internet of Things}
\newacronym{iiot}{IIoT}{Industrial Internet of Things}
\newacronym{iomt}{IoMT}{Internet of Medical Things}
\newacronym{ics}{ICS}{Industrial Control System}
\newacronym{ot}{OT}{Operational Technology}
\newacronym{can}{CAN}{Controller Area Network}
\newacronym{v2x}{V2X}{Vehicle-to-Everything}
\newacronym{vanet}{VANET}{Vehicular Ad-Hoc Network}
\newacronym{uav}{UAV}{Unmanned Aerial Vehicle}
\newacronym{sdn}{SDN}{Software-Defined Networking}
\newacronym{lan}{LAN}{Local Area Network}
\newacronym{wsn}{WSN}{Wireless Sensor Network}
\newacronym{lpwan}{LPWAN}{Low-Power Wide-Area Network}
\newacronym{isp}{ISP}{Internet Service Provider}
\newacronym{ip}{IP}{Internet Protocol}
\newacronym{dns}{DNS}{Domain Name System}
\newacronym{http}{HTTP}{Hypertext Transfer Protocol}
\newacronym{https}{HTTPS}{HTTP Secure}
\newacronym{tls}{TLS}{Transport Layer Security}
\newacronym{quic}{QUIC}{QUIC Transport Protocol}
\newacronym{mqtt}{MQTT}{Message Queuing Telemetry Transport}
\newacronym{pfcp}{PFCP}{Packet Forwarding Control Protocol}
\newacronym{pqc}{PQC}{Post-Quantum Cryptography}
\newacronym{pcap}{PCAP}{Packet Capture}
\newacronym{csv}{CSV}{Comma-Separated Values}
\newacronym{doi}{DOI}{Digital Object Identifier}
\usepackage{wasysym, pifont}
\newcommand{\NCaptures}{95\xspace}
\newcommand{\NScreened}{262\xspace}
\newcommand{\NExcluded}{28\xspace}
\newcommand{\NExclTrace}{12\xspace}

\newcommand{\NExclNoTraffic}{8\xspace}
\newcommand{\NExclUnavail}{8\xspace}
\newcommand{\NFinal}{234\xspace}

\newcommand{\NPcap}{141\xspace}

\newcommand{\NDomains}{14\xspace}

\usepackage{comment}

\specialcomment{draft}{\begingroup\color{blue}}{\endgroup}
\specialcomment{rewrite}{\begingroup\color{red}}{\endgroup}

\usepackage{todonotes}

\newcommand{\NSinceTwentyThree}{92\xspace}
\newcommand{\NLastTwoYears}{43\xspace}

\newcommand{\Obs}{\textsc{Observatory}\xspace}
\newcommand{\snapshot}{9 November 2026\xspace}
\newcommand{\release}{v1.0\xspace}
\DeclareDocumentCommand{\deriv}{g o}{%
  \smash{\raisebox{0.32em}{\tikz[baseline=0pt]{\draw[line width=0.4pt,line cap=round] (0,0) -- (0,0.58em) (-0.16em,0.40em) -- (0.16em,0.40em);}}}%
  \IfValueT{#2}{\textsuperscript{#2}}%
}

\providecommand{\symw}{1.4pt}
\DeclareRobustCommand{\cfull}{\tikz[baseline=-0.5ex]{\draw[green!60!black, line width=\symw, line cap=round, line join=round] (-0.30em,0) -- (-0.08em,-0.25em) -- (0.32em,0.30em);}}
\DeclareRobustCommand{\chalf}{\tikz[baseline=-0.5ex]{\draw[orange!85!black, line width=\symw, line cap=round] (-0.28em,0) -- (0.28em,0);}}
\DeclareRobustCommand{\cnone}{\tikz[baseline=-0.5ex]{\draw[red!70!black, line width=\symw, line cap=round] (-0.25em,-0.25em) -- (0.25em,0.25em) (-0.25em,0.25em) -- (0.25em,-0.25em);}}

\newcolumntype{Y}{>{\centering\arraybackslash}X}

\begin{document}

\title{\large The \textsc{Cybersecurity Dataset Observatory}: Which Data for AI in Cybersecurity?}

\shorttitle{The Cybersecurity Dataset Observatory}
\shortauthors{F. Cerasuolo et~al.}

\author[1]{Francesco Cerasuolo}[orcid=0009-0000-6476-8092]
\ead{francesco.cerasuolo@unina.it}
\cormark[1]

\author[1]{Gabriele Mangiacapre}
\ead{gabriele.mangiacapre@unina.it}

\author[1]{Antonio Pescap\`e}[orcid=0000-0002-0221-7444]
\ead{pescape@unina.it}

\cortext[1]{Corresponding author}

\affiliation[1]{
  organization={University of Napoli Federico II},
  country={Italy}
}


\begin{abstract}
AI is increasingly used for cybersecurity tasks such as intrusion and malware detection, in enterprise, industrial, vehicular and mobile networks.
Its effectiveness depends on data: models need representative network traffic, both to be trained and to be evaluated realistically.
Public datasets of cybersecurity network traffic, however, are scattered across laboratory pages, repositories and aggregators, and are continuously released, corrected and withdrawn: \NSinceTwentyThree{} of the \NFinal{} datasets we catalog were released in $2023$ or later.
At this pace, surveys that list them are outdated soon after publication, since they describe the field only as it was when they were written.
We present the \textsc{Cybersecurity dataset observatory}, a \emph{public} and \emph{continuously updated} catalog of cybersecurity datasets based on network traffic.
New datasets are added as they are found or submitted, every record is periodically re-checked against its original source, to which it links rather than hosting the data, and every change is recorded in a public change log.
This paper documents the first release: we define a traffic-centered scope and a seven-axis taxonomy, and review \NFinal{} datasets (204 original captures and 30 derivatives) across \NDomains{} domains.
The review shows a field that grows quickly but unevenly, where 91 datasets distribute only processed records, a few laboratories produce a large share of the data, and UAV, wireless and V2X networks have little public evidence, while 5G has few real captures.
We then relate the catalog to common uses in AI research, outline a procedure for selecting datasets, and discuss benchmark pitfalls and open gaps.
\end{abstract}



\maketitle
\input{glyphtounicode}

\section{Introduction}\label{sec:intro}
Advances in \gls{ai} have reshaped network-based security analysis: the open question is no longer whether intrusions can be detected automatically, but how to detect them reliably, at scale and under realistic conditions.
Research on \gls{ai} for network security leverages many types of traffic data as \gls{ai} model input, from packet sequences and flow statistics to protocol messages, and focus on different solution tailored for specific network environments, from enterprise networks to industrial plants, in-vehicle buses and cellular cores.
The \emph{data-driven} nature of these techniques shifts the focus 
not only onto leveraged \gls{ai} methods and evaluation protocol, but equally onto how network traffic data are collected, transformed and labeled.
This information is therefore a prerequisite for interpreting any reported result: without knowing how the traffic was obtained and prepared, it is impossible to tell whether two apparently comparable results reflect differences between detection methods or merely differences in the underlying data~\citep{sommer2010,arp2022}.

Yet, the literature has largely prioritized intrusion detection tasks over the systematic analysis and characterization of the underlying datasets.
Several comparative reviews have examined cybersecurity network traffic datasets~\citep{ring2019,goldschmidt2025,kenyon2020}, but they record the acquisition and preparation information discussed above only insofar as it matters for intrusion detection, and therefore cannot fully characterize the data required for malware-communication analysis, encrypted-traffic inspection, industrial-protocol monitoring, vehicular misbehavior detection or other tasks based on network observations.
A resource that serves all of these tasks should therefore include a dataset because it contains network traffic, not because it was designed for a given task, and record the task, the domain and the labels as properties of each dataset.

The motivations behind this work are twofold.
First, the \textit{dataset landscape is fragmented}: network traffic datasets are scattered across papers, institutional pages and generic repositories.
Moreover, few practical tools are available to help researchers identify and select the most suitable dataset for their specific research objectives and experimental requirements.
Second, the \textit{dataset landscape changes over time}: public datasets are continuously released, corrected,  re-extracted (the same captures processed again with a different feature extractor), moved or withdrawn.
Surveys are snapshots, and their tables start to age as soon as they are published.
The same holds for the dataset lists, 
which are usually checked once, when they are compiled: when we traced each entry of our own preliminary list, compiled from surveys, aggregators and exploratory searches, to its original source, we found wrong links, inconsistent names and data types, and datasets for which no public trace could be found.
A list of datasets therefore remains useful only if it is maintained.

In this work, we present the \emph{Cybersecurity Dataset Observatory} (hereinafter \Obs): 
a public and continuously updated catalog of network-traffic-based cybersecurity datasets.
It is a structured list in which every dataset is described by the same fields and linked to its original source.%
\footnote{Each record links to the original distribution: the \Obs{} never downloads, mirrors or redistributes data, so that the publisher's copy and license remain authoritative.
}
Unlike a survey table, the \Obs is not frozen at the date of this paper:
new datasets are added as they are found or submitted, sources and descriptions are re-checked, corrections and withdrawals are logged rather than silently applied, and the catalog is periodically frozen into dated, versioned releases.
The \Obs{} keeps the history of each record: because every change of a source is dated in the change log, it can show whether two works that cite the same dataset used the same release, and when a dataset was modified or withdrawn.
Therefore, this work is a citable snapshot of the field at release \release{}, with its taxonomy and analysis, while the \Obs keeps that snapshot current, providing the scientific community with an always up-to-date reference.
This reference supports researchers selecting and citing datasets for AI experiments, dataset authors announcing new releases and keeping their records accurate, and reviewers verifying which release of a dataset a submission used.

\subsection{Contributions}
\label{sec:contrib}
This work makes the following contributions:
\begin{itemize}
\item the \Obs{}, a public catalog of cybersecurity datasets that distribute network traffic, whatever security task they were built for, maintained after publication and not frozen at release. Each record links to the original source and the associated publication, and reports the dataset's lineage (real capture or derived, and from which dataset). Changes are tracked in a public changelog and in dated, versioned releases, and the community can propose datasets and corrections, which the maintainers review before publication;
\item a taxonomy describing every dataset along seven axes (domain, artifacts, acquisition, labeling, attack families, access and lineage), so that datasets from different domains and tasks can be compared;
\item a review of the \NFinal{} datasets of release \release{}, grouped into \NDomains{} domains, analyzing their distribution over time, data types, access conditions and derived versions;
\item a guide to using the datasets for AI, relating them to common uses in AI research, outlining a step-by-step selection procedure, and applying it to practical use cases;
\item a discussion of common benchmark pitfalls and of the gaps that remain open.
\end{itemize}

\subsection{Organization}
The rest of the paper is organized as follows.
Section~\ref{sec:rw} presents the background on traffic-based cybersecurity tasks and on how their datasets are built, documented and maintained, and positions the \Obs{} with respect to existing surveys and catalogs.
Section~\ref{sec:observatory} describes the \Obs{}: how it is built, updated, and verified, its public interface, and how it is maintained and extended by the community.
Section~\ref{sec:review} defines the seven-axis taxonomy, gives an overview of release \release{} and reviews its datasets domain by domain.
Section~\ref{sec:suitability} shows how to use the catalog to select datasets for \gls{ai} research.
Section~\ref{sec:pitfalls} discusses benchmark pitfalls and open gaps, and Sec.~\ref{sec:conclusion} concludes with the limitations of this release.
Finally, App.~\ref{sec:assistants} compares the catalog with the answers of three general-purpose AI assistants, App.~\ref{app:excluded} lists the datasets excluded from the first release, with the reason for each exclusion,
App.~\ref{app:acronyms} reports all the acronyms used within the manuscript.
\section{Background and Related Work}\label{sec:rw}
This section reviews the background of this work. It first discusses how network traffic datasets support cybersecurity tasks beyond intrusion detection and why their fragmentation makes them hard to discover and select (Sec.~\ref{sec: traffic_rw}), then the problems of quality, provenance and persistence that affect their reuse (Sec.~\ref{sec: dataset_rw}), and positions the \Obs{} with respect to existing surveys and catalogs (Sec.~\ref{sec:positioning}).

\subsection{From intrusion detection to traffic-centered cybersecurity}
\label{sec: traffic_rw}

Public network traffic datasets have mostly been collected to evaluate intrusion detection systems, and their history can be read as a sequence of generations. The first was built at MIT Lincoln Laboratory for the \texttt{DARPA 1998} and \texttt{DARPA 1999} evaluations~\citep{lippmann2000darpa98,lippmann2000darpa99}, from which \texttt{KDD Cup 1999}~\citep{kddcup99} and later \texttt{NSL-KDD}~\citep{tavallaee2009} were derived; despite well-known redundancy and realism problems~\citep{tavallaee2009}, these data remained the reference benchmark for more than a decade. A second generation replaced them with testbeds that generate benign traffic from user profiles and inject contemporary attacks, such as \texttt{ISCXIDS2012}~\citep{shiravi2012}, \texttt{UNSW-NB15}~\citep{moustafa2015}, \texttt{CICIDS2017} and \texttt{CSE-CIC-IDS2018}~\citep{sharafaldin2018}. These datasets distribute packet captures together with flow features computed by a specific extractor, and the feature tables, rather than the packets, became the de facto benchmark.
A third generation followed the spread of networked systems beyond enterprise IP networks. \gls{iot} datasets such as \texttt{N-BaIoT}~\citep{meidan2018nbaiot}, \texttt{BoT-IoT}~\citep{koroniotis2019}, \texttt{IoT-23}~\citep{garcia2020iot23}, \texttt{TON\_IoT}~\citep{moustafa2021toniot} and \texttt{CICIoT2023}~\citep{neto2023ciciot} capture device-specific and malware-induced traffic; industrial testbeds such as \texttt{SWaT}~\citep{goh2017swat} and \texttt{WADI}~\citep{ahmed2017wadi} pair network data with physical process measurements; and in-vehicle, vehicular, cellular, drone and wireless datasets, such as \texttt{Car-Hacking}~\citep{seo2018gids}, \texttt{ROAD}~\citep{verma2024road}, \texttt{VeReMi}~\citep{vanderheijden2018veremi}, \texttt{5G-NIDD}~\citep{samarakoon2022fivegnidd}, \texttt{UAV-NIDD}~\citep{uavnidd2025} and \texttt{AWID3}~\citep{chatzoglou2021awid3}, record \gls{can} frames, safety messages or radio-level traffic rather than IP flows. In parallel, archives of real traffic such as \texttt{MAWI}~\citep{cho2000mawi}, \texttt{UGR'16}~\citep{macia2018ugr16} and \texttt{Kyoto 2006+}~\citep{song2011kyoto} offer operational data at scale, at the cost of weak or heuristic labels.
Over the same period, the tasks addressed with traffic data have broadened beyond intrusion detection. Five families of tasks can be distinguished, which are also used to compare existing surveys in Table~\ref{tab:related}: intrusion detection in IP networks; detection of malware, botnet and command-and-control traffic, which relies on long captures of infected hosts~\citep{garcia2014ctu13,beigi2014iscxbot}; attack detection in encrypted traffic, where payloads are hidden and only timing, sizes and handshake metadata remain~\citep{hikari2021,tqhc22026}; attack detection in industrial and vehicular networks, whose traffic consists of industrial-protocol messages or \gls{can} frames rather than IP flows~\citep{goh2017swat,seo2018gids}; and detection of attacks on DNS and web services, often based on queries or requests rather than packets~\citep{gimenez2010csic,montazeri2020dohbrw,mahdavifar2021dnsexf}.
These tasks require different labels, temporal granularity and features: a packet trace recorded during a controlled attack and a table of labeled flows are not interchangeable, even when they come from the same dataset family. Existing surveys have documented this evolution~\citep{ring2019,kenyon2020,goldschmidt2025}, and earlier work has warned that the gap between benchmark data and operational networks limits what \gls{ai} results on these datasets can show~\citep{sommer2010}.

\subsection{Dataset quality, provenance and lifecycle management}
\label{sec: dataset_rw}
Dataset availability alone does not guarantee experimental reliability. 
Errors and methodological limitations in widely reused benchmarks have been documented in the literature.
\citet{tavallaee2009} analyzed the redundancy of \texttt{KDD'99} and introduced \texttt{NSL-KDD} as an alternative benchmark. 
\citet{engelen2021} documented labeling and feature-extraction problems in \texttt{CICIDS2017} and released a corrected version. 
\citet{sarhan2022standard} re-extracted several benchmarks using a common NetFlow feature set to improve comparability across datasets.
These examples highlight the importance of distinguishing dataset families from their individual releases and derivatives. Corrections, re-extractions, changes in feature representation, and revised labels may produce materially different data artifacts under the same family name. 
Consequently, a dataset catalog is useful only if it identifies the specific release being described, documents its relationship to previous versions, distinguishes independent datasets from derived representations, and keeps its records verifiable over time.
Without this information, nominally identical dataset references may conceal differences that affect feature availability, label semantics, and experimental reproducibility. Citing only a family name is therefore insufficient to establish which data were used in an experiment.
Reliable dataset discovery also depends on the quality and persistence of dataset documentation. The FAIR principles~\citep{wilkinson2016fair} and datasheets for datasets~\citep{gebru2021datasheets} provide complementary foundations for describing data resources through identifiable releases, structured metadata, provenance, licensing information, and documentation of collection procedures and intended uses. These principles are particularly relevant to cybersecurity traffic datasets, whose interpretation often depends on collection environments, attack scenarios, labeling procedures, feature extraction, and restrictions on redistribution.
The last requirement is the hardest to maintain. Web references decay over time~\citep{klein2014rot}, and for a dataset catalog a broken link is more than a documentation problem: it can prevent researchers from checking a record, obtaining the data or establishing which data a published experiment used. 
AI-assisted search adds a further risk. Asking a general-purpose AI assistant for the datasets of a given domain is a natural alternative to a catalog, and its answer is a useful lead. 
However, Large Language Models (LLMs) can return plausible but non-existent references~\citep{walters2023fabrication,ji2023hallucination}, so datasets found this way may not exist or may not match their description; moreover, the answer is not verified, changes from one assistant to another, and does not describe the datasets along common fields.
Both problems are addressed only if each record is checked against its original distribution and re-checked over time.

\begin{table*}[!t]
\centering
\footnotesize
\setlength{\tabcolsep}{3pt}
\caption[Comparison with related surveys and registries]{Comparison with related surveys and registries.
}
\label{tab:related}
\begin{adjustbox}{max width=\textwidth}
\begin{threeparttable}
\begin{tabular}{lccccccc}
\toprule
\textbf{Dimension} & \textbf{Netresec~\citep{netresec}} & \textbf{Ring~\citep{ring2019}} & \textbf{Goldschmidt~\citep{goldschmidt2025}} & \textbf{COMIDDS~\citep{comidds2024}} & \textbf{Pinto~\citep{pinto2025}} & \textbf{Ashraf~\citep{ashraf2026}} & \textbf{Ours} \\
 & \textit{PCAP list} & \textit{Survey} & \textit{Survey} & \textit{Registry} & \textit{Survey} & \textit{\begin{tabular}[t]{@{}c@{}}Survey + Repo. \end{tabular}} & \textit{Living catalog} \\
\midrule
\rowcolor[HTML]{f0f0f0}
\multicolumn{8}{l}{\textit{Coverage}}\\
Datasets covered & $75$ & $34$ & $89$ & $61$ & $42$ & $124$ & \NFinal \\
Intrusion detection (IP) & \chalf & \cfull & \cfull & \cfull & \cfull & \cfull & \cfull \\
Malware, botnet and C2 & \cfull & \chalf & \cfull & \chalf & \chalf & \chalf & \cfull \\
Encrypted traffic & \cnone & \cnone & \cnone & \cnone & \chalf & \chalf & \cfull \\
Industrial and (non-IP) vehicular networks & \chalf & \cnone & \chalf & \cnone & \chalf & \chalf & \cfull \\
DNS and web attacks & \chalf & \chalf & \chalf & \chalf & \chalf & \chalf & \cfull \\
Cross-domain & \chalf & \chalf & \chalf & \chalf & \chalf & \cfull & \cfull \\
\midrule
\rowcolor[HTML]{f0f0f0}
\multicolumn{8}{l}{\textit{Maintenance}}\\
Updated after publication & \cfull & \cnone & \chalf & \cfull & \cnone & \cfull & \cfull \\
Documented update process & \cnone & \cnone & \chalf & \chalf & \cnone & \chalf & \cfull \\
Change log of corrections & \cnone & \cnone & \cnone & \chalf & \cnone & \cnone & \cfull \\
\midrule
\rowcolor[HTML]{f0f0f0}
\multicolumn{8}{l}{\textit{Information for AI use and Data Provenance}}\\
Unified taxonomy & \chalf & \cfull & \cfull & \cfull & \cfull & \cfull & \cfull \\
Guidance for AI use & \cnone & \chalf & \chalf & \chalf & \chalf & \cfull & \cfull \\
Link to original source & \cfull & \chalf & \cfull & \cfull & \chalf & \cfull & \cfull \\
Release identity & \cnone & \chalf & \chalf & \chalf & \chalf & \chalf & \cfull \\
Lineage & \cnone & \chalf & \chalf & \chalf & \chalf & \chalf & \cfull \\
\midrule
\rowcolor[HTML]{f0f0f0}
\multicolumn{8}{l}{\textit{Reliability}}\\
Known limitations discussed & \cnone & \cfull & \cfull & \chalf & \cfull & \cfull & \cfull \\
Audit of catalog records & \cnone & \cnone & \chalf & \chalf & \cnone & \chalf & \cfull \\
\bottomrule
\end{tabular}
\begin{tablenotes}[flushleft]
\item \textbf{NB:} The second header row gives the type of each work: a list of public packet captures, a survey, a registry, or a survey with an associated online repository. \cfull{} explicit, \chalf{} partial, \cnone{} not documented. For the task rows, \cfull{} means a dedicated section or table, \chalf{} a few datasets mentioned, \cnone{} none. For living catalogs, the number of datasets refers to \snapshot{}.
\end{tablenotes}
\end{threeparttable}
\end{adjustbox}
\end{table*}
\subsection{Positioning with respect to existing surveys and catalogs}
\label{sec:positioning}

Existing surveys provide an essential foundation for understanding the network intrusion detection dataset landscape. Studies by~\citet{ring2019},~\citet{kenyon2020},~\citet{goldschmidt2025},~\citet{pinto2025}, and~\citet{ashraf2026} review public datasets and identify properties relevant to their comparison and use. Other works focus on specific settings, such as industrial networks~\citep{dobler2025}, or address the selection of datasets for intrusion detection and prevention systems~\citep{appsci2026selection}. 
Complementary online resources support dataset discovery beyond the publication of an individual survey: registries, including COMIDDS~\citep{comidds2024,comiddsrepo} and the supplementary resources of~\citet{xgoldyrepo}, and curated lists of public packet captures, of which Netresec~\citep{netresec} is the most widely used.%
\footnote{
Generic data platforms (e.g., Kaggle, Hugging Face) and the dataset pages of individual laboratories (e.g., the Canadian Institute for Cybersecurity, the Stratosphere Laboratory) are not included: they host data, often as third-party copies, and do not aim to catalog the field.
}
Table~\ref{tab:related} compares representative surveys and registries with the \Obs{} along four groups of dimensions: the coverage of cybersecurity tasks and network domains, the maintenance of the list after publication, the information provided for AI use and data provenance, and the reliability of the records.
A capability is marked as explicit (\cfull) when it is documented for every record, partial (\chalf) when it is documented only for some records or only in the text, and absent (\cnone) when no documentation of it was found in the publication or its associated resource as of \snapshot{}.

The comparison in Table~\ref{tab:related} shows that the \Obs{} extends existing work along four groups of dimensions, which correspond to the blocks of the table.

\noindent
\textbf{Coverage.} Existing surveys and registries are built around intrusion detection in IP networks. \citet{ring2019} and COMIDDS focus on enterprise traffic, with other domains appearing only as isolated entries, typically AWID for 802.11. \citet{goldschmidt2025} and \citet{pinto2025} add environment-specific datasets (IoT, industrial, SDN), and \citet{ashraf2026} group datasets by application domain, but all three select them as intrusion detection benchmarks. Netresec lists captures from more settings, including malware and industrial protocols, but as pointers to sources. In all these works, malware and C2 captures, generic encrypted traffic and non-IP communications are absent or covered only in part.
The \Obs{} instead includes any dataset of network communications, whatever its original task, and release \release{} covers \NFinal{} datasets across \NDomains{} domains, including malware and \gls{c2} captures, encrypted traffic, and non-IP communications such as \gls{can} buses and \gls{v2x} messages.

\noindent
\textbf{Continuous update.} New and corrected releases keep appearing: \NSinceTwentyThree{} of the \NFinal{} datasets in release \release{} date from $2023$ onward, \NLastTwoYears{} from $2025$--$2026$ alone, so a static list loses coverage quickly.
COMIDDS, the online repository of~\citet{ashraf2026} and the supplementary list of~\citet{goldschmidt2025} are also updated after publication, but without a documented procedure for adding, correcting and withdrawing records or a readable history of changes. Each release of the \Obs{} is versioned and dated, and new entries are added through the contribution process described in Sec.~\ref{sec:observatory}.

\noindent
\textbf{Information for AI use and data provenance.} Most surveys describe datasets through a common set of properties, and some relate labels and splits to supervised or unsupervised learning; the \Obs{} relates its columns to common uses in AI research (Sec.~\ref{sec:suitability}). Few works, however, record which release is meant and how releases derive from one another: derivations such as \texttt{DARPA}, \texttt{KDD Cup 1999} and \texttt{NSL-KDD}, or \texttt{CICIDS2017} and its corrected versions, are discussed in the text but not encoded per record. The \Obs{} identifies each release and records its lineage, so that a re-extraction, such as a common-feature NetFlow version, is not counted as an independent dataset.

\noindent
\textbf{Reliability.} Most surveys, and the \Obs{} itself, discuss the known limitations of the datasets they review. The accuracy of the catalog records is a separate question: a sound dataset may be described wrongly, or a record may point to a dataset that cannot be found. Where existing works check their records, they do so once: \citet{ring2019} checked their repositories in February 2019, \citet{goldschmidt2025} verified their links in early 2025, and \citet{pinto2025} attempted to retrieve each dataset at the time of writing, with no later re-check. The \Obs{} traces every record to its original distribution before inclusion and re-checks it at each update.

\noindent
Finally, the \Obs{} applies the FAIR principles~\citep{wilkinson2016fair} to the description of the data rather than to the data themselves.
Each record has a stable entry and a link to the original distribution, and can be searched and filtered on every axis (\textit{findable}); the access axis states whether the data are open, require registration or are released on request (\textit{accessible}); all records are described by the same fields and vocabularies, such as acquisition, labeling and attack families (\textit{interoperable}); and lineage, acquisition and labeling document how the data were produced and how they relate to other releases (\textit{reusable}).

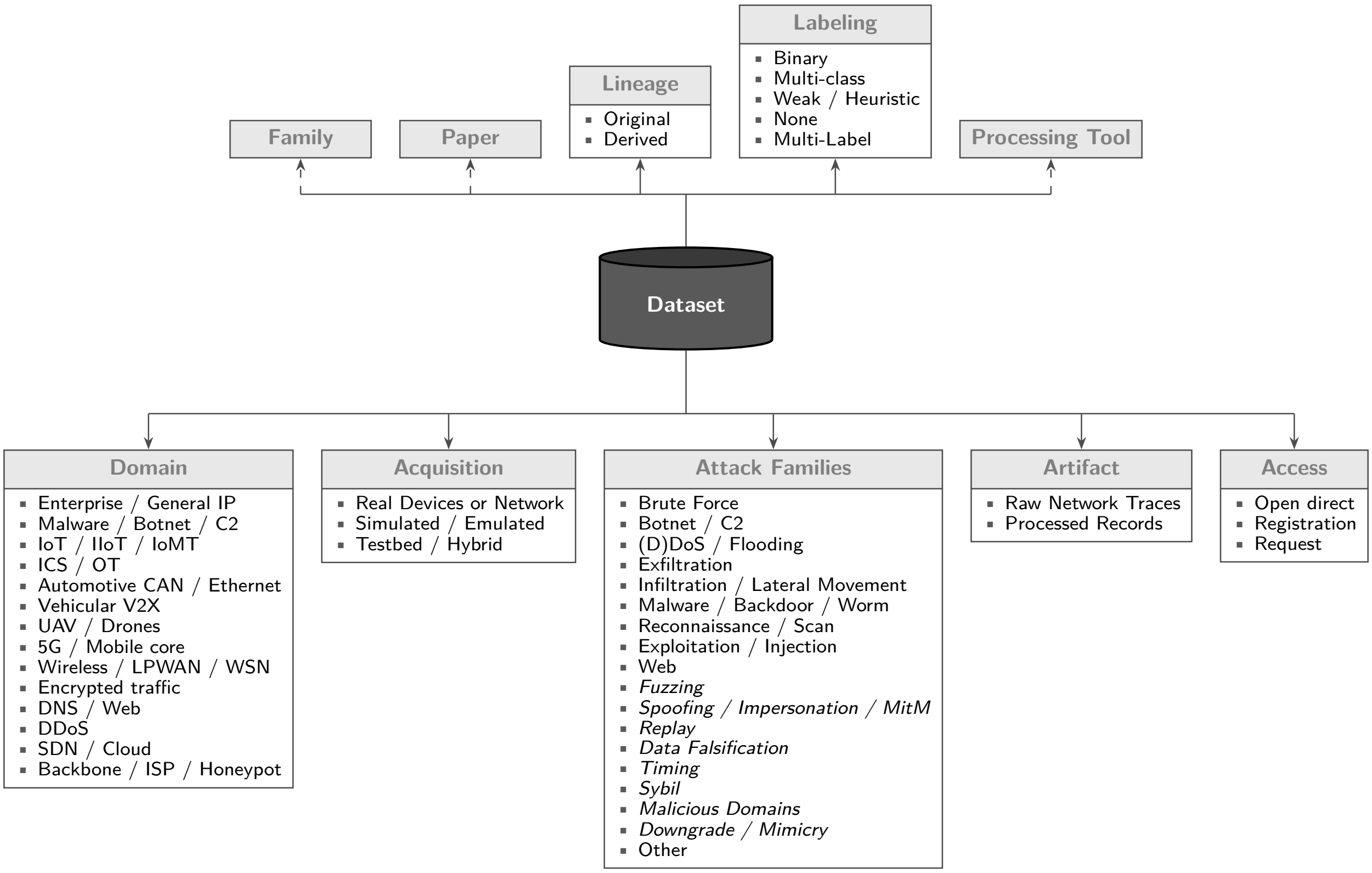
\begin{figure*}[!t]
\centering
\newcommand{\itm}{\textcolor{gray!70!black}{\textbullet}~}
\newcommand{\cl}[1]{{\renewcommand{\arraystretch}{0.92}%
  \begin{tabular}{@{}l@{}}#1\end{tabular}}}
\newcommand{\hd}[1]{\textbf{\footnotesize\strut #1}}
\begin{adjustbox}{max width=\textwidth}
\begin{tikzpicture}[
  font=\scriptsize,
  card/.style={rectangle split, rectangle split parts=2,
    rectangle split part fill={gray!18,white},
    rectangle split part align={center,left},
    draw=gray!60!black, line width=0.5pt,
    inner xsep=4pt, inner ysep=2pt, minimum width=1.8cm},
  head/.style={draw=gray!60!black, line width=0.5pt, fill=gray!18,
    inner xsep=4pt, inner ysep=2pt, minimum width=1.8cm},
  core/.style={draw=black, cylinder, shape border rotate=90,
    aspect=0.25, minimum width=2.2cm, minimum height=1.3cm,
    inner sep=0pt, cylinder uses custom fill,
    cylinder body fill=gray!70!black, cylinder end fill=gray!45!black,
    text=white, line width=0.8pt, font=\footnotesize\bfseries},
  bus/.style={gray!60!black, line width=0.5pt},
  rel/.style={bus, -{Stealth[length=1.6mm]}},
  opt/.style={rel, dashed},
]

\node[core] (ds) at (0,0) {Dataset};

\matrix (toprow) [column sep=3.5mm, anchor=south, inner sep=0pt] at (0,1.85) {
  \node[head, anchor=south] (fam) {\hd{Family}}; &
  \node[head, anchor=south] (pap) {\hd{Paper}}; &
  \node[card, anchor=south] (lin) {\hd{Lineage}
    \nodepart{two}\cl{\itm Original\\ \itm Derived}}; &
  \node[card, anchor=south] (lab) {\hd{Labeling}
    \nodepart{two}\cl{\itm Binary\\ \itm Multi-class\\ \itm Weak / Heuristic\\\itm None\\\itm Multi-Label}}; &
  \node[head, anchor=south] (der) {\hd{Processing Tool}}; \\
};

\matrix (botrow) [column sep=3.5mm, anchor=north, inner sep=0pt] at (0,-1.85) {
\node[card, anchor=north] (dom) {\hd{Domain}
    \nodepart{two}\cl{\itm Enterprise / General IP\\ \itm Malware / Botnet / C2\\
      \itm IoT / IIoT / IoMT\\ \itm ICS / OT\\ \itm Automotive CAN / Ethernet\\
      \itm Vehicular V2X\\ \itm UAV / Drones\\ \itm 5G / Mobile core\\
      \itm Wireless / LPWAN / WSN\\ \itm Encrypted traffic\\ \itm DNS / Web\\
      \itm DDoS\\ \itm SDN / Cloud\\ \itm Backbone / ISP / Honeypot}}; &
  \node[card, anchor=north] (cap) {\hd{Acquisition}
    \nodepart{two}\cl{\itm Real Devices or Network\\ \itm Simulated / Emulated\\ \itm Testbed / Hybrid}}; &
  \node[card, anchor=north] (atk) {\hd{Attack Families}
    \nodepart{two}\cl{\itm Brute Force\\ \itm Botnet / C2\\ \itm (D)DoS / Flooding\\
      \itm Exfiltration\\ \itm Infiltration / Lateral Movement\\
      \itm Malware / Backdoor / Worm\\
      \itm Reconnaissance / Scan\\ \itm Exploitation / Injection\\ \itm Web\\
      \itm \textit{Fuzzing}\\ \itm \textit{Spoofing / Impersonation / MitM}\\
      \itm \textit{Replay}\\ \itm \textit{Data Falsification}\\
      \itm \textit{Timing}\\ \itm \textit{Sybil}\\
      \itm \textit{Malicious Domains}\\ \itm \textit{Downgrade / Mimicry}\\
      \itm Other}}; &
  \node[card, anchor=north] (art) {\hd{Artifact}
    \nodepart{two}\cl{\itm Raw Network Traces\\ \itm Processed Records}}; &
  \node[card, anchor=north] (acc) {\hd{Access}
    \nodepart{two}\cl{\itm Open direct\\ \itm Registration\\ \itm Request}}; \\
};

\coordinate (bt) at (0, 1.4);
\coordinate (bb) at (0,-1.4);
\draw[bus] (ds.north) -- (bt);
\draw[bus] (ds.south) -- (bb);
\draw[bus] (fam.south |- bt) -- (der.south |- bt);
\draw[bus] (dom.north |- bb) -- (acc.north |- bb);
\foreach \n/\s in {fam/opt, pap/opt, lin/rel, lab/rel, der/opt}
  \draw[\s] (\n.south |- bt) -- (\n.south);
\foreach \n in {dom, cap, atk, art, acc}
  \draw[rel] (\n.north |- bb) -- (\n.north);

\end{tikzpicture}
\end{adjustbox}
\caption{Conceptual model of an \Obs{} record. Each dataset is linked to the papers that introduce or use it, its lineage (original or derived from another dataset) and, for derivatives, the family of its source, its labeling scheme, the processing tool used to derive its artifacts (when applicable), its application domain, its acquisition details, the attack families it covers, the artifacts it distributes, and its access link. Dashed arrows represent optional associations.
}
\label{fig:entity}
\end{figure*}

\section{The Cybersecurity Dataset Observatory}\label{sec:observatory}

This section defines the \Obs{} and how it can be accessed, and describes how its records are collected, selected and verified, 
and how it is maintained, governed and extended by the community.

\paragraph{Cybersecurity Dataset Observatory.}
The \Obs is a public catalog of network-traffic datasets for cybersecurity, 
continuously maintained and updated to support \gls{ai} research in network security.
The \Obs{} aims to include every publicly described dataset that distributes recorded network communications, or features demonstrably derived from them, whatever security task it was designed for; release \release{} covers the datasets found through the sources and searches described below.
Each record describes one dataset, or one separately published version of it. Every record is classified along the seven taxonomy axes (Sec.~\ref{sec:taxonomy})
and carries
($i$) the link to the original distribution,
($ii$) the associated paper, when one exists, and
($iii$) the date of the snapshot in which it was last checked.
Each record states the type of artifacts it distributes (raw network traces, processed records, or both), and each derivative points to the records it was produced from, which in turn list their derivatives (Fig.~\ref{fig:entity}).
The \Obs is maintained by the \textsc{Traffic} group of the \textit{University of Napoli Federico II}.
It does not download, mirror or host any dataset.
Every record points to the publisher's page or to a persistent identifier, and the publisher's copy remains the only authoritative version.
This choice respects the access conditions set by dataset authors: 71 of the 234 datasets of release \release{} are distributed only after registration (49) or on request (22), which a mirrored copy would bypass. 
The catalog is published on the \Obs{} website (\url{https://traffic-arclab.github.io/observatory/}). Each record can be searched and filtered along every axis of the taxonomy, and its detail view links to the download page and the paper and shows the record's lineage, that is, the dataset it derives from and the datasets derived from it.

\paragraph{Sources.}\label{sec:method}
Candidates for \Obs were mainly identified from five sources: ($i$) existing registries~\citep{comiddsrepo,xgoldyrepo}; ($ii$) datasets reviewed in recent surveys~\citep{goldschmidt2025,pinto2025,ashraf2026,dobler2025}; ($iii$) repositories Zenodo, IEEE DataPort and Mendeley Data; and ($iv$) searches on Google Scholar and IEEE Xplore for datasets released in $2023$--$2026$;%
\footnote{The query was (``dataset'' OR ``data set'') AND (``intrusion'' OR ``attack'' OR ``malware'' OR ``anomaly'') AND (``network traffic'' OR ``pcap'' OR ``netflow''), repeated with each protocol name (Modbus, DNP3, IEC~104, CAN, MQTT, ...) in place of the traffic terms.}
($v$) generative AI assistants.
The last source was used differently from the others. The assistants suggested candidate datasets and drafted record descriptions, but their output was treated only as a lead: no candidate was included unless it could be traced to its original distribution and associated paper, and every field of a record was checked against that source by the authors. Suggested candidates that could not be traced in this way were discarded, which is the risk discussed in Sec.~\ref{sec: dataset_rw}.

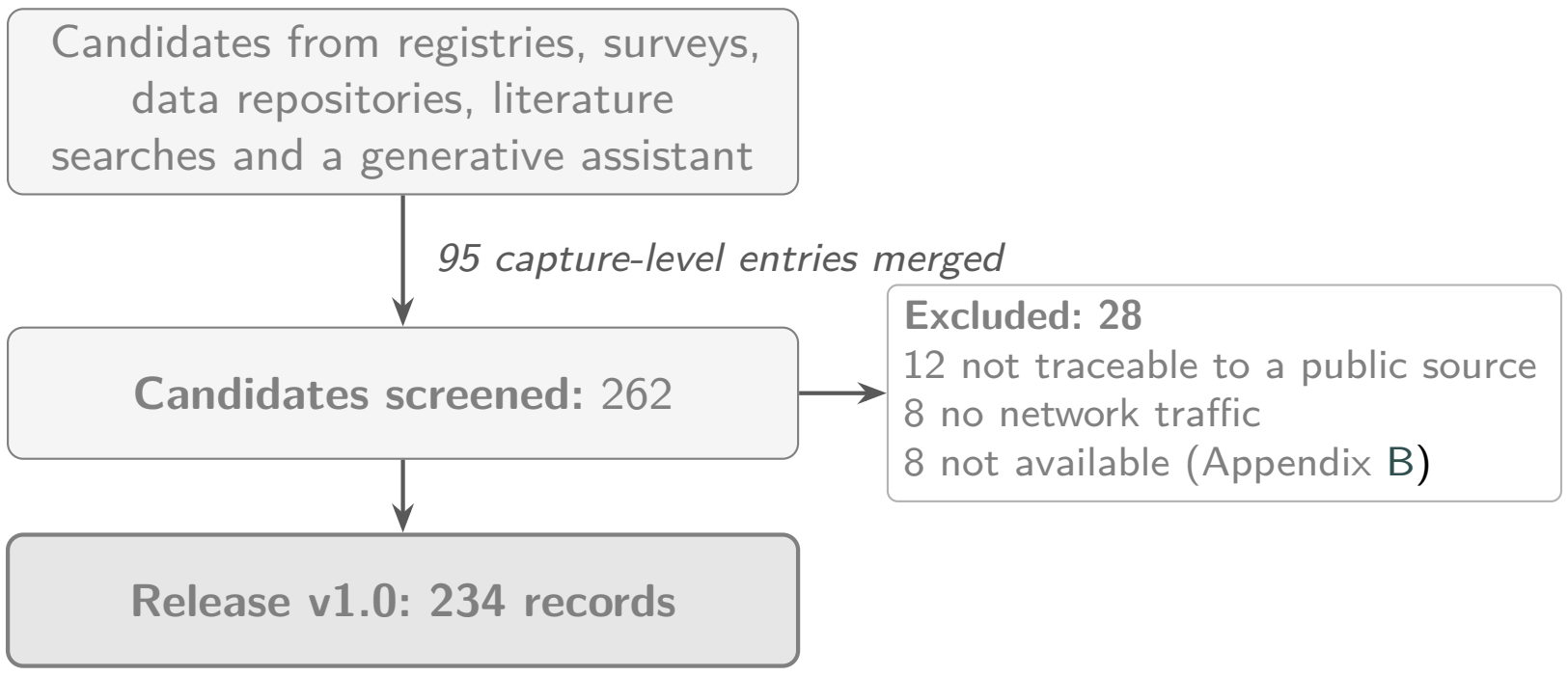
\begin{figure}[!t]
\centering
\begin{tikzpicture}[
  node distance=0.5cm,
  flowstep/.style={draw, rounded corners=3pt, fill=gray!8, text width=5.2cm, minimum height=0.9cm, align=center, font=\small},
  flowout/.style={draw, thick, rounded corners=3pt, fill=gray!20, text width=5.2cm, minimum height=0.9cm, align=center, font=\small\bfseries},
  flowexc/.style={draw=gray!60, rounded corners=2pt, fill=white, text width=4.4cm, align=left, font=\footnotesize},
  selarr/.style={-{Stealth[length=2mm]}, thick, gray!70!black}]

\node[flowstep] (src) {Candidates from registries, surveys, data repositories, literature searches and a generative assistant};
\node[flowstep, below=0.9cm of src] (scr) {\textbf{Candidates screened:} \NScreened};
\node[flowout, below=of scr] (fin) {Release \release{}: \NFinal{} records};
\node[flowexc, right=0.6cm of scr] (exc) {\textbf{Excluded: \NExcluded}\\
\NExclTrace{} not traceable to a public source\\
\NExclNoTraffic{} no network traffic\\
\NExclUnavail{} not available (Appendix~\ref{app:excluded})};

\draw[selarr] (src) -- node[right, font=\footnotesize\itshape, xshift=1mm] {\NCaptures{} capture-level entries merged} (scr);
\draw[selarr] (scr) -- (fin);
\draw[selarr] (scr.east) -- (exc.west);
\end{tikzpicture}
\caption{Construction of release \release{} (\snapshot): candidates screened against the inclusion criteria of Section~\ref{sec:scope}, excluded candidates by reason, and records in the release.}
\label{fig:flow}
\end{figure}

\paragraph{Scope and inclusion criteria.}\label{sec:scope}
The unit of the catalog is the \emph{record}: a dataset, or a separately published version of it, that can be cited and obtained on its own. Individual captures that belong to one collection, such as the \texttt{CTU malware} captures~\citep{stratosphereMCF}, do not have their own records and are listed under the collection.
A candidate becomes a record when it meets four criteria:
\begin{itemize}
  \item \textbf{Scope.} It was built for, or is documented for, a network security task, such as intrusion, attack, malware or anomaly detection, whatever its labels.
  \item \textbf{Network traffic.} It contains recorded network communications (packets, frames or bus messages) or records demonstrably derived from them (flows, logs, feature tables); host logs, malware binaries and generic sensor data are excluded.
  \item \textbf{Accessibility.} It can be obtained by third parties at the time of the release, openly, after registration or on request. Datasets kept private by their authors, withdrawn, or announced but not yet released are excluded from the release and listed in App.~\ref{app:excluded}; the website keeps them, marked as not available, so that they re-enter the catalog if their distribution is restored or published.
  \item \textbf{Consistency.} It can be traced to its original distribution (publisher, authors' repository or persistent identifier), and its name, paper and declared content agree with each other, so that the record identifies a single dataset; candidates that cannot be traced are excluded (App.~\ref{app:excluded}).
\end{itemize}
Re-extractions, relabelings and augmentations that meet these criteria are kept but marked as derivatives of their source (\deriv{}).
The searches were run between September and November~2026 by the authors. 
After merging \NCaptures{} capture-level entries into their parent collections, \NScreened{} candidates were screened against these criteria and \NExcluded{} were excluded: \NExclTrace{} could not be traced to any public source, \NExclNoTraffic{} did not contain network traffic and \NExclUnavail{} could not be obtained at the time of the release. The remaining \NFinal{} records form release \release{} (Figure~\ref{fig:flow}).

\paragraph{Verification.}
Verification was performed manually by the authors for every record. They checked that the source page exists and is reachable, that the dataset name, paper and data type match the description, and whether access is open or requires registration or a request. In the same step they collected, from the source page and the associated paper, the values of the seven taxonomy axes (Sec.~\ref{sec:taxonomy}); a value that neither source documents is left empty rather than inferred. Since the \Obs{} links to the original source and never downloads the data, verification confirms the source, not the files.


\paragraph{Maintenance and governance.}\label{sec:maintenance}
This paper documents 
release \release{} of the \Obs{} 
(counts, tables and figures refer to that release). 
After release, the catalog is maintained through three separate activities:
\begin{itemize}
  \item \textbf{Discovery.} New candidates come from two channels: the searches described above, repeated periodically, and proposals sent through a public form on the website, open to dataset authors and to any other user. A proposal must include a link to the primary source and, if available, the associated paper, and may describe the dataset along the axes of the taxonomy to speed up its review;
  \item \textbf{Validation.} Every candidate, whatever its channel, is checked manually by the maintainers against the inclusion criteria and with the verification procedure described above. Once accepted, it appears in the next update. Validation is kept manual on purpose: AI assistants can suggest candidates, but their answers are incomplete, inconsistent with each other and must be verified against the source (App.~\ref{sec:assistants});
  \item \textbf{Periodic check.} Records already in the catalog are re-checked periodically by the maintainers: whether the source page is still reachable, whether its access conditions have changed, and whether a new or corrected version has been published.
\end{itemize}
Every change resulting from these activities is dated and recorded in a public change log (\url{https://traffic-arclab.github.io/observatory/changelog.html}) that lists, for every added, corrected or withdrawn record, the date and the type of change. Versioned releases, which freeze the catalog at a given date, are published periodically on the website. Records are never deleted: a dataset whose distribution disappears is kept on the website, marked as not available and excluded from the counts of the next release (App.~\ref{app:excluded}), so that published results obtained on it remain traceable.

\paragraph{How to contribute and stay informed.}\label{sec:contribute}
The \Obs{} distinguishes three roles. \textbf{Users} consult the catalog and can subscribe by email to the domains they follow, to be notified when datasets are added. They can also propose datasets that are missing, even if they are not their authors, and send feedback on the \Obs{} itself, such as missing features, unclear descriptions or problems with the website. \textbf{Data providers} are the authors of a dataset: they can submit their own releases, including those the maintainers have not yet found, and report corrections to the corresponding records. \textbf{Maintainers}, currently the authors of this paper, check every proposal, whatever its origin, against the scope and the verification procedure described above, and run the periodic check. Trusted contributors, such as authors of several records, may become maintainers in later releases.



\section{A Unified Taxonomy and Review of Traffic-Based Cybersecurity Datasets}\label{sec:review}
This section introduces the taxonomy used to describe each dataset (Sec.~\ref{sec:taxonomy}) and applies it to the \NFinal{} datasets of release \release{}. It first gives an overview of the catalog as a whole (Sec.~\ref{sec:overall}), and then reviews the datasets domain by domain, following the columns of the domain tables (Secs.~\ref{sec:rev-iot}--\ref{sec:rev-encrypted}).

\begin{table}[!htbp]\centering\footnotesize
\caption{Eligible records by application domain: distributed artifacts, datasets recorded in real networks or installations, derivatives of earlier captures and datasets released since $2023$. 
}
\label{tab:domain}
\begin{threeparttable}
\begin{tabular}{lcccccccc}\toprule
& & \multicolumn{3}{c}{\textbf{Distributed Artifacts}} & & & \\ \cmidrule(lr){3-5}
\multirow{-2}{*}{\textbf{Domain}} & \multirow{-2}{*}{\textbf{Number}} & Raw+Proc & Raw & Proc. & 
\multirow{-2}{*}{\textbf{Open}} & 
\multirow{-2}{*}{\textbf{Real}} & \multirow{-2}{*}{\textbf{Derived \deriv{}}} & \multirow{-2}{*}{\textbf{2023+}} \\ \midrule
IoT / IIoT / IoMT & 50 & 28 & 4 & 18 & 32 & 2 & 8 & 27 \\
Malware / botnet / C2 & 15 & 7 & 3 & 5 & 10 & 1 & 2 & 0 \\
ICS / OT & 22 & 10 & 5 & 7 & 14 & 2 & 0 & 6 \\
Automotive CAN/Ethernet & 24 & 2 & 20 & 2 & 19 & 12 & 0 & 11 \\
Vehicular V2X & 6 & 0 & 6 & 0 & 5 & 0 & 0 & 2 \\
UAV / drones & 5 & 3 & 0 & 2 & 3 & 0 & 0 & 5 \\
5G / mobile core & 13 & 6 & 3 & 4 & 10 & 0 & 0 & 11 \\
Wireless / LPWAN / WSN & 5 & 2 & 0 & 3 & 3 & 0 & 1 & 2 \\
DNS / web & 10 & 2 & 0 & 6\tnote{a} & 3 & 2 & 2 & 2 \\
DDoS & 7 & 2 & 2 & 3 & 3 & 2 & 0 & 2 \\
SDN / cloud & 5 & 2 & 1 & 2 & 3 & 1 & 0 & 2 \\
Backbone / ISP / Honeypot & 11 & 2 & 2 & 7 & 9 & 9 & 1 & 2 \\
Enterprise / general IP & 57 & 17 & 9 & 31 & 46 & 6 & 16 & 18 \\
Encrypted traffic & 4 & 2 & 1 & 1 & 3 & 0 & 0 & 2 \\
\midrule
Total & 234 & 85 & 56 & 91 & 163 & 37 & 30 & 92 \\
\bottomrule\end{tabular}
\begin{tablenotes}[flushleft]
\scriptsize
\item \textbf{N.B.:} 
\textbf{Domain:}
Command and Control (C2),
Internet of Things (IoT), Industrial Internet of Things (IIoT), Internet of Medical Things (IoMT),
Industrial Control Systems (ICS), Operational Technology (OT),
Controller Area Network (CAN), Vehicle-to-Everything (V2X),
Unmanned Aerial Vehicle (UAV),
Low-Power Wide-Area Network (LPWAN), Wireless Sensor Network (WSN),
Domain Name System (DNS), Distributed Denial of Service (DDoS),
Software-Defined Networking (SDN),
Internet Service Provider (ISP), Internet Protocol (IP);
\textbf{Distributed Artifacts:} \emph{Raw+Proc}: raw network traces together with processed records; \emph{Raw}: raw network traces only; \emph{Proc.}: processed records only (flows, logs or feature tables). For automotive and V2X datasets, traces are raw CAN/Ethernet frames or V2X messages rather than IP packets. \emph{Real}: datasets marked \CIRCLE{} in the \emph{Acq.} column of their domain table (recorded in a real network, installation or vehicle); \textbf{Derived\deriv{}}: derivatives of an earlier capture; \emph{2023+}: datasets whose first release is 2023 or later.
\item[a] Two datasets release only raw HTTP requests, with neither traces nor extracted records, and are not counted in the artifact columns.
\end{tablenotes}
\end{threeparttable}
\end{table}

\subsection{Dataset Taxonomy}\label{sec:taxonomy}
Existing surveys usually classify datasets by application domain or by the attacks they contain. This is not enough to decide whether a dataset can support a given experiment: two \gls{iot} datasets may differ in whether they ship packets or only feature tables, in whether their traffic was recorded on real devices or simulated, and in whether their labels are binary, multiclass or only heuristic.
The \Obs{} therefore describes every record along seven axes: the \emph{domain}, which assigns each dataset to one of the tables of this section, and six axes
that are the same for every domain: 
\begin{itemize}
  \item \textbf{Artifact} distinguishes datasets that release raw network traces (packets, frames or messages) from those that release only processed records such as flows, logs or feature tables. The difference is substantial: packet traces preserve timing, ordering and, subject to capture settings, payload, whereas processed records fix the choices of the extractor, recorded in \emph{Processing Tool} (Proc. Tool in tables), and cannot be turned back into packets;
  \item \textbf{Acquisition} (\emph{Acq.} in tables) records where the traffic comes from: a real network or installation, a testbed or hybrid setup that combines physical devices with simulated or synthetic components, or an emulated or simulated environment; ``realistic'' is thus not a binary property. The value reflects the capture environment: an operational network counts as real even when the authors launch attacks during the capture, and a laboratory counts as a testbed even when it carries real users;
  \item \textbf{Labeling} (\emph{Lab.} in tables) and \emph{Classes} describe the ground truth: whether labels are binary, multiclass, weak (when the label cannot be traced to individual records, because it is assigned per file or per time interval without identifying the injected records, or derived from IDS alerts, blacklists or detectors) or absent, whether a benign class is present, and how many attack classes are distinguished;
  \item \textbf{Attack families} map the native attack names of each dataset to a common vocabulary, so that coverage can be compared without flattening the original labels. The vocabulary of each domain was built bottom-up from the attack names documented by its datasets: names describing the same kind of attack (for example \emph{SYN flood}, \emph{UDP flood} and \emph{HTTP flood}) were grouped into one family, families shared by most domains, such as (D)DoS, reconnaissance and brute force, keep the same letter everywhere, and families specific to one domain, such as fuzzing for \gls{can} or data falsification for \gls{v2x}, are added to that domain only; attacks that fit no family are recorded as \emph{Other} (O).
  Letters denote the same family across tables, with domain-specific meanings stated in the legend of each table.
  Datasets labeled with MITRE ATT\&CK tactics or techniques rather than attack names are mapped to families through them (for example, Reconnaissance and Discovery to reconnaissance), and their number of classes is marked with an asterisk.
  A family is marked as present when the documentation of the dataset describes at least one attack of that family, not from the label name alone.
  The legend of each domain table lists its families and, where a family is not self-explanatory, examples of the native names it includes.
  Families are therefore comparable across datasets, whereas the number of attack classes (\emph{Att.}) is not, since it depends on how finely each dataset splits its attacks.
  \item \textbf{Access} records whether a dataset is openly downloadable, requires registration or is released on request; datasets that cannot be obtained at the time of the release are not part of it (App.~\ref{app:excluded}).
  \item \textbf{Lineage}, shown by the derivative mark (\deriv{}), records whether a release is derived from an earlier capture. A release is considered a derivative when it is a re-extraction, which processes the same captures again with a different tool or feature set; a relabeling, which keeps the records but changes their labels, for example to correct known errors; or an augmentation, which adds synthetic records, for example to rebalance classes. Derivatives are legitimate, and some are preferable to their originals because they correct known defects, as in the revised release of \texttt{CICIDS2017}~\citep{engelen2021}. They can change the results obtained on a dataset, but none adds new observations of network behavior.
\end{itemize}
These axes correspond to the columns of each domain table (Secs.~\ref{sec:rev-iot}--\ref{sec:rev-encrypted}) and to the filters of the \Obs{} website.
They are also deliberately independent: a vehicular trace may be simulated or recorded on a physical bus, a packet capture may be open or released on request, and a flow table may be an original release or a re-extraction of an earlier capture.

\subsection{Overview}\label{sec:overall}
This subsection gives an overview of release \release{}. Table~\ref{tab:domain} reports, for each domain, the number of datasets, the artifacts they distribute, and how many were recorded in real networks, derived from earlier captures, or released since $2023$; domains are listed in the order of Secs.~\ref{sec:rev-iot}--\ref{sec:rev-encrypted}.

Each dataset is assigned to exactly one domain, with a fixed order of precedence. 
A dataset is first assigned to the specialized network environment in which it was captured: \gls{iot}, \gls{ics}, automotive, \gls{v2x}, \gls{uav}, 5G core, wireless, \gls{sdn}/cloud, or backbone, ISP and honeypot networks. Datasets captured in a generic environment are assigned to a property-based domain when their defining property is the traffic type (encrypted traffic, DNS/web) or the threat (malware/C2, DDoS), and otherwise to the enterprise and general-purpose domain, which therefore acts as the residual one and is presented last. 
For example, a DDoS dataset captured in an \gls{iot} testbed belongs to \gls{iot}, while \texttt{CIC-DDoS2019}, captured in a generic testbed, belongs to \gls{ddos}. Because each dataset appears in one domain only, per-domain counts understate cross-cutting topics; the website therefore also tags each record with its protocol and threat, and these tags can be combined with the domain filter.
The domains differ not only in scale and technology, but also in what counts as observable network behavior: enterprise benchmarks rely on IP packet captures and flow features, \gls{iot} collections expose device-specific and malware-induced communication, industrial datasets must be read together with protocol and process semantics, and automotive and vehicular traces consist of \gls{can} frames or safety messages rather than IP traffic.

\emph{Coverage is very uneven.} Enterprise networks ($57$ datasets) and \gls{iot} ($50$) account for nearly half of the catalog, followed by automotive networks ($24$), industrial control systems ($22$) and malware and botnets ($15$). 
Other domains, such as V2X, UAV and encrypted traffic, are the subject of growing research interest, yet they are among those with the fewest public datasets.
\emph{As for the released artifact, the picture is equally varied}. Raw traces are the norm for automotive and \gls{v2x} datasets ($22$ of $24$ and all six) and are released by about two thirds of the industrial and \gls{iot} ones, but more than half of the enterprise datasets ($31$ of $57$) and six of the ten DNS/web datasets release only processed records, and two further DNS/web datasets only raw HTTP requests.
\emph{Access is mostly open}: $163$ of the $234$ datasets can be downloaded directly, $49$ require registration and $22$ are released only on request, among them some of the most cited testbeds, such as SWaT, WADI, EPIC and AWID. DNS/web is the most restrictive domain, with only three of its ten datasets openly available.
\emph{Datasets captured in an operational environment remain rare}: $37$ of $234$. Only backbone/honeypot has a majority of them ($9$ of $11$), followed by automotive ($12$ of $24$).
\emph{The number of datasets overstates the number of independent captures}: $30$ of the $234$ records are re-extractions, relabelings or augmentations of 14 earlier captures, so the catalog contains $204$ original datasets. Derivation is concentrated in the enterprise and IoT groups ($16$ and $8$ derivatives), and three captures alone, \texttt{CSE-CIC-IDS2018}, \texttt{CICIDS2017} and \texttt{UNSW-NB15}, account for $14$ of the $30$ derivatives.
Finally, \emph{the field also changes quickly}. Since $2023$ alone, $18$ of the $57$ enterprise datasets, $27$ of the $50$ \gls{iot} datasets and all \gls{uav} datasets have been released. A catalog compiled before $2023$ would therefore miss about a third of the enterprise datasets and more than half of the \gls{iot} ones.
\emph{Growth is not uniform}: the malware group has received no new release since $2020$, and recent malware traffic appears instead in other domains, such as \texttt{CIC-YNU-IoTMal2026} (IoT), \texttt{DEDALE} (enterprise) and \texttt{TQH-C2} (encrypted traffic), or in continuously updated collections such as \texttt{Malware-Traffic-Analysis.net}.

The following subsections discuss each domain through the columns of its table, emphasizing the properties that affect reuse and comparative evaluation rather than describing every dataset. 

\begin{table}[!htbp]
\centering\scriptsize\setlength{\tabcolsep}{2pt}
\providecommand{\acqR}{\CIRCLE}
\providecommand{\acqH}{\LEFTcircle}
\providecommand{\acqE}{\Circle}
\providecommand{\deriv}{\smash{\raisebox{0.32em}{\tikz[baseline=0pt]{\draw[line width=0.4pt,line cap=round] (0,0) -- (0,0.58em) (-0.16em,0.40em) -- (0.16em,0.40em);}}}}
\providecommand{\cmark}{\textcolor{green!55!black}{\ding{51}}}
\providecommand{\xmark}{\textcolor{red!55!black}{\ding{55}}}
\caption{IoT / IIoT / IoMT datasets in the \Obs.
}
\label{tab:rev-iot}
\begin{tabular}{llcccccccccccccccccc}
\toprule
\multirow{2}{*}{\textbf{Dataset}} & 
\multirow{2}{*}{\textbf{Year$\downarrow$}} &
\multicolumn{3}{c}{\textbf{Artifact}} & 
\multirow{2}{*}{\textbf{Access}}
&
\multirow{2}{*}{\textbf{Acq.}}
&
\multirow{2}{*}{\textbf{Lab.}}
& \multicolumn{2}{c}{\textbf{Classes}} & \multicolumn{10}{c}{\textbf{Attack families}}
\\
\cmidrule(lr){3-5}\cmidrule(lr){9-10}\cmidrule(lr){11-20}
 &  & \textbf{Raw} & \textbf{Proc.} & \textbf{Proc. Tool} & &  &  & \textbf{Ben.} & \textbf{Att.} & \textbf{B} & \textbf{C} & \textbf{D} & \textbf{E} & \textbf{L} & \textbf{M} & \textbf{R} & \textbf{V} & \textbf{W} & \textbf{O} \\
\midrule
\texttt{Kitsune}~\cite{mirsky2018kitsune} & 2018 & \cmark & \cmark & AfterImage & \faLockOpen & \acqH & B & \cmark & 9 & \xmark&\cmark&\cmark&\xmark&\xmark&\xmark&\cmark&\xmark&\xmark&\cmark \\
\rowcolor[HTML]{f0f0f0}
\texttt{N-BaIoT}~\cite{meidan2018nbaiot} & 2018 & \xmark & \cmark & AfterImage & \faLockOpen & \acqH & M & \cmark & 10 & \xmark&\cmark&\cmark&\xmark&\xmark&\xmark&\cmark&\xmark&\xmark&\xmark \\
\texttt{UNSW IoT Attack Traces}~\cite{hamza2019mud} & 2018 & \cmark & \cmark & custom & \faLockOpen & \acqH & W & \cmark & 8 & \xmark&\xmark&\cmark&\xmark&\xmark&\xmark&\xmark&\xmark&\xmark&\cmark \\
\rowcolor[HTML]{f0f0f0}
\texttt{BoT-IoT}~\cite{koroniotis2019} & 2018 & \cmark & \cmark & Argus & \faLockOpen & \acqE & M & \cmark & 4 & \xmark&\xmark&\cmark&\cmark&\xmark&\xmark&\cmark&\xmark&\xmark&\xmark \\
\texttt{UNSW IoT Traffic Traces}~\cite{sivanathan2019iot} & 2018 & \cmark & \cmark & custom & \faLockOpen & \acqH & -- & \cmark & 0 & \xmark&\xmark&\xmark&\xmark&\xmark&\xmark&\xmark&\xmark&\xmark&\xmark \\
\rowcolor[HTML]{f0f0f0}
\texttt{IoT Network Intrusion Dataset}~\cite{kang2019iotnid} & 2019 & \cmark & \xmark & -- & \faUser & \acqH & M & \cmark & 8 & \cmark&\cmark&\cmark&\xmark&\xmark&\xmark&\cmark&\xmark&\xmark&\cmark \\
\texttt{DAD}~\cite{dad2020} & 2020 & \cmark & \xmark & -- & \faLockOpen & \acqE & M & \cmark & 3 & \xmark&\xmark&\xmark&\xmark&\xmark&\xmark&\xmark&\xmark&\xmark&\cmark \\
\rowcolor[HTML]{f0f0f0}
\texttt{IoT-23}~\cite{garcia2020iot23} & 2020 & \cmark & \cmark & Zeek & \faLockOpen & \acqH & M & \cmark & 9 & \xmark&\cmark&\cmark&\xmark&\xmark&\cmark&\cmark&\xmark&\xmark&\xmark \\
\texttt{TON\_IoT}~\cite{moustafa2021toniot} & 2020 & \cmark & \cmark & Zeek & \faLockOpen & \acqH & M & \cmark & 9 & \cmark&\xmark&\cmark&\xmark&\xmark&\cmark&\cmark&\cmark&\cmark&\cmark \\
\rowcolor[HTML]{f0f0f0}
\texttt{ECU-IoHT}~\cite{ahmed2021ecuioht} & 2020 & \cmark & \cmark & tshark & \faLockOpen & \acqH & M & \cmark & 4 & \xmark&\xmark&\cmark&\xmark&\xmark&\xmark&\cmark&\xmark&\xmark&\cmark \\
\texttt{IoTID20}~\cite{ullah2020iotid20}\ \deriv{}[\cite{kang2019iotnid}] & 2020 & \xmark & \cmark & CFM & \faLockOpen & \acqH & M & \cmark & 8 & \cmark&\cmark&\cmark&\xmark&\xmark&\xmark&\cmark&\xmark&\xmark&\cmark \\
\rowcolor[HTML]{f0f0f0}
\texttt{MedBIoT}~\cite{guerra2020medbiot} & 2020 & \cmark & \cmark & AfterImage & \faLockOpen & \acqH & M & \cmark & 3 & \xmark&\cmark&\xmark&\xmark&\xmark&\xmark&\cmark&\xmark&\xmark&\xmark \\
\texttt{MQTT-IoT-IDS2020}~\cite{hindy2021mqttiotids} & 2020 & \cmark & \cmark & custom & \faUser & \acqE & M & \cmark & 4 & \cmark&\xmark&\xmark&\xmark&\xmark&\xmark&\cmark&\xmark&\xmark&\xmark \\
\rowcolor[HTML]{f0f0f0}
\texttt{MQTTset}~\cite{vaccari2020mqttset} & 2020 & \cmark & \cmark & tshark & \faLockOpen & \acqE & M & \cmark & 5 & \cmark&\xmark&\cmark&\xmark&\xmark&\xmark&\xmark&\xmark&\xmark&\cmark \\
\texttt{WUSTL-EHMS-2020}~\cite{hady2020wustlehms} & 2020 & \xmark & \cmark & Argus & \faLockOpen & \acqH & B & \cmark & 2 & \xmark&\xmark&\xmark&\xmark&\xmark&\xmark&\xmark&\xmark&\xmark&\cmark \\
\rowcolor[HTML]{f0f0f0}
\texttt{NF-BoT-IoT-v2}~\cite{sarhan2022standard}\ \deriv{}[\cite{koroniotis2019}] & 2021 & \xmark & \cmark & nProbe & \faLockOpen & \acqE & M & \cmark & 4 & \xmark&\xmark&\cmark&\cmark&\xmark&\xmark&\cmark&\xmark&\xmark&\xmark \\
\texttt{NF-ToN-IoT-v2}~\cite{sarhan2022standard}\ \deriv{}[\cite{moustafa2021toniot}] & 2021 & \xmark & \cmark & nProbe & \faLockOpen & \acqH & M & \cmark & 9 & \cmark&\xmark&\cmark&\xmark&\xmark&\cmark&\cmark&\cmark&\cmark&\cmark \\
\rowcolor[HTML]{f0f0f0}
\texttt{X-IIoTID}~\cite{alhawawreh2022xiiotid} & 2021 & \xmark & \cmark & custom & \faUser & \acqH & M & \cmark & 18 & \cmark&\cmark&\cmark&\cmark&\cmark&\cmark&\cmark&\cmark&\xmark&\cmark \\
\texttt{WUSTL-IIoT-2021}~\cite{wustliiot2021} & 2021 & \xmark & \cmark & Argus & \faLockOpen & \acqH & M & \cmark & 4 & \xmark&\xmark&\cmark&\xmark&\xmark&\cmark&\cmark&\cmark&\xmark&\xmark \\
\rowcolor[HTML]{f0f0f0}
\texttt{CICIoT2022}~\cite{dadkhah2022ciciot2022} & 2022 & \cmark & \cmark & CFM & \faUser & \acqH & M & \cmark & 2 & \cmark&\xmark&\cmark&\xmark&\xmark&\xmark&\xmark&\xmark&\xmark&\xmark \\
\texttt{Edge-IIoTset}~\cite{ferrag2022edgeiiotset} & 2022 & \cmark & \cmark & Zeek, tshark & \faLockOpen & \acqH & M & \cmark & 14 & \xmark&\xmark&\cmark&\xmark&\xmark&\cmark&\cmark&\cmark&\cmark&\cmark \\
\rowcolor[HTML]{f0f0f0}
\texttt{LATAM-DDoS-IoT}~\cite{almaraz2022latam} & 2022 & \cmark & \cmark & Argus & \faUser & \acqH & M & -- & -- & \xmark&\xmark&\cmark&\xmark&\xmark&\xmark&\xmark&\xmark&\xmark&\xmark \\
\texttt{RT-IoT2022}~\cite{sharmila2023rtiot} & 2022 & \xmark & \cmark & Zeek & \faLockOpen & \acqH & M & \cmark & 9 & \cmark&\xmark&\cmark&\xmark&\xmark&\xmark&\cmark&\xmark&\xmark&\cmark \\
\rowcolor[HTML]{f0f0f0}
\texttt{ACI-IoT-2023}~\cite{bastian2023aciiot} & 2023 & \cmark & \cmark & CFM & \faUser & \acqH & M & \cmark & 11 & \cmark&\xmark&\cmark&\xmark&\xmark&\xmark&\cmark&\xmark&\xmark&\cmark \\
\texttt{DoS/DDoS-MQTT-IoT}~\cite{alatram2023dosmqtt} & 2023 & \cmark & \cmark & tshark & \faLockOpen & \acqH & M & \cmark & 10 & \xmark&\xmark&\cmark&\xmark&\xmark&\xmark&\xmark&\xmark&\xmark&\xmark \\
\rowcolor[HTML]{f0f0f0}
\texttt{CICIoT2023}~\cite{neto2023ciciot} & 2023 & \cmark & \cmark & custom & \faUser & \acqH & M & \cmark & 33 & \cmark&\cmark&\cmark&\xmark&\xmark&\xmark&\cmark&\xmark&\cmark&\cmark \\
\texttt{IoMT-TrafficData}~\cite{areia2024iomt} & 2023 & \cmark & \cmark & Zeek, tshark & \faLockOpen & \acqH & M & \cmark & 8 / 2 & \xmark&\xmark&\cmark&\xmark&\xmark&\xmark&\cmark&\xmark&\xmark&\cmark \\
\rowcolor[HTML]{f0f0f0}
\texttt{ROSIDS23}~\cite{rosids23} & 2023 & \cmark & \cmark & CFM & \faLockOpen & \acqH & M & -- & -- & \xmark&\xmark&\cmark&\xmark&\xmark&\xmark&\xmark&\xmark&\xmark&\cmark \\
\texttt{CICAPT-IIoT2024}~\cite{ghiasvand2024cicapt} & 2024 & \cmark & \cmark & SPADE & \faUser & \acqH & M & \cmark & 8\textsuperscript{*} & \xmark&\cmark&\xmark&\cmark&\cmark&\xmark&\cmark&\xmark&\xmark&\cmark \\
\rowcolor[HTML]{f0f0f0}
\texttt{CICEVSE2024}~\cite{buedi2024cicevse} & 2024 & \cmark & \cmark & NFStream & \faUser & \acqH & M & -- & -- & \xmark&\xmark&\cmark&\xmark&\xmark&\cmark&\cmark&\xmark&\xmark&\cmark \\
\texttt{IoT-DH}~\cite{iotdh2024} & 2024 & \cmark & \cmark & Argus & \faLockOpen & \acqR & W & -- & -- & \xmark&\xmark&\cmark&\xmark&\xmark&\xmark&\xmark&\xmark&\xmark&\xmark \\
\rowcolor[HTML]{f0f0f0}
\texttt{CICIoMT2024}~\cite{dadkhah2024ciciomt} & 2024 & \cmark & \cmark & CFM & \faUser & \acqH & M & \cmark & 18 & \xmark&\xmark&\cmark&\xmark&\xmark&\xmark&\cmark&\xmark&\xmark&\cmark \\
\texttt{ROSPaCe}~\cite{rospace2024} & 2024 & \xmark & \cmark & tshark, custom & \faLockOpen & \acqH & M & \cmark & 6 & \xmark&\xmark&\cmark&\xmark&\xmark&\xmark&\cmark&\xmark&\xmark&\xmark \\
\rowcolor[HTML]{f0f0f0}
\texttt{CIC IoT-DIAD 2024}~\cite{ciciotdiad2024}\ \deriv{}[\cite{neto2023ciciot}] & 2024 & \xmark & \cmark & CFM & \faUser & \acqH & M & \cmark & 33 & \cmark&\cmark&\cmark&\xmark&\xmark&\xmark&\cmark&\xmark&\cmark&\cmark \\
\texttt{TabularIoTAttack-2024}~\cite{tabulariot2024}\ \deriv{}[\cite{neto2023ciciot}] & 2024 & \xmark & \cmark & CFM & \faUser & \acqH & M & -- & -- & \xmark&\cmark&\cmark&\cmark&\xmark&\cmark&\xmark&\cmark&\xmark&\cmark \\
\rowcolor[HTML]{f0f0f0}
\texttt{BCCC-IoT-IDS-Zwave-2025}~\cite{bcccZwave2025} & 2025 & \cmark & \cmark & NFStream & \faEnvelope & \acqH & M & -- & 81 / 24 & \multicolumn{10}{c}{--} \\
\texttt{BCCC-IoT-MQTT-IDS-2025}~\cite{bcccMqtt2025} & 2025 & \cmark & \cmark & MFL & \faEnvelope & \acqH & M & -- & -- & \cmark&\xmark&\cmark&\xmark&\xmark&\xmark&\cmark&\xmark&\xmark&\cmark \\
\rowcolor[HTML]{f0f0f0}
\texttt{NIMS IoT 2025-1}~\cite{nims2025a} & 2025 & \cmark & \xmark & -- & \faUser & \acqH & W & -- & 4 & \xmark&\xmark&\cmark&\xmark&\xmark&\xmark&\cmark&\xmark&\xmark&\xmark \\
\texttt{DataSense}~\cite{datasense2025} & 2025 & \cmark & \cmark & CFM & \faUser & \acqH & M & \cmark & 49 & \cmark&\xmark&\cmark&\xmark&\xmark&\cmark&\cmark&\xmark&\cmark&\cmark \\
\rowcolor[HTML]{f0f0f0}
\texttt{Gotham}~\cite{gotham2025} & 2025 & \cmark & \cmark & tshark & \faLockOpen & \acqE & M & \cmark & 9 & \cmark&\cmark&\cmark&\xmark&\xmark&\xmark&\cmark&\cmark&\xmark&\cmark \\
\texttt{MQTTEEB-D}~\cite{mqtteebd2025} & 2025 & \xmark & \cmark & custom & \faLockOpen & \acqR & M & \cmark & 5 & \cmark&\xmark&\cmark&\xmark&\xmark&\xmark&\xmark&\cmark&\xmark&\xmark \\
\rowcolor[HTML]{f0f0f0}
\texttt{MQTT\_UAD}~\cite{mqttuad2025} & 2025 & \xmark & \cmark & tshark, custom & \faLockOpen & -- & B & \cmark & 3 & \xmark&\xmark&\cmark&\xmark&\xmark&\xmark&\xmark&\xmark&\xmark&\cmark \\
\texttt{NF-ToN-IoT-v3}~\cite{luay2025nfv3}\ \deriv{}[\cite{moustafa2021toniot}] & 2025 & \xmark & \cmark & nProbe & \faLockOpen & \acqH & M & \cmark & 9 & \cmark&\xmark&\cmark&\xmark&\xmark&\cmark&\cmark&\cmark&\cmark&\cmark \\
\rowcolor[HTML]{f0f0f0}
\texttt{NF-BoT-IoT-v3}~\cite{luay2025nfv3}\ \deriv{}[\cite{koroniotis2019}] & 2025 & \xmark & \cmark & nProbe & \faLockOpen & \acqE & M & \cmark & 4 & \xmark&\xmark&\cmark&\cmark&\xmark&\xmark&\cmark&\xmark&\xmark&\xmark \\
\texttt{UNSW-IoTraffic}~\cite{unswiotraffic2025} & 2025 & \cmark & \cmark & custom & \faLockOpen & \acqH & -- & \cmark & 0 & \xmark&\xmark&\xmark&\xmark&\xmark&\xmark&\xmark&\xmark&\xmark&\xmark \\
\rowcolor[HTML]{f0f0f0}
\texttt{NIMS IoT 2025-2}~\cite{nims2025b} & 2026 & \cmark & \xmark & -- & \faUser & \acqH & W & -- & 9 & \cmark&\xmark&\cmark&\xmark&\xmark&\xmark&\cmark&\xmark&\xmark&\cmark \\
\texttt{CyberFlowIoT-GICAP}~\cite{cyberflowiot2026} & 2026 & \cmark & \cmark & NFStream & \faLockOpen & \acqH & B & \cmark & 1 & \multicolumn{10}{c}{--} \\
\rowcolor[HTML]{f0f0f0}
\texttt{IoT-WinRL}~\cite{iotwinrl2026}\ \deriv{}[\cite{neto2023ciciot}] & 2026 & \xmark & \cmark & custom & \faLockOpen & \acqH & M & \cmark & 2 & \cmark&\cmark&\cmark&\xmark&\xmark&\xmark&\cmark&\xmark&\cmark&\cmark \\
\texttt{IoT-SFC v2}~\cite{iotsfc2026} & 2026 & \xmark & \cmark & Zeek & \faLockOpen & \acqH & M & \cmark & 14 & \cmark&\cmark&\cmark&\xmark&\xmark&\xmark&\cmark&\xmark&\xmark&\cmark \\
\rowcolor[HTML]{f0f0f0}
\texttt{CIC-YNU-IoTMal2026}~\cite{cicynu2026} & 2026 & \xmark & \cmark & tshark & \faUser & \acqE & M & \cmark & 7 & \xmark&\cmark&\xmark&\xmark&\xmark&\cmark&\xmark&\xmark&\xmark&\xmark \\
\bottomrule
\end{tabular}

\smallskip
\parbox{\linewidth}{\scriptsize\raggedright
\textbf{N.B:}
\textbf{Artifact:} Raw Network traces (Raw), Processed Records (Proc.);
\textbf{Processing Tool (Proc. Tool):} tool that produced the processed records; CICFlowMeter (CFM), including its modified/fixed versions, MQTTFlowLyzer (MFL), Zeek includes its former name Bro, custom = authors' own extractor, including scripts built on parsing libraries such as dpkt or PyShark;
\textbf{Access:} \faLockOpen~open; \faUser~registration; \faEnvelope~on request;
\textbf{Acquisition (Acq.):} \acqR~real devices; \acqH~real and simulated devices; \acqE~emulated/simulated; \acqH = merge of sources;
\textbf{Labeling (Lab.):} Binary (B), Multiclass (M), Weak/heuristic (W), None (--);
\textbf{Classes:} Benign (Ben.), Attack (Att.: number of attack types; \textsuperscript{*}~number of MITRE ATT\&CK tactics or techniques used as labels, mapped to the attack families);
\textbf{Attack Families:}
Brute force (B), Botnet/C2 (C), (D)DoS (D), Exfiltration (E), Infiltration/lateral movement (L), Malware/backdoor/worm (M), Reconnaissance/scan (R), Exploitation (V), Web (W), Other, incl.\ spoofing/MitM and protocol-specific attacks (O);
\deriv{}~derivative of an earlier capture, followed by the reference of its source;
\cmark~present, \xmark~not present, --~not specified or not documented.
}
\end{table}
\subsection{General, Industrial and Medical Internet of Things}\label{sec:rev-iot}

Table~\ref{tab:rev-iot} lists the $50$ IoT, IIoT and IoMT datasets in the \Obs{}. \emph{The group is younger and is growing faster}:
$27$ datasets were released from $2023$ onward. Its subdomains (industrial, medical, MQTT, Z-Wave) are increasingly covered by dedicated releases.
In this domain, \emph{derivatives are few}, and half of them are NetFlow re-extractions of \texttt{BoT-IoT} and \texttt{TON\_IoT}. 
However, \emph{the tools used to extract processed records from the raw traces vary widely} (\emph{Proc. Tool} column): about ten different extractors are used, with \texttt{tshark} and CICFlowMeter the most frequent, so processed records from different datasets rarely share a feature space.
As for acquisition (\textit{Acq.} column), \emph{most datasets come from testbeds that combine physical devices with simulated components}. This is more realistic than pure emulation, but only two datasets observe real deployments.
\emph{Labels are mostly multiclass} (\textit{Lab.} column), but the benign class is undocumented for one dataset in five, and the number of attacks ranges from $0$ to $81$, again because some releases count fine-grained, device-specific attack variants (\textit{Classes} columns).
\emph{The attack families reflect the IoT threat model}: (D)DoS appears in most datasets, followed by reconnaissance and protocol-specific attacks such as spoofing or MitM (O), while lateral movement and ATT\&CK-labeled campaigns are almost absent (\textit{Attack families} columns). 
Overall, most datasets in the group release raw traffic and few are derivatives, but its heterogeneity in tools, labels and devices makes cross-dataset evaluation difficult.

\providecommand{\deriv}{\smash{\raisebox{0.32em}{\tikz[baseline=0pt]{\draw[line width=0.4pt,line cap=round] (0,0) -- (0,0.58em) (-0.16em,0.40em) -- (0.16em,0.40em);}}}}
\begin{table}[!htbp]
\centering\scriptsize\setlength{\tabcolsep}{2pt}
\providecommand{\acqR}{\CIRCLE}
\providecommand{\acqH}{\LEFTcircle}
\providecommand{\acqE}{\Circle}
\providecommand{\cmark}{\textcolor{green!55!black}{\ding{51}}}
\providecommand{\xmark}{\textcolor{red!55!black}{\ding{55}}}
\caption{Malware / botnet / C2 datasets in the \Obs.}
\label{tab:rev-malware}
\begin{tabular}{llcccccccccccccccccc}
\toprule
\multirow{2}{*}{\textbf{Dataset}} & 
\multirow{2}{*}{\textbf{Year$\downarrow$}} &
\multicolumn{3}{c}{\textbf{Artifact}} & 
\multirow{2}{*}{\textbf{Access}}
&
\multirow{2}{*}{\textbf{Acq.}}
&
\multirow{2}{*}{\textbf{Lab.}}
& \multicolumn{2}{c}{\textbf{Classes}} & \multicolumn{10}{c}{\textbf{Attack families}}
\\
\cmidrule(lr){3-5}\cmidrule(lr){9-10}\cmidrule(lr){11-20}
 &  & \textbf{Raw} & \textbf{Proc.} & \textbf{Proc. Tool} & &  &  & \textbf{Ben.} & \textbf{Att.} & \textbf{B} & \textbf{C} & \textbf{D} & \textbf{E} & \textbf{L} & \textbf{M} & \textbf{R} & \textbf{V} & \textbf{W} & \textbf{O} \\
\midrule
\texttt{ISOT Botnet}~\cite{saad2011isot} & 2010 & \cmark & \xmark & -- & \faLockOpen & \acqH & B & \cmark & 2 & \xmark&\cmark&\xmark&\xmark&\xmark&\xmark&\xmark&\xmark&\xmark&\cmark \\
\rowcolor[HTML]{f0f0f0}
\texttt{CTU-13}~\cite{garcia2014ctu13} & 2011 & \cmark & \cmark & Argus & \faLockOpen & \acqR & M & \cmark & 1 & \xmark&\cmark&\cmark&\xmark&\xmark&\xmark&\xmark&\xmark&\xmark&\cmark \\
\texttt{Stratosphere MCF}~\cite{stratosphereMCF} & 2011--17 & \cmark & \cmark & Argus, Zeek & \faLockOpen & \acqH & W & \cmark & 1 & \xmark&\cmark&\xmark&\xmark&\xmark&\cmark&\xmark&\xmark&\xmark&\xmark \\
\rowcolor[HTML]{f0f0f0}
\texttt{VAST 2012 MC2}~\cite{vast2012} & 2012 & \xmark & \cmark & ASA, Snort & \faLockOpen & \acqE & W & -- & -- & \xmark&\cmark&\xmark&\xmark&\xmark&\xmark&\xmark&\xmark&\xmark&\xmark \\
\texttt{Malware-Traffic-Analysis.net}~\cite{mtanet} & 2013--26 & \cmark & \xmark & -- & \faLockOpen & \acqH & -- & -- & -- & \xmark&\cmark&\xmark&\xmark&\xmark&\cmark&\xmark&\xmark&\xmark&\xmark \\
\rowcolor[HTML]{f0f0f0}
\texttt{CTU-mixed-Capture}~\cite{ctumixed} & 2015--16 & \cmark & \cmark & Argus, Zeek & \faLockOpen & \acqH & W & \cmark & 1 & \xmark&\cmark&\xmark&\xmark&\xmark&\cmark&\xmark&\xmark&\xmark&\xmark \\
\texttt{CIC-AAGM2017}~\cite{lashkari2017aagm} & 2017 & \cmark & \cmark & CFM & \faUser & \acqH & M & \cmark & 10 & \xmark&\xmark&\xmark&\xmark&\xmark&\cmark&\xmark&\xmark&\xmark&\xmark \\
\rowcolor[HTML]{f0f0f0}
\texttt{CIC-AndMal2017}~\cite{lashkari2018andmal} & 2017 & \cmark & \cmark & CFM & \faUser & \acqH & M & \cmark & 4 & \xmark&\xmark&\xmark&\xmark&\xmark&\cmark&\xmark&\xmark&\xmark&\xmark \\
\texttt{ISOT HTTP Botnet}~\cite{alenazi2017isothttp} & 2017 & \cmark & \xmark & -- & \faLockOpen & \acqH & M & \cmark & 9 & \xmark&\cmark&\xmark&\xmark&\xmark&\xmark&\xmark&\xmark&\xmark&\xmark \\
\rowcolor[HTML]{f0f0f0}
\texttt{CICInvesAndMal2019}~\cite{taheri2019invesandmal}\ \deriv{}[\cite{lashkari2018andmal}] & 2019 & \xmark & \cmark & CFM & \faUser & \acqH & M & \cmark & 4 & \xmark&\xmark&\xmark&\xmark&\xmark&\cmark&\xmark&\xmark&\xmark&\xmark \\
\texttt{MTA-KDD'19}~\cite{letteri2020mtakdd}\ \deriv{}[\cite{mtanet}] & 2019 & \xmark & \cmark & custom & \faLockOpen & \acqH & B & \cmark & 1 & \xmark&\cmark&\xmark&\xmark&\xmark&\cmark&\xmark&\xmark&\xmark&\xmark \\
\rowcolor[HTML]{f0f0f0}
\texttt{CICMalDroid2020}~\cite{mahdavifar2020maldroid} & 2020 & \xmark & \cmark & CopperDroid & \faUser & \acqE & M & \cmark & 4 & \xmark&\xmark&\xmark&\xmark&\xmark&\cmark&\xmark&\xmark&\xmark&\xmark \\
\texttt{Android Mischief}~\cite{androidmischief} & 2020 & \cmark & \cmark & Zeek & \faLockOpen & \acqH & -- & \xmark & 8 & \xmark&\cmark&\xmark&\xmark&\xmark&\cmark&\xmark&\xmark&\xmark&\xmark \\
\rowcolor[HTML]{f0f0f0}
\texttt{NetML}~\cite{barut2020netml} & 2020 & \cmark & \cmark & Intel FL & \faLockOpen & \acqH & M & \cmark & 20 & \xmark&\cmark&\xmark&\xmark&\xmark&\cmark&\xmark&\xmark&\xmark&\xmark \\
\texttt{CCCS-CIC-AndMal-2020}~\cite{rahali2020andmal2020} & 2020 & \xmark & \cmark & custom & \faUser & \acqE & M & \cmark & 14 & \xmark&\xmark&\xmark&\xmark&\xmark&\cmark&\xmark&\xmark&\xmark&\xmark \\
\bottomrule
\end{tabular}
\smallskip
\parbox{\linewidth}{\scriptsize\raggedright
\textbf{N.B:}
\textbf{Artifact:} Raw Network traces (Raw), Processed Records (Proc.);
\textbf{Processing Tool (Proc. Tool):} tool that produced the processed records; CICFlowMeter (CFM), including its modified/fixed versions, Intel flow library (Intel FL), Cisco ASA firewall logs (ASA), Zeek includes its former name Bro, custom = authors' own extractor;
\textbf{Access:} \faLockOpen~open; \faUser~registration;
\textbf{Acquisition (Acq.):} \acqR~operational network; \acqH~real and simulated devices in a controlled environment; \acqE~emulated/simulated;
\textbf{Labeling (Lab.):} Binary (B), Multiclass (M), Weak/heuristic (W), None (--);
\textbf{Classes:} Benign (Ben.), Attack (Att.: number of malicious classes or malware families);
\textbf{Attack Families:}
Brute force (B), Botnet/C2 (C), (D)DoS (D), Exfiltration (E), Infiltration/lateral movement (L), Malware, incl.\ backdoor, worm, RAT, adware, ransomware (M), Reconnaissance/scan (R), Exploitation (V), Web (W), Other, e.g.\ spam, click fraud (O);
\deriv{}~derivative of an earlier capture, followed by the reference of its source;
\cmark~present, \xmark~not present, --~not specified or not documented.
}
\end{table}

\subsection{Malware, botnet and command-and-control traffic}\label{sec:rev-malware}

Table~\ref{tab:rev-malware} lists the 15 malware, botnet and C2 datasets in the \Obs{}. Unlike the IoT groups, this one has received no new release after $2020$, although malware traffic still appears in datasets assigned to other domains (Secs.~\ref{sec:rev-iot}, \ref{sec:rev-enterprise}, and~\ref{sec:rev-encrypted}), and the only recent announcement, \texttt{MalwareTraffic}, has not yet been published (App.~\ref{app:excluded}). 
Its most current sources are the continuously extended collections, \texttt{Stratosphere MCF} and \texttt{Malware-Traffic-Analysis.net}, rather than one-off releases.
\emph{Raw traces are widely available}: ten of the $15$ datasets release packet traces (\emph{Artifact} column), which is essential for malware traffic, whose value lies in C2 timing and payload behavior that flow features only partially preserve.
\emph{Malware is mostly executed in controlled environments}: on real and simulated devices or in emulators (\emph{Acq.} column); only \texttt{CTU-13} records infections in an operational network.
\emph{Class counts are not comparable across datasets}: they range from a single malicious class to $20$ malware classes (\emph{Att.} column).
\emph{The focus on attacks is narrow}: botnet/C2 and malware dominate, with only a few datasets covering (D)DoS or other attacks such as spam and click fraud (\emph{Attack families} columns).
\emph{Android traffic is a distinct subgroup}, mostly from CIC, whose records are in some cases produced by sandboxes or app execution rather than by network capture.
Overall, the group offers rich traces of real malware, but it is aging and its labels are hard to compare. Models trained on it may learn the behavior of malware families that are no longer active~\cite{jordaney2017transcend}.


\begin{table}[!htbp]
\centering\scriptsize\setlength{\tabcolsep}{2pt}
\providecommand{\acqR}{\CIRCLE}
\providecommand{\acqH}{\LEFTcircle}
\providecommand{\acqE}{\Circle}
\providecommand{\deriv}{\smash{\raisebox{0.32em}{\tikz[baseline=0pt]{\draw[line width=0.4pt,line cap=round] (0,0) -- (0,0.58em) (-0.16em,0.40em) -- (0.16em,0.40em);}}}}
\providecommand{\cmark}{\textcolor{green!55!black}{\ding{51}}}
\providecommand{\xmark}{\textcolor{red!55!black}{\ding{55}}}
\caption{ICS/OT datasets in the \Obs.}
\label{tab:rev-ics}
\begin{tabular}{llcccccccccccccccccc}
\toprule
\multirow{2}{*}{\textbf{Dataset}} & 
\multirow{2}{*}{\textbf{Year$\downarrow$}} &
\multicolumn{3}{c}{\textbf{Artifact}} & 
\multirow{2}{*}{\textbf{Access}}
&
\multirow{2}{*}{\textbf{Acq.}}
&
\multirow{2}{*}{\textbf{Lab.}}
& \multicolumn{2}{c}{\textbf{Classes}} & \multicolumn{10}{c}{\textbf{Attack families}}
\\
\cmidrule(lr){3-5}\cmidrule(lr){9-10}\cmidrule(lr){11-20}
 &  & \textbf{Raw} & \textbf{Proc.} & \textbf{Proc. Tool} & &  &  & \textbf{Ben.} & \textbf{Att.} & \textbf{B} & \textbf{C} & \textbf{D} & \textbf{E} & \textbf{L} & \textbf{M} & \textbf{R} & \textbf{V} & \textbf{W} & \textbf{O} \\
\midrule
\texttt{MSU Gas Pipeline \& Water Tank}~\cite{morris2014ics} & 2014 & \xmark & \cmark & custom & \faLockOpen & \acqH & M & \cmark & 7 & \xmark&\xmark&\cmark&\xmark&\xmark&\xmark&\cmark&\cmark&\xmark&\xmark \\
\rowcolor[HTML]{f0f0f0}
\texttt{MSU/ORNL Power System}~\cite{pan2015power} & 2014 & \xmark & \cmark & OpenPDC & \faLockOpen & \acqH & M & \cmark\, & 28 & \xmark&\xmark&\xmark&\xmark&\xmark&\xmark&\xmark&\cmark&\xmark&\cmark \\
\texttt{4SICS}~\cite{fourSICS} & 2015 & \cmark & \xmark & -- & \faLockOpen & \acqH & -- & -- & -- & \multicolumn{10}{c}{--} \\
\rowcolor[HTML]{f0f0f0}
\texttt{SWaT}~\cite{goh2017swat} & 2015 & \cmark & \cmark & custom & \faEnvelope & \acqH & B & \cmark & 1 & \xmark&\xmark&\xmark&\xmark&\xmark&\xmark&\xmark&\cmark&\xmark&\cmark \\
\texttt{Lemay Modbus}~\cite{lemay2016modbus} & 2016 & \cmark & \cmark & custom & \faLockOpen & \acqE & B & \cmark & 1 & \xmark&\cmark&\xmark&\xmark&\xmark&\xmark&\cmark&\cmark&\xmark&\cmark \\
\rowcolor[HTML]{f0f0f0}
\texttt{WADI}~\cite{ahmed2017wadi} & 2017 & \cmark & \cmark & custom & \faEnvelope & \acqH & B & \cmark & 1 & \xmark&\xmark&\xmark&\xmark&\xmark&\xmark&\xmark&\cmark&\xmark&\cmark \\
\texttt{2017QUT\_S7comm}~\cite{rodofile2017qut} & 2017 & \cmark & \cmark & tshark & \faLockOpen & \acqH & M & \cmark & 13 & \xmark&\xmark&\cmark&\xmark&\xmark&\xmark&\xmark&\cmark&\xmark&\xmark \\
\rowcolor[HTML]{f0f0f0}
\texttt{2017QUT\_DNP3}~\cite{rodofile2017qut} & 2017 & \cmark & \cmark & tshark & \faLockOpen & \acqE & M & \cmark & 5 & \xmark&\xmark&\cmark&\xmark&\xmark&\xmark&\xmark&\cmark&\xmark&\cmark \\
\texttt{EPIC}~\cite{epic2018} & 2018 & \cmark & \cmark & custom & \faEnvelope & \acqH & -- & \cmark & 0 & \xmark&\xmark&\xmark&\xmark&\xmark&\xmark&\xmark&\xmark&\xmark&\xmark \\
\rowcolor[HTML]{f0f0f0}
\texttt{WUSTL-IIoT-2018}~\cite{teixeira2018wustl} & 2018 & \xmark & \cmark & Argus & \faLockOpen & \acqH & B & \cmark & 1 & \xmark&\xmark&\xmark&\xmark&\xmark&\xmark&\cmark&\cmark&\xmark&\xmark \\
\texttt{Electra}~\cite{perales2019electra} & 2019 & \xmark & \cmark & Scapy & \faEnvelope & \acqR & M & \cmark & 6 & \xmark&\xmark&\xmark&\xmark&\xmark&\xmark&\cmark&\cmark&\xmark&\cmark \\
\rowcolor[HTML]{f0f0f0}
\texttt{M2M using OPC UA}~\cite{pinto2020opcua} & 2020 & \xmark & \cmark & PyShark & \faUser & \acqH & M & \cmark & 3 & \xmark&\xmark&\cmark&\xmark&\xmark&\xmark&\xmark&\xmark&\xmark&\cmark \\
\texttt{HIL WDT}~\cite{faramondi2021wdt} & 2021 & \xmark & \cmark & Wireshark & \faUser & \acqH & M & \cmark & 4 & \xmark&\xmark&\cmark&\xmark&\xmark&\xmark&\cmark&\xmark&\xmark&\cmark \\
\rowcolor[HTML]{f0f0f0}
\texttt{DNP3 Intrusion Detection}~\cite{dnp3ids2022} & 2022 & \cmark & \cmark & CFM, custom & \faLockOpen & \acqH & M & \cmark & 9 & \xmark&\xmark&\cmark&\xmark&\xmark&\xmark&\xmark&\cmark&\xmark&\cmark \\
\texttt{Hybrid IoT/OT Honeypots}~\cite{dtuhoneypots2022} & 2022 & \cmark & \xmark & -- & \faEnvelope & \acqR & W & \xmark & 2 & \xmark&\xmark&\xmark&\xmark&\xmark&\xmark&\cmark&\xmark&\xmark&\cmark \\
\rowcolor[HTML]{f0f0f0}
\texttt{IEC 104 Intrusion Detection}~\cite{iec104ids2022} & 2022 & \cmark & \cmark & CFM, custom & \faLockOpen & \acqH & M & \cmark & 14 & \xmark&\xmark&\cmark&\xmark&\xmark&\xmark&\xmark&\cmark&\xmark&\cmark \\
\texttt{CICModbus2023}~\cite{cicmodbus2023} & 2023 & \cmark & \xmark & -- & \faUser & \acqE & M & \cmark & 9 & \cmark&\xmark&\cmark&\xmark&\xmark&\xmark&\cmark&\cmark&\xmark&\cmark \\
\rowcolor[HTML]{f0f0f0}
\texttt{ICS-Flow}~\cite{dehlaghi2023icsflow} & 2023 & \cmark & \cmark & ICSFG & \faLockOpen & \acqE & M & \cmark & 4 & \xmark&\xmark&\cmark&\xmark&\xmark&\xmark&\cmark&\xmark&\xmark&\cmark \\
\texttt{MARSIM}~\cite{spravil2023detecting} & 2023 & \cmark & \xmark & -- & \faLockOpen & \acqE & B & \cmark & 1 & \xmark&\xmark&\xmark&\xmark&\xmark&\xmark&\xmark&\xmark&\xmark&\cmark \\
\rowcolor[HTML]{f0f0f0}
\texttt{Westermo}~\cite{westermo2023} & 2023 & \cmark & \cmark & ICSFG & \faLockOpen & \acqH & M & \cmark & 5 & \cmark&\xmark&\xmark&\xmark&\xmark&\xmark&\cmark&\xmark&\xmark&\cmark \\
\texttt{Cyber4OT}~\cite{cyber4ot2025} & 2025 & \cmark & \xmark & -- & \faLockOpen & \acqH & -- & -- & -- & \xmark&\xmark&\xmark&\xmark&\xmark&\xmark&\cmark&\xmark&\xmark&\cmark \\
\rowcolor[HTML]{f0f0f0}
\texttt{ICSCASD-MPLC}~\cite{icscasd2025} & 2025 & \xmark & \cmark & Wireshark, custom & \faLockOpen & \acqH & M & \cmark & 7 & \xmark&\xmark&\cmark&\xmark&\xmark&\xmark&\cmark&\cmark&\xmark&\cmark \\
\bottomrule
\end{tabular}

\smallskip
\parbox{\linewidth}{\scriptsize\raggedright
\textbf{N.B:}
\textbf{Artifact:} Raw Network traces (Raw), Processed Records (Proc.);
\textbf{Proc. Tool (Processing Tool):} tool that produced the processed records; CICFlowMeter (CFM), including its modified/fixed versions, ICSFlowGenerator (ICSFG) (Zeek includes its former name Bro, custom = authors' own extractor, (mod.) = modified version);
\textbf{Access:} \faLockOpen~open; \faUser~registration; \faEnvelope~ on request;
\textbf{Acquisition (Acq.):} \acqR~real network or plant (incl.\ Internet-exposed honeypots); \acqH~physical testbed or hardware-in-the-loop; \acqE~emulated/simulated;
\textbf{Labeling (Lab.):} Binary (B), Multiclass (M); Weak/heuristic (W); None (--);
\textbf{Classes:} Benign (Ben.; in brackets: number of benign/no-event subclasses), Attack (Att.);
\textbf{Attack Families:}
Brute force (B), Botnet/C2/covert channel (C), (D)DoS (D), Exfiltration (E), Infiltration/lateral movement (L), Malware/backdoor/worm (M), Reconnaissance/scan (R), Exploitation/injection, incl.\ false data injection (V), Web (W), Other, incl.\ spoofing/MitM, replay, misconfiguration, physical faults (O);
\deriv{}~derivative of an earlier capture;
\cmark~present, \xmark~not present, --~not specified or not documented.
}
\end{table}

\subsection{Industrial Control Systems and Operational Technology}\label{sec:rev-ics}

Table~\ref{tab:rev-ics} lists the $22$ ICS and OT datasets in the \Obs{}. 
\emph{The group keeps growing and has no derivatives}: nine of its $22$ datasets were released since $2022$, and each is an independent capture, usually built around a single industrial protocol.
These datasets must be interpreted in relation to the processes and protocols that generate their traffic: Modbus/TCP, DNP3, IEC 60870-5-104, S7 and OPC UA messages encode control operations and timing relationships that many generic IP-flow feature sets do not preserve.
\emph{Raw traces are available for most datasets}: $15$ of the $22$ release them, while seven distribute only processed records (\emph{Artifact} column).
\textit{Access} is a real obstacle for some of the best-known testbeds: \texttt{SWaT}, \texttt{WADI} and \texttt{EPIC} are released only on request, as are \texttt{Electra} and a honeypot collection, which makes it hard to reproduce studies that use them.
\emph{Labeling is mostly multiclass but tied to the process}: $13$ of the $22$ datasets are multiclass, with up to $28$ scenarios, while five, including the widely used \texttt{SWaT} and \texttt{WADI}, use binary labels with a single attack class (\emph{Lab.} and \emph{Classes} columns). In all cases label semantics depend on the interaction between network messages and process state, since an attack may manifest itself through a process transition, an abnormal command sequence or a deviation from periodic communication patterns~\citep{dobler2025}.
\emph{The attack families reflect the ICS threat model}: spoofing, replay and MitM (O, $17$ datasets), injection (V, $13$), reconnaissance ($11$) and (D)DoS ($10$) dominate, while malware, exfiltration, lateral movement and web attacks never appear (\emph{Attack families} columns).
Overall, the group offers independent and protocol-specific data, but it is fragmented across protocols, often hard to access, and does not cover multi-stage attacks.

\begin{table}[!htbp]
\centering\scriptsize\setlength{\tabcolsep}{2pt}
\providecommand{\acqR}{\CIRCLE}
\providecommand{\acqH}{\LEFTcircle}
\providecommand{\acqE}{\Circle}
\providecommand{\deriv}{\smash{\raisebox{0.32em}{\tikz[baseline=0pt]{\draw[line width=0.4pt,line cap=round] (0,0) -- (0,0.58em) (-0.16em,0.40em) -- (0.16em,0.40em);}}}}
\providecommand{\cmark}{\textcolor{green!55!black}{\ding{51}}}
\providecommand{\xmark}{\textcolor{red!55!black}{\ding{55}}}
\caption{Automotive CAN/Ethernet datasets in the \Obs.}
\label{tab:rev-automotive}
\begin{tabular}{llccccccccccccc}
\toprule
\multirow{2}{*}{\textbf{Dataset}} & 
\multirow{2}{*}{\textbf{Year$\downarrow$}} &
\multicolumn{3}{c}{\textbf{Artifact}} & 
\multirow{2}{*}{\textbf{Access}}
&
\multirow{2}{*}{\textbf{Acq.}}
&
\multirow{2}{*}{\textbf{Lab.}}
& \multicolumn{2}{c}{\textbf{Classes}} & \multicolumn{5}{c}{\textbf{Attack families}}
\\
\cmidrule(lr){3-5}\cmidrule(lr){9-10}\cmidrule(lr){11-15}
 &  & \textbf{Raw} & \textbf{Proc.} & \textbf{Proc. Tool} & &  &  & \textbf{Ben.} & \textbf{Att.} & \textbf{D} & \textbf{F} & \textbf{S} & \textbf{P} & \textbf{O} \\
\midrule
\texttt{OTIDS}~\cite{lee2017otids} & 2017 & \cmark & \xmark & -- & \faLockOpen & \acqR & W & \cmark & 3 & \cmark&\cmark&\cmark&\xmark&\xmark \\
\rowcolor[HTML]{f0f0f0}
\texttt{HCRL Survival Analysis}~\cite{han2018survival} & 2018 & \cmark & \xmark & -- & \faLockOpen & \acqR & B & \cmark & 3 & \cmark&\cmark&\cmark&\xmark&\xmark \\
\texttt{HCRL Car-Hacking}~\cite{seo2018gids} & 2018 & \cmark & \xmark & -- & \faLockOpen & \acqR & B & \cmark & 4 & \cmark&\cmark&\cmark&\xmark&\xmark \\
\rowcolor[HTML]{f0f0f0}
\texttt{TU/e CAN v2}~\cite{dupont2019tuecan} & 2019 & \cmark & \xmark & -- & \faLockOpen & \acqH & W & \cmark & 5 & \cmark&\cmark&\xmark&\cmark&\cmark \\
\texttt{SynCAN}~\cite{hanselmann2020syncan} & 2019 & \xmark & \cmark & CAN-D & \faLockOpen & \acqE & B & \cmark & 5 & \cmark&\xmark&\cmark&\cmark&\cmark \\
\rowcolor[HTML]{f0f0f0}
\texttt{UNIMORE Bus-Off}~\cite{unimorebusoff} & 2019 & \cmark & \xmark & -- & \faLockOpen & \acqH & -- & \cmark & 2 & \xmark&\xmark&\xmark&\xmark&\cmark \\
\texttt{Car Hacking Challenge 2020}~\cite{carhackchallenge2020} & 2020 & \cmark & \xmark & -- & \faUser & \acqR & M & \cmark & 4 & \cmark&\cmark&\cmark&\cmark&\xmark \\
\rowcolor[HTML]{f0f0f0}
\texttt{ROAD}~\cite{verma2024road} & 2020 & \cmark & \cmark & CAN-D & \faLockOpen & \acqR & M & \cmark & 13 & \xmark&\cmark&\cmark&\xmark&\cmark \\
\texttt{AVTP Ethernet Intrusion}~\cite{jeong2021avtp} & 2021 & \cmark & \xmark & -- & \faUser & \acqH & W & \cmark & 1 & \xmark&\xmark&\xmark&\cmark&\xmark \\
\rowcolor[HTML]{f0f0f0}
\texttt{SOME/IP IDS}~\cite{alkhatib2021someip} & 2021 & \xmark & \cmark & custom & \faLockOpen & \acqE & M & \cmark & 4 & \xmark&\xmark&\xmark&\xmark&\cmark \\
\texttt{HCRL CAN-FD}~\cite{hcrlcanfd2022} & 2022 & \cmark & \xmark & -- & \faLockOpen & \acqR & B & \cmark & 3 & \cmark&\cmark&\xmark&\xmark&\cmark \\
\rowcolor[HTML]{f0f0f0}
\texttt{TOW-IDS}~\cite{han2023towids} & 2022 & \cmark & \xmark & -- & \faUser & \acqH & M & \cmark & 5 & \cmark&\xmark&\cmark&\cmark&\cmark \\
\texttt{UNIMORE DAGA}~\cite{unimoredaga} & 2022 & \cmark & \xmark & -- & \faLockOpen & \acqR & W & \cmark & 5 & \cmark&\cmark&\xmark&\cmark&\xmark \\
\rowcolor[HTML]{f0f0f0}
\texttt{can-train-and-test}~\cite{lampe2024cantrain} & 2023 & \cmark & \xmark & -- & \faLockOpen & \acqR & B & \cmark & 9 & \cmark&\cmark&\cmark&\xmark&\cmark \\
\texttt{CrySyS CAN}~\cite{gazdag2023crysys} & 2023 & \cmark & \xmark & -- & \faLockOpen & \acqH & M & \cmark & 24 & \xmark&\xmark&\cmark&\cmark&\cmark \\
\rowcolor[HTML]{f0f0f0}
\texttt{HCRL M-CAN}~\cite{hcrlmcan2023} & 2023 & \cmark & \xmark & -- & \faLockOpen & \acqH & B & \cmark & 2 & \cmark&\cmark&\xmark&\xmark&\xmark \\
\texttt{HCRL B-CAN}~\cite{hcrlbcan2023} & 2023 & \cmark & \xmark & -- & \faLockOpen & \acqH & B & \cmark & 2 & \cmark&\cmark&\xmark&\xmark&\xmark \\
\rowcolor[HTML]{f0f0f0}
\texttt{CAN-MIRGU}~\cite{rajapaksha2024canmirgu} & 2024 & \cmark & \xmark & -- & \faLockOpen & \acqR & -- & \cmark & 6 & \cmark&\cmark&\cmark&\cmark&\cmark \\
\texttt{CICIoV2024}~\cite{neto2024ciciov} & 2024 & \cmark & \xmark & -- & \faUser & \acqH & M & \cmark & 5 & \cmark&\xmark&\cmark&\xmark&\xmark \\
\rowcolor[HTML]{f0f0f0}
\texttt{X-CANIDS}~\cite{jeong2024xcanids} & 2024 & \cmark & \cmark & DBC & \faUser & \acqH & B & \cmark & 5 & \xmark&\cmark&\cmark&\cmark&\cmark \\
\texttt{SAE J1939}~\cite{lee2024j1939} & 2024 & \cmark & \xmark & -- & \faLockOpen & \acqE & B & \cmark & 11 & \cmark&\xmark&\cmark&\xmark&\cmark \\
\rowcolor[HTML]{f0f0f0}
\texttt{GEM-CAN v3}~\cite{gemcan2026} & 2026 & \cmark & \xmark & -- & \faLockOpen & \acqR & B & \cmark & 2 & \cmark&\xmark&\cmark&\xmark&\xmark \\
\texttt{CarDS}~\cite{hellemans2025cards} & 2026 & \cmark & \xmark & -- & \faLockOpen & \acqR & -- & \cmark & -- & \xmark&\xmark&\cmark&\xmark&\cmark \\
\rowcolor[HTML]{f0f0f0}
\texttt{AutoHack}~\cite{song2026autohack} & 2026 & \cmark & \xmark & -- & \faLockOpen & \acqR & M & \cmark & 5 & \cmark&\cmark&\cmark&\cmark&\cmark \\
\bottomrule
\end{tabular}

\smallskip
\parbox{\linewidth}{\scriptsize\raggedright
\textbf{N.B:}
\textbf{Artifact:} Raw Network Traces, i.e.\ raw CAN/Ethernet frames (Raw), Processed Records, e.g.\ decoded signals (Proc.);
\textbf{Processing Tool (Proc.):} tool that produced the processed records, CAN-D (CAN-D signal translation), DBC = signals decoded with the vehicle's DBC file, custom = authors' own parser;
\textbf{Access:} \faLockOpen~open; \faUser~registration; 
\textbf{Acquisition (Acq.):}    \acqR~real vehicle (on road, stationary or on a dynamometer) with injected attacks; \acqH~real ECUs on a bench, or real benign traffic with synthetically inserted attacks; \acqE~synthetic, incl.\ hardware bus simulators and software packet generators;
\textbf{Labeling (Lab.):} Binary, per message (B), Multiclass (M); Weak, per file or attack interval (W); None (--);
\textbf{Classes:} Benign (Ben.), Attack (Att.);
\textbf{Attack Families:}
DoS/flooding (D), Fuzzing (F), Spoofing/fabrication/masquerade/impersonation (S), Replay (P), Other, e.g.\ suspension, bus-off, malfunction, PTP/MAC attacks, transport-protocol memory attacks, SOME/IP protocol errors, scans (O);
\cmark~present, \xmark~not present, --~not specified or not documented.
}
\end{table}
\subsection{Automotive CAN and Ethernet}\label{sec:rev-automotive}
Table~\ref{tab:rev-automotive} lists the $24$ automotive datasets in the \Obs{}, $11$ of them released since $2023$. 
\emph{They do not contain IP packets between hosts}: in-vehicle networks carry frames on buses such as \gls{can}, CAN-FD and SAE J1939, the variant used in heavy vehicles, or on automotive Ethernet. Four datasets cover the latter: \texttt{TOW-IDS}, the \texttt{AVTP} dataset, \texttt{SOME/IP IDS} and \texttt{CarDS}, which records both the \gls{can} and Ethernet buses of the same vehicle.
Their protocols, prediction targets and evaluation assumptions differ substantially, and none is interchangeable with IP captures.
\emph{Almost all datasets release raw frames}: $22$ of the $24$ do so, and only two distribute processed records alone, decoded signals in \texttt{SynCAN} and parsed SOME/IP messages in \texttt{SOME/IP IDS} (\emph{Artifact} column). \texttt{ROAD} and \texttt{X-CANIDS} add a signal-level view to their frames.
\emph{Half of the datasets were recorded on real vehicles}: $12$ of the $24$, while nine use real ECUs on a bench or insert attacks into real traffic, and three are synthetic (\texttt{SynCAN}, \texttt{SOME/IP IDS} and the SAE J1939 dataset, generated with hardware bus simulators) (\emph{Acq.} column). In all cases, however, the attacks are injected by the authors rather than observed in the wild.
\emph{Labeling is the weakest point}: only seven datasets have multiclass labels, ten use per-message binary labels, four use weak per-file or per-interval labels, and three have none or do not document them (\emph{Lab.} column).
\emph{The attack families are few and specific to the bus}: DoS/flooding ($17$ datasets), spoofing ($16$) and fuzzing ($14$) dominate, followed by other attacks such as suspension, bus-off or transport-protocol abuse ($14$) and by replay ($10$) (\emph{Attack families} columns).
Overall, the group offers raw frames from real vehicles, but with a narrow set of injected attacks and inconsistent labels. Moreover, \gls{can} identifiers and signals are vehicle-specific, so results rarely transfer across datasets.
\begin{table}[!htbp]
\centering\scriptsize\setlength{\tabcolsep}{2pt}
\providecommand{\acqR}{\CIRCLE}
\providecommand{\acqH}{\LEFTcircle}
\providecommand{\acqE}{\Circle}
\providecommand{\deriv}{\smash{\raisebox{0.32em}{\tikz[baseline=0pt]{\draw[line width=0.4pt,line cap=round] (0,0) -- (0,0.58em) (-0.16em,0.40em) -- (0.16em,0.40em);}}}}
\providecommand{\cmark}{\textcolor{green!55!black}{\ding{51}}}
\providecommand{\xmark}{\textcolor{red!55!black}{\ding{55}}}
\caption{Vehicular V2X datasets in the \Obs.}
\label{tab:rev-v2x}
\begin{tabular}{llcccccccccccccc}
\toprule
\multirow{2}{*}{\textbf{Dataset}} & 
\multirow{2}{*}{\textbf{Year$\downarrow$}} &
\multicolumn{3}{c}{\textbf{Artifact}} & 
\multirow{2}{*}{\textbf{Access}}
&
\multirow{2}{*}{\textbf{Acq.}}
&
\multirow{2}{*}{\textbf{Lab.}}
& \multicolumn{2}{c}{\textbf{Classes}} & \multicolumn{6}{c}{\textbf{Attack families}}
\\
\cmidrule(lr){3-5}\cmidrule(lr){9-10}\cmidrule(lr){11-16}
 &  & \textbf{Raw} & \textbf{Proc.} & \textbf{Proc. Tool} & &  &  & \textbf{Ben.} & \textbf{Att.} & \textbf{D} & \textbf{F} & \textbf{P} & \textbf{T} & \textbf{Y} & \textbf{O} \\
\midrule
\texttt{VeReMi}~\cite{vanderheijden2018veremi} & 2018 & \cmark & \xmark & -- & \faLockOpen & \acqE & M & \cmark & 5 & \xmark&\cmark&\xmark&\xmark&\xmark&\cmark \\
\rowcolor[HTML]{f0f0f0}
\texttt{V2X Security Threats}~\cite{goncalves2020v2x} & 2020 & \cmark & \xmark & -- & \faLockOpen & \acqE & M & -- & -- & \cmark&\cmark&\xmark&\xmark&\xmark&\xmark \\
\texttt{VeReMi Extension}~\cite{kamel2020veremiext} & 2020 & \cmark & \xmark & -- & \faLockOpen & \acqE & M & \cmark & 19 & \cmark&\cmark&\cmark&\cmark&\cmark&\cmark \\
\rowcolor[HTML]{f0f0f0}
\texttt{VANET DDoS}~\cite{yang2022vanetddos} & 2022 & \cmark & \xmark & -- & \faLockOpen & \acqE & -- & \cmark & 4 & \cmark&\cmark&\xmark&\xmark&\xmark&\xmark \\
\texttt{MisbehaviorX}~\cite{misbehaviorx2024} & 2024 & \cmark & \xmark & -- & \faUser & \acqE & M & \cmark & 68 & \multicolumn{6}{c}{--} \\
\rowcolor[HTML]{f0f0f0}
\texttt{VeReMi NextGen}~\cite{veremings2026} & 2026 & \cmark & \xmark & -- & \faLockOpen & \acqE & B & \cmark & 15 & \cmark&\cmark&\cmark&\cmark&\xmark&\cmark \\
\bottomrule
\end{tabular}

\smallskip
\parbox{\linewidth}{\scriptsize\raggedright
\textbf{N.B:}
\textbf{Artifact:} Raw Network traces, i.e.\ logged V2X messages (Raw), Processed Records (Proc.);
\textbf{Processing Tool (Proc. Tool):} tool that produced the processed records;
\textbf{Access:} \faLockOpen~open; \faUser~registration;
\textbf{Acquisition (Acq.):} \acqE~simulated (VEINS/F2MD, Eclipse MOSAIC, SUMO + OMNeT++);
\textbf{Labeling (Lab.):} Binary, attacker 0/1 per message with attack type given by the subset (B), Multiclass (M); None (--);
\textbf{Classes:} Benign (Ben.), Attack/misbehaviour types (Att.);
\textbf{Attack Families:}
DoS/flooding (D), Data falsification of position, speed, acceleration or heading (F), Replay (P), Timing, i.e.\ delayed messages (T), Sybil (Y), Other, e.g.\ eventual stop, faulty sensors (O);
\cmark~present, \xmark~not present, --~not specified or not documented.
}
\end{table}

\subsection{Cooperative vehicular communication}\label{sec:rev-v2x}

Table~\ref{tab:rev-v2x} lists the six V2X datasets in the \Obs{}. Like the automotive group, \emph{they do not contain IP packets between hosts} but safety messages exchanged between vehicles and roadside units, such as cooperative awareness messages.
Three of the six belong to the \texttt{VeReMi} family~\citep{vanderheijden2018veremi}, which models misbehavior such as falsified position or speed reports in cooperative awareness messages; the others cover synthesized attacks and \gls{ddos} against vehicular networks.
\emph{Every dataset is simulated} (\emph{Acq.} column): none was recorded on real vehicles or roadside infrastructure. Labels and message fields therefore reflect the simulator and the message format that produced them, and results obtained on these datasets may not hold for real V2X traffic.
\emph{Labels are mostly multiclass}: four of the six datasets, while \texttt{VeReMi NextGen} uses binary labels with the attack type given by the subset and \texttt{VANET DDoS} does not document its labels (\emph{Lab.} column).
\emph{Class counts are not comparable across datasets}: they range from $4$ attack types in \texttt{VANET DDoS} to $68$ misbehavior types in \texttt{MisbehaviorX} (\emph{Att.} column).
\emph{Data falsification is the dominant attack family}: it appears in all five datasets that document their attacks, followed by DoS in four (\emph{Attack families} columns).
Overall, the group offers labeled data of simulated V2X traffic, but without any real V2X traffic it cannot show that a detector works on deployed vehicles.

\begin{table}[!htbp]
\centering\scriptsize\setlength{\tabcolsep}{2pt}
\providecommand{\acqR}{\CIRCLE}
\providecommand{\acqH}{\LEFTcircle}
\providecommand{\acqE}{\Circle}
\providecommand{\deriv}{\smash{\raisebox{0.32em}{\tikz[baseline=0pt]{\draw[line width=0.4pt,line cap=round] (0,0) -- (0,0.58em) (-0.16em,0.40em) -- (0.16em,0.40em);}}}}
\providecommand{\cmark}{\textcolor{green!55!black}{\ding{51}}}
\providecommand{\xmark}{\textcolor{red!55!black}{\ding{55}}}
\caption{UAV / drones datasets in the \Obs.}
\label{tab:rev-uav}
\begin{tabular}{llccccccccccccccc}
\toprule
\multirow{2}{*}{\textbf{Dataset}} & 
\multirow{2}{*}{\textbf{Year$\downarrow$}} &
\multicolumn{3}{c}{\textbf{Artifact}} & 
\multirow{2}{*}{\textbf{Access}}
&
\multirow{2}{*}{\textbf{Acq.}}
&
\multirow{2}{*}{\textbf{Lab.}}
& \multicolumn{2}{c}{\textbf{Classes}} & \multicolumn{7}{c}{\textbf{Attack families}}
\\
\cmidrule(lr){3-5}\cmidrule(lr){9-10}\cmidrule(lr){11-17}
 &  & \textbf{Raw} & \textbf{Proc.} & \textbf{Proc. Tool} & &  &  & \textbf{Ben.} & \textbf{Att.} & \textbf{B} & \textbf{D} & \textbf{P} & \textbf{R} & \textbf{S} & \textbf{V} & \textbf{O} \\
\midrule
\texttt{UAV Cyber-Physical}~\cite{uavcyberphys2023} & 2023 & \xmark & \cmark & CFM & \faUser & \acqH & M & \cmark & 4 & \xmark&\cmark&\cmark&\xmark&\cmark&\cmark&\xmark \\
\rowcolor[HTML]{f0f0f0}
\texttt{ISOT Drone}~\cite{isotdrone2024} & 2024 & \cmark & \cmark & custom & \faLockOpen & \acqH & B & \cmark & 5 & \xmark&\xmark&\cmark&\xmark&\cmark&\cmark&\xmark \\
\texttt{UAV-NIDD}~\cite{uavnidd2025} & 2025 & \cmark & \cmark & CFM & \faLockOpen & \acqH & M & \cmark & 10 & \cmark&\cmark&\cmark&\cmark&\cmark&\xmark&\cmark \\
\rowcolor[HTML]{f0f0f0}
\texttt{UAVIDS-2025}~\cite{uavids2025} & 2025 & \xmark & \cmark & custom & \faUser & \acqE & M & \cmark & 4 & \xmark&\cmark&\xmark&\xmark&\xmark&\xmark&\cmark \\
\texttt{GCS-NIDD}~\cite{gcsnidd2026} & 2026 & \cmark & \cmark & CFM, Zeek & \faLockOpen & \acqH & M & \cmark & 9 & \cmark&\cmark&\cmark&\cmark&\cmark&\xmark&\cmark \\
\bottomrule
\end{tabular}

\smallskip
\parbox{\linewidth}{\scriptsize\raggedright
\textbf{N.B:}
\textbf{Artifact:} Raw Network Traces (Raw), Processed Records (Proc.);
\textbf{Processing Tool (Proc. Tool):} tool that produced the processed records; CICFlowMeter (CFM), custom = authors' own extractor;
\textbf{Access:} \faLockOpen~open; \faUser~registration;
\textbf{Acquisition (Acq.):} \acqH~physical testbed with real UAVs, access points and ground control stations; \acqE~simulated (NS-3);
\textbf{Labeling (Lab.):} Binary, benign vs anomalous (B), Multiclass (M);
\textbf{Classes:} Benign (Ben.), Attack (Att.);
\textbf{Attack Families:}
Brute force (B), (D)DoS/flooding/de-authentication (D), Replay (P), Reconnaissance/scan (R), Spoofing/MitM/evil twin, incl.\ GPS and IP spoofing, data interception (S), False data injection, incl.\ command injection, payload manipulation (V), Other, e.g.\ fake landing, GPS jamming, blackhole, wormhole, Sybil (O);
\cmark~present, \xmark~not present, --~not specified or not documented.}
\end{table}

\subsection{Unmanned Aerial Vehicle and drone networks}\label{sec:rev-uav}

Table~\ref{tab:rev-uav} lists the five \gls{uav} datasets in the \Obs{}. 
\emph{The group is one of the smallest and the youngest of the catalog}, with all datasets released since $2023$; several \gls{uav} and LoRaWAN candidates could not be traced to any public source and were excluded.
These datasets mix network traffic between drones and ground control stations with flight telemetry, and only their traffic component falls within the scope of this review.
\emph{Most datasets come from physical testbeds}: four of the five were captured with real drones, access points and ground stations, and one is simulated (\emph{Acq.} column).
\emph{Raw traces are not always available}: two of the five datasets distribute processed records only (\emph{Artifact} column). 
\emph{The attack families reflect the wireless link between drone and ground station}: DoS and de-authentication, replay, and spoofing (including GPS spoofing) each appear in four of the five datasets, while other, drone-specific attacks such as fake landing or GPS jamming (O) appear in three (\emph{Attack families} columns).
Overall, the group is too small and too recent to support general conclusions: each dataset comes from a single testbed or simulation, and none has yet been reused or re-extracted.

\begin{table}[!htbp]
\centering\scriptsize\setlength{\tabcolsep}{2pt}
\providecommand{\acqR}{\CIRCLE}
\providecommand{\acqH}{\LEFTcircle}
\providecommand{\acqE}{\Circle}
\providecommand{\deriv}{\smash{\raisebox{0.32em}{\tikz[baseline=0pt]{\draw[line width=0.4pt,line cap=round] (0,0) -- (0,0.58em) (-0.16em,0.40em) -- (0.16em,0.40em);}}}}
\providecommand{\cmark}{\textcolor{green!55!black}{\ding{51}}}
\providecommand{\xmark}{\textcolor{red!55!black}{\ding{55}}}
\caption{5G / mobile core datasets in the \Obs.}
\label{tab:rev-fiveg}
\begin{tabular}{llccccccccccccccc}
\toprule
\multirow{2}{*}{\textbf{Dataset}} &
\multirow{2}{*}{\textbf{Year$\downarrow$}} &
\multicolumn{3}{c}{\textbf{Artifact}} &
\multirow{2}{*}{\textbf{Access}}
&
\multirow{2}{*}{\textbf{Acq.}}
&
\multirow{2}{*}{\textbf{Lab.}}
& \multicolumn{2}{c}{\textbf{Classes}} & \multicolumn{7}{c}{\textbf{Attack families}}
\\
\cmidrule(lr){3-5}\cmidrule(lr){9-10}\cmidrule(lr){11-17}
 &  & \textbf{Raw} & \textbf{Proc.} & \textbf{Proc. Tool} & &  &  & \textbf{Ben.} & \textbf{Att.} & \textbf{B} & \textbf{D} & \textbf{F} & \textbf{P} & \textbf{R} & \textbf{S} & \textbf{O} \\
\midrule
\texttt{5G-NIDD}~\cite{samarakoon2022fivegnidd} & 2022 & \cmark & \cmark & Argus & \faLockOpen & \acqH & M & \cmark & 8 & \xmark&\cmark&\xmark&\xmark&\cmark&\xmark&\xmark \\
\rowcolor[HTML]{f0f0f0}
\texttt{5GAD-2022}~\cite{fivegad2022} & 2022 & \cmark & \xmark & -- & \faLockOpen & \acqE & M & \cmark & 10 & \xmark&\xmark&\xmark&\xmark&\cmark&\cmark&\cmark \\
\texttt{5G Core IDS}~\cite{sheikhi2023ddos} & 2023 & \xmark & \cmark & tshark & \faLockOpen & \acqE & B & \cmark & 3 & \xmark&\cmark&\xmark&\xmark&\xmark&\cmark&\cmark \\
\rowcolor[HTML]{f0f0f0}
\texttt{5GC PFCP}~\cite{pfcp2023} & 2023 & \cmark & \cmark & CFM, custom & \faLockOpen & \acqE & M & \cmark & 4 & \xmark&\cmark&\xmark&\xmark&\xmark&\xmark&\cmark \\
\texttt{5G Slicing DoS/DDoS}~\cite{khan2023slicing} & 2023 & \cmark & \cmark & CFM & \faUser & \acqE & -- & \cmark & -- & \xmark&\cmark&\xmark&\xmark&\xmark&\xmark&\xmark \\
\rowcolor[HTML]{f0f0f0}
\texttt{5G Flow-based}~\cite{wehbe2024fiveg} & 2024 & \xmark & \cmark & CFM & \faUser & \acqE & M & \cmark & 6 & \xmark&\cmark&\xmark&\xmark&\xmark&\xmark&\cmark \\
\texttt{WUSTL-HDRL-2024}~\cite{wustlhdrl2024} & 2024 & \xmark & \cmark & Argus & \faLockOpen & \acqE & M & \cmark & 4 & \xmark&\cmark&\xmark&\xmark&\xmark&\cmark&\cmark \\
\rowcolor[HTML]{f0f0f0}
\texttt{5G Core GTP-U}~\cite{fivegcgad2025} & 2025 & \cmark & \cmark & CFM & \faLockOpen & \acqE & M & \cmark & 5 & \cmark&\cmark&\cmark&\xmark&\xmark&\xmark&\cmark \\
\texttt{5G Security \& Resource Alloc.}~\cite{fivegsra2025} & 2025 & \cmark & \cmark & 5GPP & \faLockOpen & \acqH & B & \cmark & 8 & \cmark&\cmark&\cmark&\cmark&\xmark&\cmark&\xmark \\
\rowcolor[HTML]{f0f0f0}
\texttt{5GCID}~\cite{hu2026fivegcid} & 2025 & \cmark & \xmark & -- & \faLockOpen & \acqE & W & \cmark & 3 & \xmark&\cmark&\xmark&\xmark&\xmark&\xmark&\xmark \\
\texttt{NetsLab-5GORAN-IDD}~\cite{civciss2025netslab} & 2025 & \cmark & \cmark & Zeek & \faLockOpen & \acqH & W & \cmark & 18 & \cmark&\cmark&\xmark&\xmark&\cmark&\xmark&\cmark \\
\rowcolor[HTML]{f0f0f0}
\texttt{5GCoreSec}~\cite{sheikhi2026fivegcoresec} & 2026 & \xmark & \cmark & tshark & \faLockOpen & \acqE & M & \cmark & 8 & \cmark&\cmark&\xmark&\xmark&\xmark&\cmark&\cmark \\
\texttt{CVE5G}~\cite{rosa2026cve5g} & 2026 & \cmark & \xmark & -- & \faUser & \acqE & M & \cmark & 5 & \xmark&\cmark&\cmark&\xmark&\cmark&\xmark&\xmark \\
\bottomrule
\end{tabular}

\smallskip
\parbox{\linewidth}{\scriptsize\raggedright
\textbf{N.B:}
\textbf{Artifact:} Raw Network Traces (Raw), Processed Records (Proc.);
\textbf{Processing Tool (Proc. Tool):} tool that produced the processed records; CICFlowMeter (CFM), 5GPacketParser (5GPP), custom = authors' own extractor;
\textbf{Access:} \faLockOpen~open; \faUser~registration; 
\textbf{Acquisition (Acq.):} \acqH~testbed with real radio equipment (5G Test Network Oulu with Nokia small cells; Amarisoft); \acqE~emulated or simulated 5G core (free5GC, Open5GS, virtual UE/gNB with UERANSIM; Simu5G);
\textbf{Labeling (Lab.):} Binary (B), Multiclass (M), Weak (W; one label per capture file), not documented (--);
\textbf{Classes:} Benign (Ben.), Attack (Att.);
\textbf{Attack Families:}
Brute force, incl.\ GTP-U TEID guessing (B), DoS/flooding, incl.\ HTTP/2 rapid reset, slow-rate (D), Fuzzing, malformed NAS/GTP (F), Replay (P), Reconnaissance/scan, incl.\ NF discovery (R), Spoofing/impersonation, fake AMF/SMF, MitM (S), Other, e.g.\ PFCP session modification/redirection, NF reconfiguration, GTP encapsulation, HTTP/2 stream multiplexing, ransomware, SQL injection (O);
\cmark~present, \xmark~not present, --~not specified or not documented.
}
\end{table}

\subsection{5G and mobile core networks}\label{sec:rev-5g}
Table~\ref{tab:rev-fiveg} lists the 13 5G and mobile-core datasets in the \Obs{}, all released since $2022$ and six of them since $2025$.
In cellular and virtualized networks, the monitoring location relative to control-plane functions, data-plane elements or orchestration infrastructure determines the meaning of a trace. Accordingly, the datasets observe different interfaces of the 5G system, from user-plane traffic to the \gls{pfcp}, NGAP and GTP-U protocols of the core.
\emph{Most datasets release packet traces}: nine of the $13$ do so, six of them together with flow or feature records, while four distribute processed records only (\emph{Artifact} column).
\emph{The extractors are rarely designed for 5G}: when a tool is documented, it is most often CICFlowMeter, a generic IP-flow extractor that does not parse 5G protocol fields; only the PFCP dataset and the resource-allocation dataset add a dedicated parser (\emph{Proc. Tool} column).
\emph{Most 5G cores are emulated}: ten datasets emulate or simulate the core with open-source software (free5GC, Open5GS, OpenAirInterface, UERANSIM, Simu5G), and only three use real radio equipment, namely \texttt{5G-NIDD}, the resource-allocation dataset and \texttt{NetsLab-5GORAN-IDD} (\emph{Acq.} column).
\emph{DoS and flooding dominate the attack families}: they are present in $12$ of the $13$ datasets, followed by attacks specific to the 5G core or to the services it carries, such as PFCP session modification, NF reconfiguration and GTP encapsulation (O, eight datasets) (\emph{Attack families} columns).
Overall, the group is growing quickly but remains dominated by emulated cores, whose behavior may differ from operator deployments, and by DoS scenarios, so it does not yet cover the variety of attacks that a 5G core may face.

\begin{table}[!htbp]
\centering\scriptsize\setlength{\tabcolsep}{2pt}
\providecommand{\acqR}{\CIRCLE}
\providecommand{\acqH}{\LEFTcircle}
\providecommand{\acqE}{\Circle}
\providecommand{\deriv}{\smash{\raisebox{0.32em}{\tikz[baseline=0pt]{\draw[line width=0.4pt,line cap=round] (0,0) -- (0,0.58em) (-0.16em,0.40em) -- (0.16em,0.40em);}}}}
\providecommand{\cmark}{\textcolor{green!55!black}{\ding{51}}}
\providecommand{\xmark}{\textcolor{red!55!black}{\ding{55}}}
\caption{Wireless 802.11/LPWAN/WSN datasets in the \Obs.}
\label{tab:rev-wireless}
\begin{tabular}{llccccccccccccccccc}
\toprule
\multirow{2}{*}{\textbf{Dataset}} & 
\multirow{2}{*}{\textbf{Year$\downarrow$}} &
\multicolumn{3}{c}{\textbf{Artifact}} & 
\multirow{2}{*}{\textbf{Access}}
&
\multirow{2}{*}{\textbf{Acq.}}
&
\multirow{2}{*}{\textbf{Lab.}}
& \multicolumn{2}{c}{\textbf{Classes}} & \multicolumn{8}{c}{\textbf{Attack families}}
\\
\cmidrule(lr){3-5}\cmidrule(lr){9-10}\cmidrule(lr){11-18}
 &  & \textbf{Raw} & \textbf{Proc.} & \textbf{Proc. Tool} & &  &  & \textbf{Ben.} & \textbf{Att.} & \textbf{B} & \textbf{C} & \textbf{D} & \textbf{M} & \textbf{S} & \textbf{V} & \textbf{W} & \textbf{O} \\
\midrule
\texttt{AWID2}~\cite{kolias2016} & 2015 & \xmark & \cmark & tshark & \faEnvelope & \acqH & M & \cmark & 3 / 15 & \xmark&\xmark&\cmark&\xmark&\cmark&\cmark&\xmark&\xmark \\
\rowcolor[HTML]{f0f0f0}
\texttt{WSN-DS}~\cite{almomani2016wsnds} & 2016 & \xmark & \cmark & NS-2 & \faLockOpen & \acqE & M & \cmark & 4 & \xmark&\xmark&\cmark&\xmark&\xmark&\xmark&\xmark&\cmark \\
\texttt{AWID3}~\cite{chatzoglou2021awid3} & 2021 & \cmark & \cmark & tshark & \faEnvelope & \acqH & M & \cmark & 13 & \cmark&\cmark&\cmark&\cmark&\cmark&\cmark&\cmark&\xmark \\
\rowcolor[HTML]{f0f0f0}
\texttt{LoRaWAN Jamming}~\cite{lorawanjam2026}\ \deriv{}[\cite{povalac2023lorawan}] & 2026 & \xmark & \cmark & custom & \faLockOpen & \acqE & M & \cmark & 2 & \xmark&\xmark&\cmark&\xmark&\xmark&\xmark&\xmark&\cmark \\
\texttt{WPA3 Attacks}~\cite{wpa3attacks2026} & 2026 & \cmark & \cmark & Wireshark & \faLockOpen & \acqH & -- & \cmark & 12 & \xmark&\xmark&\cmark&\xmark&\cmark&\cmark&\xmark&\cmark \\
\bottomrule
\end{tabular}

\smallskip
\parbox{\linewidth}{\scriptsize\raggedright
\textbf{N.B:}
\textbf{Artifact:} Raw Network Traces (Raw), Processed Records (Proc.);
\textbf{Processing Tool (Proc. Tool):} tool that produced the processed records;
\textbf{Access:} \faLockOpen~open; \faUser~registration; \faEnvelope~ on request;
\textbf{Acquisition (Acq.):} \acqH~physical Wi-Fi lab \acqE~simulated (NS-2, LEACH), or real LoRaWAN traces with semi-synthetic attack augmentation;
\textbf{Labeling (Lab.):} Multiclass (M); None (--);
\textbf{Classes:} Benign (Ben.), Attack (Att.; for AWID2, 3 major classes / 15 individual attacks);
\textbf{Attack Families:}
Brute force (B), Botnet (C), DoS/flooding/de-authentication/jamming, incl.\ WSN blackhole, grayhole, flooding, scheduling (D), Malware (M), Impersonation/evil twin/rogue AP/spoofing (S), Injection/exploitation, incl.\ Krack, Kr00k (V), Web, SQL injection (W), Other, e.g.\ battery depletion, WSN routing attacks, WPA3 downgrade (O);
\deriv{}~derivative of an earlier capture;
\cmark~present, \xmark~not present, --~not specified or not documented.
}
\end{table}

\subsection{Wireless Local Area, Sensor and Low-Power Wide-Area Networks}\label{sec:rev-wireless}

Table~\ref{tab:rev-wireless} lists the five wireless datasets in the \Obs{}, covering IEEE 802.11, wireless sensor networks and LPWAN. 
\emph{These datasets introduce observation layers for which an IP-centric taxonomy is insufficient}: the AWID family~\citep{kolias2016} contains 802.11 security observations whose radio- and link-layer semantics cannot be inferred from TCP/IP connection fields, and AWID3 extends it to more recent 802.11 security mechanisms. WSN-DS covers attacks on wireless sensor networks, and a semi-synthetic LoRaWAN jamming dataset is the only LPWAN record in the catalog.
\emph{Raw traces are rare}: only \texttt{AWID3} and the \texttt{WPA3} dataset release packet traces, while the other three distribute processed records only, extracted with tshark, produced directly by the NS-2 simulator or obtained with a custom extractor (\emph{Artifact} and \emph{Proc. Tool} columns).
\emph{The reference benchmarks are hard to obtain}: both \texttt{AWID} releases are available only on request (\emph{Access} column).
\emph{Link-layer attacks dominate}: the DoS family, which includes de-authentication and jamming, appears in all five datasets, followed by impersonation and injection in three each; only \texttt{AWID3} also includes higher-layer attacks such as brute force, botnet, malware and web (\emph{Attack families} columns).
Overall, the group is small and centered on a single family: \texttt{AWID} remains the reference for 802.11, while sensor and \gls{lpwan} networks are covered by one dataset each.

\begin{table}[!htbp]
\centering\scriptsize\setlength{\tabcolsep}{2pt}
\providecommand{\acqR}{\CIRCLE}
\providecommand{\acqH}{\LEFTcircle}
\providecommand{\acqE}{\Circle}
\providecommand{\deriv}{\smash{\raisebox{0.32em}{\tikz[baseline=0pt]{\draw[line width=0.4pt,line cap=round] (0,0) -- (0,0.58em) (-0.16em,0.40em) -- (0.16em,0.40em);}}}}
\providecommand{\cmark}{\textcolor{green!55!black}{\ding{51}}}
\providecommand{\xmark}{\textcolor{red!55!black}{\ding{55}}}
\caption{DNS / web datasets in the \Obs.}
\label{tab:rev-dnsweb}
\begin{tabular}{llcccccccccccccc}
\toprule
\multirow{2}{*}{\textbf{Dataset}} & 
\multirow{2}{*}{\textbf{Year$\downarrow$}} &
\multicolumn{3}{c}{\textbf{Artifact}} & 
\multirow{2}{*}{\textbf{Access}}
&
\multirow{2}{*}{\textbf{Acq.}}
&
\multirow{2}{*}{\textbf{Lab.}}
& \multicolumn{2}{c}{\textbf{Classes}} & \multicolumn{6}{c}{\textbf{Attack families}}
\\
\cmidrule(lr){3-5}\cmidrule(lr){9-10}\cmidrule(lr){11-16}
 &  & \textbf{Raw} & \textbf{Proc.} & \textbf{Proc. Tool} & &  &  & \textbf{Ben.} & \textbf{Att.} & \textbf{W} & \textbf{D} & \textbf{C} & \textbf{E} & \textbf{M} & \textbf{O} \\
\midrule
\texttt{ECML/PKDD 2007}~\cite{raissi2007ecml} & 2007 & \xmark & \xmark & -- & \faLockOpen & \acqE & M & \cmark & 7 & \cmark&\xmark&\xmark&\xmark&\xmark&\xmark \\
\rowcolor[HTML]{f0f0f0}
\texttt{CSIC 2010}~\cite{gimenez2010csic} & 2010 & \xmark & \xmark & -- & \faLockOpen & \acqE & B & \cmark & 1 & \cmark&\xmark&\xmark&\xmark&\xmark&\xmark \\
\texttt{Biblio-US17}~\cite{diazverdejo2024biblious17} & 2017 & \xmark & \cmark & custom & \faEnvelope & \acqR & W & \cmark & 1 & \cmark&\xmark&\xmark&\xmark&\xmark&\xmark \\
\rowcolor[HTML]{f0f0f0}
\texttt{CIRA-CIC-DoHBrw-2020}~\cite{montazeri2020dohbrw} & 2020 & \cmark & \cmark & DoHMeter & \faUser & \acqH & M & \cmark & 1 & \xmark&\xmark&\xmark&\cmark&\xmark&\xmark \\
\texttt{SR-BH 2020}~\cite{riera2022srbh} & 2020 & \xmark & \cmark & ModSecurity & \faLockOpen & \acqR & ML & \cmark & 12 & \cmark&\xmark&\xmark&\xmark&\xmark&\xmark \\
\rowcolor[HTML]{f0f0f0}
\texttt{CIC-Bell-DNS2021}~\cite{mahdavifar2021dns} & 2021 & \xmark & \cmark & custom & \faUser & \acqH & M & \cmark & 3 & \xmark&\xmark&\xmark&\xmark&\cmark&\xmark \\
\texttt{CIC-Bell-DNS-EXF-2021}~\cite{mahdavifar2021dnsexf} & 2021 & \xmark & \cmark & custom & \faUser & \acqH & M & \cmark & 2 & \xmark&\xmark&\xmark&\cmark&\xmark&\xmark \\
\rowcolor[HTML]{f0f0f0}
\texttt{H23Q}~\cite{chatzoglou2023h23q} & 2022 & \cmark & \cmark & tshark & \faEnvelope & \acqH & M & \cmark & 4 & \cmark&\cmark&\xmark&\xmark&\xmark&\cmark \\
\texttt{BCCC-CIRA-CIC-DoHBrw-2020}~\cite{bcccdohbrw2023}\ \deriv{}[\cite{montazeri2020dohbrw}] & 2023 & \xmark & \cmark & SMOTE & \faEnvelope & \acqH & B & \cmark & 1 & \xmark&\xmark&\xmark&\cmark&\xmark&\xmark \\
\rowcolor[HTML]{f0f0f0}
\texttt{BCCC-CIC-Bell-DNS-2024}~\cite{bcccbelldns2024}\ \deriv{}[\cite{mahdavifar2021dns}] & 2024 & \xmark & \cmark & ALFlowLyzer & \faEnvelope & \acqH & M & -- & 6 & \xmark&\xmark&\xmark&\cmark&\cmark&\xmark \\
\bottomrule
\end{tabular}

\smallskip
\parbox{\linewidth}{\scriptsize\raggedright
\textbf{N.B:}
\textbf{Artifact:} Raw Network Traces (Raw), Processed Records, e.g.\ HTTP requests, DNS records, flows (Proc.);
\textbf{Processing Tool (Proc. Tool):} tool that produced the processed records, custom = authors' own extractor;
\textbf{Access:} \faLockOpen~open; \faUser~registration; \faEnvelope~ on request;
\textbf{Acquisition (Acq.):} \acqR~real production server or Internet-exposed web server; \acqH~testbed with real browsers, servers, DNS resolvers or cloud servers; \acqE~synthetically generated HTTP requests;
\textbf{Labeling (Lab.):} Binary (B), Multiclass (M), Multi-label, CAPEC (ML); Weak, IDS alerts with confidence levels (W);
\textbf{Classes:} Benign (Ben.), Attack (Att.);
\textbf{Attack families:}
Web attacks, e.g.\ SQLi, XSS, path traversal, injection, request smuggling (W), (D)DoS/slow DoS (D), Botnet/C2 (C), DNS tunnelling/exfiltration (E), Malicious domains, i.e.\ domains serving malware, phishing or spam, counted with malware (M), Other, e.g.\ QUIC/HTTP-3 protocol attacks, fuzzing (O);
\deriv{}~derivative of an earlier capture, followed by the reference of its source;
\cmark~present, \xmark~not present, --~not specified or not documented.
}
\end{table}
\subsection{DNS and Web traffic}\label{sec:rev-dnsweb}

Table~\ref{tab:rev-dnsweb} lists the ten datasets in the \Obs{} that contain attacks carried over DNS or the web.
\emph{They mostly record application-layer objects}, HTTP requests or DNS exchanges, rather than full network captures. Classic web-attack corpora such as \texttt{ECML/PKDD 2007} and \texttt{CSIC 2010} contain labeled HTTP requests against web applications, while more recent releases cover DNS over HTTPS, malicious domains and DNS exfiltration.
\emph{Only two datasets support packet-level models}: \texttt{CIRA-CIC-DoHBrw-2020} and \texttt{H23Q} release packet traces, the other six distribute records produced by an extractor, and the two classic corpora release only anonymized raw HTTP requests in XML, with no extraction software involved (\emph{Artifact} and \emph{Proc. Tool} columns). In \texttt{BCCC-CIRA-CIC-DoHBrw-2020}, part of the records are synthetic samples generated with SMOTE to rebalance the classes, rather than observed traffic.
\emph{Access is restricted for most datasets}: only three are open, three require registration and four are available on request (\emph{Access} column).
\emph{Real traffic comes with indirect labels}: \texttt{Biblio-US17} and \texttt{SR-BH 2020} record real web servers, but their labels come from IDS or WAF alerts rather than ground truth (\emph{Acq.} column).
\emph{Labeling schemes vary widely}: binary, multiclass, weak labels with confidence levels and multi-label CAPEC annotations coexist (\emph{Lab.} column).
\emph{The attack families split the group in two}: the five web datasets cover web attacks such as SQL injection and XSS, the five DNS datasets cover exfiltration and malicious domains, and only \texttt{H23Q} adds denial of service and protocol attacks on HTTP/3 and QUIC; no dataset covers C2 traffic (\emph{Attack families} columns).
Overall, the group is useful for application-layer detection and its real-traffic datasets inherit the limits of the tools that labeled them.

\begin{table}[!htbp]
\centering\scriptsize\setlength{\tabcolsep}{2pt}
\providecommand{\acqR}{\CIRCLE}
\providecommand{\acqH}{\LEFTcircle}
\providecommand{\acqE}{\Circle}
\providecommand{\deriv}{\smash{\raisebox{0.32em}{\tikz[baseline=0pt]{\draw[line width=0.4pt,line cap=round] (0,0) -- (0,0.58em) (-0.16em,0.40em) -- (0.16em,0.40em);}}}}
\providecommand{\cmark}{\textcolor{green!55!black}{\ding{51}}}
\providecommand{\xmark}{\textcolor{red!55!black}{\ding{55}}}
\caption{DDoS datasets in the \Obs.}
\label{tab:rev-ddos}
\begin{tabular}{llcccccccc*{3}{>{\centering\arraybackslash}p{2.1em}}}
\toprule
\multirow{2}{*}{\textbf{Dataset}} & 
\multirow{2}{*}{\textbf{Year$\downarrow$}} &
\multicolumn{3}{c}{\textbf{Artifact}} & 
\multirow{2}{*}{\textbf{Access}}
&
\multirow{2}{*}{\textbf{Acq.}}
&
\multirow{2}{*}{\textbf{Lab.}}
& \multicolumn{2}{c}{\textbf{Classes}} & \multicolumn{3}{c}{\textbf{Attack families}}
\\
\cmidrule(lr){3-5}\cmidrule(lr){9-10}\cmidrule(lr){11-13}
 &  & \textbf{Raw} & \textbf{Proc.} & \textbf{Proc. Tool} & &  &  & \textbf{Ben.} & \textbf{Att.} & \textbf{D} & \textbf{R} & \textbf{V} \\
\midrule
\texttt{DARPA 2000 LLDOS}~\cite{darpa2000} & 2000 & \cmark & \xmark & -- & \faLockOpen & \acqE & W & -- & 5 & \cmark&\cmark&\cmark \\
\rowcolor[HTML]{f0f0f0}
\texttt{CAIDA DDoS Attack 2007}~\cite{caida2007ddos} & 2007 & \cmark & \xmark & -- & \faEnvelope & \acqR & -- & \xmark & 1 & \cmark&\xmark&\xmark \\
\texttt{CIC-DDoS2019}~\cite{sharafaldin2019ddos} & 2019 & \cmark & \cmark & CFM & \faUser & \acqH & M & \cmark & 13 & \cmark&\cmark&\xmark \\
\rowcolor[HTML]{f0f0f0}
\texttt{BOUN DDoS}~\cite{erhan2020boun} & 2020 & \xmark & \cmark & Wireshark & \faUser & \acqR & W & \cmark & 2 & \cmark&\xmark&\xmark \\
\texttt{USB-IDS-1}~\cite{catillo2021usbids} & 2021 & \cmark & \cmark & CFM & \faLockOpen & \acqH & B & \cmark & 4 & \cmark&\xmark&\xmark \\
\rowcolor[HTML]{f0f0f0}
\texttt{USB-IDS-TC}~\cite{usbidstc2024} & 2025 & \xmark & \cmark & CFM & \faLockOpen & \acqH & M & \cmark & 4 & \cmark&\xmark&\xmark \\
\texttt{BCCC-UDP-QUIC-IDS-2025}~\cite{bcccudpquic2025} & 2025 & \xmark & \cmark & UDP/QUICFlowLyzer & \faEnvelope & \acqH & M & \cmark & 7 & \cmark&\xmark&\xmark \\
\bottomrule
\end{tabular}

\smallskip
\parbox{\linewidth}{\scriptsize\raggedright
\textbf{N.B:}
\textbf{Artifact:} Raw Network Traces (Raw), Processed Records (Proc.);
\textbf{Processing Tool (Proc. Tool):} tool that produced the processed records, CFM = CICFlowMeter;
\textbf{Access:} \faLockOpen~open; \faUser~registration; \faEnvelope~on request;
\textbf{Acquisition (Acq.):} \acqR~real network (CAIDA backbone telescope trace; university campus backbone with injected floods); \acqH~testbed with synthetic benign profiles; \acqE~emulated (DARPA 2000);
\textbf{Labeling (Lab.):} Binary (B); Multiclass (M); Weak, per-phase ground truth or attack packets identified by the victim IP (W); None, attack-only trace (--);
\textbf{Classes:} Benign (Ben.), Attack (Att.; DARPA 2000: attack phases);
\textbf{Attack families:}
(D)DoS, incl.\ reflection/amplification, flooding, slow application-layer DoS (D), Reconnaissance/scan (R), Exploitation, e.g.\ sadmind (V);
\cmark~present, \xmark~not present, --~not specified or not documented.
}
\end{table}

\subsection{Denial-of-service traffic}\label{sec:rev-ddos}

Table~\ref{tab:rev-ddos} lists the seven datasets in the \Obs{} built around DDoS attacks, from volumetric floods to reflection and amplification and slow application-layer attacks. 
They range from historical traces (the \texttt{DARPA 2000 LLDOS} scenarios and the \texttt{CAIDA 2007} attack trace) to purpose-built laboratory benchmarks such as \texttt{CIC-DDoS2019} and protocol-specific releases such as \texttt{BCCC-UDP-QUIC-IDS-2025}.
\emph{Raw traces are available for four datasets}: the other three (\texttt{BOUN DDoS}, \texttt{USB-IDS-TC} and \texttt{BCCC-UDP-QUIC-IDS-2025}) distribute processed records only, produced by Wireshark, CICFlowMeter or protocol-specific flow extractors (\emph{Artifact} and \emph{Proc. Tool} columns).
\emph{Real networks are rare}: four of the seven datasets come from testbeds with synthetic benign profiles and \texttt{DARPA 2000} is emulated, while only two observe real networks, the \texttt{CAIDA} trace, which contains attack traffic only, and \texttt{BOUN DDoS}, where floods are injected into a campus backbone (\emph{Acq.} column). In laboratory captures, attack volume, rate and duration are set by the generation tools, so results on these datasets say more about the separability of tool-generated floods than about detection in operational networks.
\emph{Labeling is weak or absent in three datasets}: \texttt{DARPA 2000} provides per-phase ground truth, \texttt{BOUN DDoS} identifies attack packets by the victim address, and \texttt{CAIDA} has no labels (\emph{Lab.} column). The others distinguish from $1$ to $13$ classes, mostly variants of reflection, amplification and flooding (\emph{Att.} column).
\emph{The attack families are almost only (D)DoS, by construction}: only \texttt{DARPA 2000} and \texttt{CIC-DDoS2019} include the reconnaissance that precedes an attack, and only \texttt{DARPA 2000} includes exploitation (\emph{Attack families} columns).
Overall, the group offers plenty of attack traffic but little realistic benign traffic, so a high detection rate on these datasets is easy to obtain and hard to transfer.

\begin{table}[!htbp]
\centering\scriptsize\setlength{\tabcolsep}{2pt}
\providecommand{\acqR}{\CIRCLE}
\providecommand{\acqH}{\LEFTcircle}
\providecommand{\acqE}{\Circle}
\providecommand{\deriv}{\smash{\raisebox{0.32em}{\tikz[baseline=0pt]{\draw[line width=0.4pt,line cap=round] (0,0) -- (0,0.58em) (-0.16em,0.40em) -- (0.16em,0.40em);}}}}
\providecommand{\cmark}{\textcolor{green!55!black}{\ding{51}}}
\providecommand{\xmark}{\textcolor{red!55!black}{\ding{55}}}
\caption{SDN / cloud datasets in the \Obs.}
\label{tab:rev-sdn}
\begin{tabular}{llcccccccccccccccc}
\toprule
\multirow{2}{*}{\textbf{Dataset}} & 
\multirow{2}{*}{\textbf{Year$\downarrow$}} &
\multicolumn{3}{c}{\textbf{Artifact}} & 
\multirow{2}{*}{\textbf{Access}}
&
\multirow{2}{*}{\textbf{Acq.}}
&
\multirow{2}{*}{\textbf{Lab.}}
& \multicolumn{2}{c}{\textbf{Classes}} & \multicolumn{8}{c}{\textbf{Attack families}}
\\
\cmidrule(lr){3-5}\cmidrule(lr){9-10}\cmidrule(lr){11-18}
 &  & \textbf{Raw} & \textbf{Proc.} & \textbf{Proc. Tool} & &  &  & \textbf{Ben.} & \textbf{Att.} & \textbf{B} & \textbf{C} & \textbf{D} & \textbf{M} & \textbf{R} & \textbf{V} & \textbf{W} & \textbf{O} \\
\midrule
\texttt{ISOT-CID}~\cite{aldribi2020isotcid} & 2018 & \cmark & \xmark & -- & \faEnvelope & \acqR & B & \cmark & 1 & \cmark&\xmark&\cmark&\cmark&\cmark&\xmark&\cmark&\cmark \\
\rowcolor[HTML]{f0f0f0}
\texttt{SDN DDoS}~\cite{ahuja2020sdnddos} & 2020 & \xmark & \cmark & Ryu & \faLockOpen & \acqE & B & \cmark & 1 & \xmark&\xmark&\cmark&\xmark&\xmark&\xmark&\xmark&\xmark \\
\texttt{InSDN}~\cite{elsayed2020insdn} & 2020 & \cmark & \cmark & CFM & \faLockOpen & \acqE & M & \cmark & 7 & \cmark&\cmark&\cmark&\xmark&\cmark&\cmark&\cmark&\xmark \\
\rowcolor[HTML]{f0f0f0}
\texttt{UNR-IDD}~\cite{das2023unridd} & 2023 & \xmark & \cmark & ONOS & \faLockOpen & \acqE & M & \cmark & 5 & \xmark&\xmark&\cmark&\xmark&\cmark&\xmark&\xmark&\cmark \\
\texttt{BCCC-cPacket-Cloud-DDoS-2024}~\cite{bccccpacket2024} & 2024 & \cmark & \cmark & NTLFlowLyzer & \faEnvelope & \acqH & M & \cmark & 17 & \xmark&\xmark&\cmark&\xmark&\xmark&\xmark&\xmark&\xmark \\
\bottomrule
\end{tabular}

\smallskip
\parbox{\linewidth}{\scriptsize\raggedright
\textbf{N.B:}
\textbf{Artifact:} Raw Network Traces (Raw), Processed Records (Proc.);
\textbf{Processing Tool (Proc. Tool):} tool that produced the processed records, CFM = CICFlowMeter;
\textbf{Access:} \faLockOpen~open; \faUser~registration; \faEnvelope~ on request;
\textbf{Acquisition (Acq.):} \acqR~real cloud environment; \acqH~cloud testbed; \acqE~emulated SDN (Mininet with Ryu or ONOS controller, virtual machines);
\textbf{Labeling (Lab.):} Binary (B; ISOT-CID: per packet, from the list of benign hosts), Multiclass (M);
\textbf{Classes:} Benign (Ben.), Attack (Att.);
\textbf{Attack families:}
Brute force (B), Botnet (C), (D)DoS/flooding (D), Malware, e.g.\ trojan, backdoor, cryptomining (M), Reconnaissance/scan (R), Exploitation, U2R (V), Web attacks (W), Other, e.g.\ fuzzing, stepping stone, flow-table overflow, blackhole, traffic diversion (O);
\cmark~present, \xmark~not present, --~not specified or not documented.
}
\end{table}
\subsection{Software-Defined and Cloud networks}\label{sec:rev-sdn}

Table~\ref{tab:rev-sdn} lists the five \gls{sdn} and cloud datasets in the \Obs{}. These datasets capture traffic in virtualized environments, where the monitoring point (controller, virtual switch or tenant network) determines what can be observed.
\emph{Some datasets use representations that exist only in \gls{sdn}}: two of the three \gls{sdn} datasets, \texttt{SDN DDoS} and \texttt{UNR-IDD}, do not export flows from packets but statistics collected by the controller itself (Ryu, ONOS), which cannot be compared with IP-flow features (\emph{Proc. Tool} column).
\emph{Most datasets are emulated}: the three \gls{sdn} datasets are emulated in Mininet, \texttt{BCCC-cPacket-Cloud-DDoS-2024} comes from a cloud testbed, and only \texttt{ISOT-CID} records a real cloud environment (\emph{Acq.} column). That dataset, however, labels packets as binary from the list of benign hosts, merging about $16$ attack types into a single class (\emph{Lab.} and \emph{Att.} columns).
\emph{(D)DoS dominates the attack families}: it is present in every dataset and is the only family in two of them, while the Other family, which includes \gls{sdn}-specific attacks such as flow-table overflow or traffic diversion, appears only in \texttt{ISOT-CID} and \texttt{UNR-IDD} (\emph{Attack families} columns).
Overall, the group is small and mostly emulated, and its only real-traffic dataset has the coarsest labels.

\begin{table}[!htbp]
\centering\scriptsize\setlength{\tabcolsep}{2pt}
\providecommand{\acqR}{\CIRCLE}
\providecommand{\acqH}{\LEFTcircle}
\providecommand{\acqE}{\Circle}
\providecommand{\deriv}{\smash{\raisebox{0.32em}{\tikz[baseline=0pt]{\draw[line width=0.4pt,line cap=round] (0,0) -- (0,0.58em) (-0.16em,0.40em) -- (0.16em,0.40em);}}}}
\providecommand{\cmark}{\textcolor{green!55!black}{\ding{51}}}
\providecommand{\xmark}{\textcolor{red!55!black}{\ding{55}}}
\caption{Backbone, ISP and honeypot datasets in the \Obs.}
\label{tab:rev-backbone}\label{tab:rev-honeypot}
\begin{tabular}{llcccccccccccccccccc}
\toprule
\multirow{2}{*}{\textbf{Dataset}} &
\multirow{2}{*}{\textbf{Year$\downarrow$}} &
\multicolumn{3}{c}{\textbf{Artifact}} &
\multirow{2}{*}{\textbf{Access}} &
\multirow{2}{*}{\textbf{Acq.}} &
\multirow{2}{*}{\textbf{Lab.}} &
\multicolumn{2}{c}{\textbf{Classes}} &
\multicolumn{9}{c}{\textbf{Attack families}} \\
\cmidrule(lr){3-5}\cmidrule(lr){9-10}\cmidrule(lr){11-19}
 & & \textbf{Raw} & \textbf{Proc.} & \textbf{Proc. Tool} & & & & \textbf{Ben.} & \textbf{Att.} & \textbf{B} & \textbf{C} & \textbf{D} & \textbf{L} & \textbf{M} & \textbf{R} & \textbf{V} & \textbf{W} & \textbf{O} \\
\midrule
\texttt{MAWI samplepoint-F}~\cite{cho2000mawi} & 2006 & \cmark & \xmark & -- & \faLockOpen & \acqR & -- & -- & -- & \multicolumn{9}{c}{--} \\
\rowcolor[HTML]{f0f0f0}
\texttt{Kyoto 2006+}~\cite{song2011kyoto} & 2006 & \xmark & \cmark & Zeek & \faLockOpen & \acqR & W & \cmark & 2 & \multicolumn{9}{c}{--} \\
\texttt{CAIDA Anonymized Traces}~\cite{caidaanon} & 2008--19 & \cmark & \xmark & -- & \faEnvelope & \acqR & -- & -- & -- & \multicolumn{9}{c}{--} \\
\rowcolor[HTML]{f0f0f0}
\texttt{Twente Labeled}~\cite{sperotto2009twente} & 2008 & \xmark & \cmark & softflowd & \faEnvelope & \acqR & W & \xmark & 3 & \cmark&\xmark&\xmark&\xmark&\xmark&\cmark&\xmark&\cmark&\xmark \\
\texttt{MAWILab}~\cite{fontugne2010mawilab}\ \deriv{}[\cite{cho2000mawi}] & 2010 & \cmark & \cmark & SCANN & \faLockOpen & \acqR & W & \cmark & 3 & \xmark&\xmark&\cmark&\xmark&\xmark&\cmark&\xmark&\xmark&\cmark \\
\rowcolor[HTML]{f0f0f0}
\texttt{UGR'16}~\cite{macia2018ugr16} & 2016 & \xmark & \cmark & nfdump & \faLockOpen & \acqR & W & \cmark & 8 & \xmark&\cmark&\cmark&\xmark&\xmark&\cmark&\xmark&\xmark&\cmark \\
\texttt{LITNET-2020}~\cite{damasevicius2020litnet} & 2020 & \xmark & \cmark & nfdump & \faLockOpen & \acqR & M & \cmark & 12 & \xmark&\xmark&\cmark&\xmark&\cmark&\cmark&\xmark&\xmark&\cmark \\
\rowcolor[HTML]{f0f0f0}
\texttt{RoEduNet-SIMARGL2021}~\cite{mihailescu2021simargl} & 2021 & \xmark & \cmark & nProbe & \faLockOpen & \acqH & M & \cmark & 3 & \xmark&\xmark&\cmark&\xmark&\xmark&\cmark&\xmark&\xmark&\xmark \\
\texttt{Hornet 15}~\cite{valeros2022hornet15} & 2021 & \xmark & \cmark & Argus & \faLockOpen & \acqR & -- & -- & -- & \multicolumn{9}{c}{--} \\
\rowcolor[HTML]{f0f0f0}
\texttt{FLNET2023}~\cite{flnet2023} & 2023 & \cmark & \cmark & CFM & \faLockOpen & \acqE & M & \cmark & 10 & \xmark&\xmark&\cmark&\cmark&\xmark&\xmark&\xmark&\cmark&\xmark \\
\texttt{CTU Hornet 65 Niner}~\cite{hornet65niner2024} & 2024 & \xmark & \cmark & Zeek & \faLockOpen & \acqR & -- & -- & -- & \multicolumn{9}{c}{--} \\
\bottomrule
\end{tabular}

\smallskip
\parbox{\linewidth}{\scriptsize\raggedright
\textbf{N.B:}
\textbf{Artifact:} Raw Network Traces (Raw), Processed Records (Proc.);
\textbf{Proc. Tool:} tool that produced the processed records, CFM = CICFlowMeter; Zeek includes its former name Bro;
\textbf{Access:} \faLockOpen~open; \faUser~registration; \faEnvelope~on request;
\textbf{Acquisition (Acq.):} \acqR~real traffic: backbone, ISP or campus links (UGR'16: real ISP traffic with attacks injected in the last 30 days) or Internet-facing honeypots (Kyoto University; University of Twente; 8 cloud honeypots in 8 cities for Hornet 15, 9 for CTU Hornet 65 Niner); \acqH~real network traffic with attacks from a lab setup; \acqE~emulated ISP-like topology;
\textbf{Labeling (Lab.):} Multiclass (M); Weak/heuristic (W; MAWILab: detector consensus;   UGR'16: part of the labels derived from blacklists and anomaly detectors; Kyoto: IDS, antivirus and shellcode-detector alerts; Twente: IDS alerts semi-automatically correlated to flows); None (--; for honeypots, all traffic is unsolicited);
\textbf{Classes:} Benign (Ben.), Attack/anomaly classes (Att.; Kyoto: known and unknown attack);
\textbf{Attack families:}
Brute force, e.g.\ SSH/FTP (B), Botnet (C), (D)DoS/flooding (D), Infiltration (L), Worms (M), Scan (R), Exploitation (V), Web attacks (W), Other, e.g.\ spam, blacklisted IPs, fragmentation, land attack, other MAWILab anomaly categories (O);
\deriv{}~derivative of an earlier capture, followed by the reference of its source;
\cmark~present, \xmark~not present, --~not specified, not documented or no attack labels.
}
\end{table}
\subsection{Backbone, ISP and honeypot traffic}\label{sec:rev-backbone}

Table~\ref{tab:rev-backbone} lists the datasets collected on backbone and ISP links and on Internet-facing honeypots.
\emph{Almost all were recorded in real networks}: nine of the $11$, with \texttt{MAWI} and \texttt{CAIDA} as long-running archives; the only exceptions are \texttt{RoEduNet-SIMARGL2021}, whose attacks were generated in a lab, and \texttt{FLNET2023}, which emulates an ISP-like topology (\emph{Acq.} column).
\emph{Realism comes at the cost of ground truth}: \texttt{MAWI}, \texttt{CAIDA} and the two Hornet collections carry no labels, while \texttt{MAWILab}, \texttt{UGR'16}, \texttt{Kyoto 2006+} and the \texttt{Twente dataset} are labeled weakly, by detector consensus, IDS alerts or the separation between background and injected attacks. 
Only \texttt{LITNET-2020}, \texttt{RoEduNet-SIMARGL2021} and \texttt{FLNET2023} provide multiclass labels, and two of them do not come from a fully real network (\emph{Lab.} column).
\emph{Most datasets release only flows or logs}: packet traces are available for \texttt{MAWI}, \texttt{CAIDA} (on request), \texttt{MAWILab} and \texttt{FLNET2023}, whereas the others release flows or connection logs produced by nfdump, nProbe, softflowd, Argus or Zeek, so the unit of analysis is fixed by the extractor and differs from one dataset to another (\emph{Artifact} and \emph{Proc. Tool} columns). \texttt{MAWILab}, moreover, is not a new capture but a labeling layer over the \texttt{MAWI} archive.
\emph{Where labels exist, flooding and scanning dominate}: (D)DoS and reconnaissance each appear in five datasets, while other families appear in one or two, such as brute force and web attacks in the \texttt{Twente dataset}, botnet in \texttt{UGR'16}, malware in \texttt{LITNET-2020}, and spam and blacklisted hosts (O) in \texttt{MAWILab}, \texttt{UGR'16} and \texttt{LITNET-2020} (\emph{Attack families} columns).
Overall, these collections are of limited use as labeled attack benchmarks, but they are among the few sources of operational and long-running traffic. This makes them a natural basis for self-supervised pretraining, drift analysis and the assessment of false positives on realistic background traffic, as long as anonymization, truncation and the available timing and host metadata are taken into account.

{\scriptsize\setlength{\tabcolsep}{1pt}
\providecommand{\acqR}{\CIRCLE}
\providecommand{\acqH}{\LEFTcircle}
\providecommand{\acqE}{\Circle}
\DeclareDocumentCommand{\deriv}{g o}{%
  \smash{\raisebox{0.32em}{\tikz[baseline=0pt]{\draw[line width=0.4pt,line cap=round] (0,0) -- (0,0.58em) (-0.16em,0.40em) -- (0.16em,0.40em);}}}%
  \IfValueT{#2}{\ifmmode^{#2}\else\textsuperscript{#2}\fi}%
}
\providecommand{\cmark}{\textcolor{green!55!black}{\ding{51}}}
\providecommand{\xmark}{\textcolor{red!55!black}{\ding{55}}}

\centering
\begin{longtable}{llcccccccccccccccccccc}
\caption{Enterprise and general IP datasets in the \Obs.}\label{tab:rev-enterprise}\\
\toprule
\multirow{2}{*}{\textbf{Dataset}} & 
\multirow{2}{*}{\textbf{Year$\downarrow$}} &
\multicolumn{3}{c}{\textbf{Artifact}} & 
\multirow{2}{*}{\textbf{Access}} &
\multirow{2}{*}{\textbf{Acq.}} &
\multirow{2}{*}{\textbf{Lab.}} &
\multicolumn{2}{c}{\textbf{Classes}} & \multicolumn{10}{c}{\textbf{Attack families}} \\
\cmidrule(lr){3-5}\cmidrule(lr){9-10}\cmidrule(lr){11-20}
 &  & \textbf{Raw} & \textbf{Proc.} & \textbf{Proc. Tool} & &  &  & \textbf{Ben.} & \textbf{Att.} & \textbf{B} & \textbf{C} & \textbf{D} & \textbf{E} & \textbf{L} & \textbf{M} & \textbf{R} & \textbf{V} & \textbf{W} & \textbf{O} \\
\midrule
\endfirsthead

\multicolumn{20}{l}{\scriptsize\textit{Table~\thetable{} (continued)}}\\
\toprule
\multirow{2}{*}{\textbf{Dataset}} & 
\multirow{2}{*}{\textbf{Year$\downarrow$}} &
\multicolumn{3}{c}{\textbf{Artifact}} & 
\multirow{2}{*}{\textbf{Access}} &
\multirow{2}{*}{\textbf{Acq.}} &
\multirow{2}{*}{\textbf{Lab.}} &
\multicolumn{2}{c}{\textbf{Classes}} & \multicolumn{10}{c}{\textbf{Attack families}} \\
\cmidrule(lr){3-5}\cmidrule(lr){9-10}\cmidrule(lr){11-20}
 &  & \textbf{Tr.} & \textbf{Der.} & \textbf{Tool} & &  &  & \textbf{Ben.} & \textbf{Att.} & \textbf{B} & \textbf{C} & \textbf{D} & \textbf{E} & \textbf{L} & \textbf{M} & \textbf{R} & \textbf{V} & \textbf{W} & \textbf{O} \\
\midrule
\endhead

\bottomrule
\endfoot

\bottomrule
\endlastfoot

\texttt{DARPA 1998}~\cite{lippmann2000darpa98} 
& 1998 & \cmark & \xmark & -- & \faLockOpen & \acqE & M & \cmark & 4 & \cmark&\xmark&\cmark&\xmark&\xmark&\xmark&\cmark&\cmark&\xmark&\xmark \\
\rowcolor[HTML]{f0f0f0}
\texttt{DARPA 1999}~\cite{lippmann2000darpa99}
& 1999 & \cmark & \xmark & -- & \faLockOpen & \acqE & M & \cmark & 4 & \cmark&\xmark&\cmark&\xmark&\xmark&\xmark&\cmark&\cmark&\xmark&\xmark \\
\texttt{KDD Cup 1999}~\cite{kddcup99} & 1999 & \xmark & \cmark & Zeek & \faLockOpen & \acqE & M & \cmark & 4 & \cmark&\xmark&\cmark&\xmark&\xmark&\xmark&\cmark&\cmark&\xmark&\xmark \\
\rowcolor[HTML]{f0f0f0}
\texttt{gureKDDCup}~\cite{perona2008gurekdd} & 2008 & \xmark & \cmark & Zeek & \faLockOpen & \acqE & M & \cmark & 4 & \cmark&\xmark&\cmark&\xmark&\xmark&\xmark&\cmark&\cmark&\xmark&\xmark \\
\texttt{ASNM Datasets}~\cite{homoliak2020asnm} & 2009 & \xmark & \cmark & custom & \faUser & \acqH & B/M & \cmark & 1 & \xmark&\xmark&\xmark&\xmark&\xmark&\xmark&\xmark&\cmark&\xmark&\cmark \\
\rowcolor[HTML]{f0f0f0}
\texttt{NSL-KDD}~\cite{tavallaee2009} & 2009 & \xmark & \cmark & Zeek & \faUser & \acqE & M & \cmark & 4 & \cmark&\xmark&\cmark&\xmark&\xmark&\xmark&\cmark&\cmark&\xmark&\xmark \\
\texttt{MACCDC}~\cite{maccdc} & $2010$--$12$ & \cmark & \xmark & -- & \faLockOpen & \acqH & -- & -- & -- & \multicolumn{10}{c}{--} \\
\rowcolor[HTML]{f0f0f0}
\texttt{VAST Challenge 2011}~\cite{vast2011} & 2011 & \cmark & \xmark & -- & \faLockOpen & \acqE & W & -- & -- & \xmark&\xmark&\cmark&\xmark&\xmark&\xmark&\cmark&\xmark&\xmark&\cmark \\
\texttt{ISCXIDS2012}~\cite{shiravi2012} & 2012 & \cmark & \cmark & custom & \faUser & \acqH & B & \cmark & 1 & \cmark&\cmark&\cmark&\xmark&\cmark&\xmark&\xmark&\xmark&\xmark&\xmark \\
\rowcolor[HTML]{f0f0f0}
\texttt{CTU Normal Captures}~\cite{ctunormal}
& 2013--17 & \cmark & \cmark & Argus, Zeek & \faLockOpen & \acqR & -- & \cmark & 0 & \xmark&\xmark&\xmark&\xmark&\xmark&\xmark&\xmark&\xmark&\xmark&\xmark \\
\texttt{LANL}~\cite{kent2016lanl} & 2015 & \xmark & \cmark & NetFlow & \faUser & \acqR & W & \cmark & 1 & \xmark&\xmark&\xmark&\xmark&\cmark&\xmark&\xmark&\xmark&\xmark&\xmark \\
\rowcolor[HTML]{f0f0f0}
\texttt{UNSW-NB15}~\cite{moustafa2015} & 2015 & \cmark & \cmark & Argus, Zeek & \faLockOpen & \acqH & M & \cmark & 9 & \xmark&\xmark&\cmark&\xmark&\xmark&\cmark&\cmark&\cmark&\xmark&\cmark \\
\texttt{NDSec-1}~\cite{beer2017ndsec} & 2016 & \cmark & \cmark & YAF & \faLockOpen & \acqE & M & \cmark & 12 & \cmark&\cmark&\cmark&\xmark&\xmark&\xmark&\cmark&\cmark&\cmark&\cmark \\
\rowcolor[HTML]{f0f0f0}
\texttt{NGIDS-DS}~\cite{haider2017ngids} & 2016 & \cmark & \cmark & custom & \faLockOpen & \acqH & M & \cmark & 7 & \xmark&\xmark&\cmark&\xmark&\xmark&\cmark&\cmark&\cmark&\xmark&\cmark \\
\texttt{CIDDS-001}~\cite{ring2017cidds001} & 2017 & \xmark & \cmark & NetFlow & \faLockOpen & \acqH & M & \cmark & 4 & \cmark&\xmark&\cmark&\xmark&\xmark&\xmark&\cmark&\xmark&\xmark&\xmark \\
\rowcolor[HTML]{f0f0f0}
\texttt{CIDDS-002}~\cite{ring2017cidds002} & 2017 & \xmark & \cmark & NetFlow & \faLockOpen & \acqE & M & \cmark & 5 & \xmark&\xmark&\xmark&\xmark&\xmark&\xmark&\cmark&\xmark&\xmark&\xmark \\
\texttt{LANL UHN}~\cite{turcotte2018uhn}
& 2017 & \xmark & \cmark & NetFlow & \faUser & \acqR & -- & \cmark & 0 & \xmark&\xmark&\xmark&\xmark&\xmark&\xmark&\xmark&\xmark&\xmark&\xmark \\
\rowcolor[HTML]{f0f0f0}
\texttt{CICIDS2017}~\cite{sharafaldin2018} & 2017 & \cmark & \cmark & CFM & \faUser & \acqH & M & \cmark & 14 & \cmark&\cmark&\cmark&\xmark&\cmark&\xmark&\cmark&\cmark&\cmark&\xmark \\
\texttt{CSE-CIC-IDS2018}~\cite{cicids2018} & 2018 & \cmark & \cmark & CFM & \faLockOpen & \acqE & M & \cmark & 14 & \cmark&\cmark&\cmark&\xmark&\cmark&\xmark&\cmark&\cmark&\cmark&\xmark \\
\rowcolor[HTML]{f0f0f0}
\texttt{OTRF-Atomic}~\cite{otrf} & 2019 & \cmark & \xmark & -- & \faLockOpen & \acqE & W & \xmark & 7\textsuperscript{*} & \multicolumn{10}{c}{--} \\
\texttt{DAPT 2020}~\cite{myneni2020dapt} & 2020 & \cmark & \cmark & CFM & \faLockOpen & \acqH & M & \cmark & 4 & \xmark&\xmark&\xmark&\cmark&\cmark&\xmark&\cmark&\cmark&\xmark&\xmark \\
\rowcolor[HTML]{f0f0f0}
\texttt{LUFlow}~\cite{mills2022luflow} & 2020 & \xmark & \cmark & Joy & \faLockOpen & \acqR & W & \cmark & 2 & \multicolumn{10}{c}{--} \\
\texttt{I-Sec-IDS}~\cite{isecids2022} & 2021 & \xmark & \cmark & CFM & \faLockOpen & \acqE & M & \cmark & 2 & \xmark&\xmark&\cmark&\xmark&\xmark&\xmark&\cmark&\xmark&\xmark&\xmark \\
\rowcolor[HTML]{f0f0f0}
\texttt{NF-CSE-CIC-IDS}~\cite{dhoogla2023nfcsecicids2018}\ \deriv{}[\cite{cicids2018}] & 2021 & \xmark & \cmark & NFS & \faLockOpen & \acqE & M & \cmark & 6 & \cmark&\cmark&\cmark&\xmark&\cmark&\xmark&\xmark&\cmark&\cmark&\xmark \\
\texttt{Improved CICIDS2017}~\cite{engelen2021}\ \deriv{}[\cite{sharafaldin2018}] & 2021 & \cmark & \cmark & CFM & \faLockOpen & \acqH & M & \cmark & 14 & \cmark&\cmark&\cmark&\xmark&\cmark&\xmark&\cmark&\cmark&\cmark&\xmark \\
\rowcolor[HTML]{f0f0f0}
\texttt{LYCOS-IDS2017}~\cite{rosay2022lycos}\ \deriv{}[\cite{sharafaldin2018}] & 2021 & \xmark & \cmark & LST & \faLockOpen & \acqH & M & \cmark & 13 & \cmark&\cmark&\cmark&\xmark&\xmark&\xmark&\cmark&\cmark&\cmark&\xmark \\
\texttt{NF-CSE-CIC-IDS2018-v2}~\cite{sarhan2022standard}\ \deriv{}[\cite{cicids2018}] & 2021 & \xmark & \cmark & nProbe & \faLockOpen & \acqE & M & \cmark & 6 & \cmark&\cmark&\cmark&\xmark&\cmark&\xmark&\xmark&\cmark&\cmark&\xmark \\
\rowcolor[HTML]{f0f0f0}
\texttt{NF-UNSW-NB15-v2}~\cite{sarhan2022standard}\ \deriv{}[\cite{moustafa2015}] & 2021 & \xmark & \cmark & nProbe & \faLockOpen & \acqH & M & \cmark & 9 & \xmark&\xmark&\cmark&\xmark&\xmark&\cmark&\cmark&\cmark&\xmark&\cmark \\
\texttt{NF-UQ-NIDS-v2}~\cite{sarhan2022standard}\ \deriv{}* & 2021 & \xmark & \cmark & nProbe & \faLockOpen & \acqH & M & \cmark & 20 & \cmark&\cmark&\cmark&\xmark&\cmark&\cmark&\cmark&\cmark&\cmark&\cmark \\
\rowcolor[HTML]{f0f0f0}
\texttt{SCVIC-APT-2021}~\cite{liu2022scvic} & 2021 & \xmark & \cmark & CFM & \faUser & \acqE & M & \cmark & 5 & \xmark&\xmark&\xmark&\cmark&\cmark&\xmark&\cmark&\cmark&\xmark&\xmark \\
\texttt{Unraveled}~\cite{myneni2023unraveled} & 2021 & \cmark & \cmark & NFS & \faLockOpen & \acqE & M & \cmark & 5 & \cmark&\cmark&\xmark&\cmark&\cmark&\xmark&\cmark&\cmark&\xmark&\xmark \\
\rowcolor[HTML]{f0f0f0}
\texttt{AIT-LDSv2}~\cite{landauer2023ait}
& 2022 & \cmark & \xmark & -- & \faLockOpen & \acqE & M & \cmark & 10 & \xmark&\xmark&\xmark&\cmark&\cmark&\xmark&\cmark&\cmark&\cmark&\xmark \\
\texttt{AnoShift}~\cite{dragoi2022anoshift}\ \deriv{}[\cite{song2011kyoto}] & 2022 & \xmark & \cmark & custom & \faLockOpen & \acqR & W & \cmark & 2 & \multicolumn{10}{c}{--} \\
\rowcolor[HTML]{f0f0f0}
\texttt{CUPID}~\cite{lawrence2022cupid} & 2022 & \cmark & \cmark & CFM & \faLockOpen & \acqH & B & \cmark & 1 & \cmark&\xmark&\cmark&\xmark&\xmark&\xmark&\cmark&\cmark&\cmark&\xmark \\
\texttt{Improved CSE-CIC-IDS2018}~\cite{liu2022error}\ \deriv{}[\cite{cicids2018}] & 2022 & \xmark & \cmark & CFM & \faLockOpen & \acqE & M & \cmark & -- & \cmark&\cmark&\cmark&\xmark&\cmark&\xmark&\cmark&\cmark&\cmark&\xmark \\
\rowcolor[HTML]{f0f0f0}
\texttt{OD-IDS2022}~\cite{odids2022} & 2022 & \xmark & \cmark & CFM & \faEnvelope & \acqE & M & \cmark & 28 & \cmark&\xmark&\cmark&\xmark&\xmark&\xmark&\xmark&\cmark&\cmark&\xmark \\
\texttt{PWNJUTSU}~\cite{berady2022pwnjutsu} & 2022 & \cmark & \xmark & -- & \faLockOpen & \acqE & -- & \xmark & -- & \xmark&\xmark&\xmark&\xmark&\cmark&\xmark&\cmark&\cmark&\xmark&\xmark \\
\rowcolor[HTML]{f0f0f0}
\texttt{UWF-ZeekData22}~\cite{bagui2023zeek22} & 2022 & \cmark & \cmark & Zeek & \faLockOpen & \acqE & M & \cmark & 10\textsuperscript{*} & \xmark&\xmark&\xmark&\cmark&\cmark&\xmark&\cmark&\cmark&\xmark&\cmark \\
\texttt{UWF-ZeekDataFall22}~\cite{bagui2023fall22} & 2022 & \xmark & \cmark & Zeek & \faLockOpen & \acqE & M & \cmark & 12\textsuperscript{*} & \xmark&\xmark&\xmark&\xmark&\xmark&\xmark&\cmark&\xmark&\xmark&\cmark \\
\rowcolor[HTML]{f0f0f0}
\texttt{APPRAISE-H2020}~\cite{appraise2023}
& 2023 & \xmark & \cmark & nProbe & \faLockOpen & \acqH & M & \cmark & 3 & \cmark&\xmark&\cmark&\xmark&\xmark&\xmark&\cmark&\xmark&\xmark&\xmark \\
\texttt{CREMEv2}~\cite{cremev2} & 2023 & \xmark & \cmark & Argus & \faLockOpen & \acqE & M & -- & 6 & \cmark&\cmark&\cmark&\xmark&\xmark&\cmark&\cmark&\cmark&\xmark&\xmark \\
\rowcolor[HTML]{f0f0f0}
\texttt{CSE-CIC-IDS2018 v2}~\cite{abluva_cse_cic_ids2018_v2}\ \deriv{}[\cite{cicids2018}] & 2023 & \xmark & \cmark & CFM & \faLockOpen & \acqE & M & \cmark & 15 & \cmark&\cmark&\cmark&\xmark&\cmark&\xmark&\cmark&\cmark&\cmark&\xmark \\
\texttt{CTU-SME-11}~\cite{bendl2023ctusme11} & 2023 & \cmark & \cmark & Zeek & \faLockOpen & \acqR & M & \cmark & 2 & \xmark&\cmark&\xmark&\xmark&\xmark&\cmark&\cmark&\xmark&\xmark&\xmark \\
\rowcolor[HTML]{f0f0f0}
\texttt{OTRF-LSASS}~\cite{otrf}
& 2023 & \cmark & \cmark & Zeek & \faLockOpen & \acqE & -- & \xmark & --\textsuperscript{*} & \xmark&\xmark&\xmark&\xmark&\xmark&\xmark&\xmark&\xmark&\xmark&\cmark \\
\texttt{TII-SSRC-23}~\cite{herzalla2023tii} & 2023 & \cmark & \cmark & CFM & \faLockOpen & \acqH & M & \cmark\,(6) & 26 & \cmark&\cmark&\cmark&\xmark&\xmark&\xmark&\cmark&\xmark&\xmark&\xmark \\
\rowcolor[HTML]{f0f0f0}
\texttt{BCCC-CIC-IDS2017}~\cite{bccccicids2017}\ \deriv{}[\cite{sharafaldin2018}] & 2024 & \xmark & \cmark & NTL & \faEnvelope & \acqH & M & \cmark & -- & \cmark&\cmark&\cmark&\xmark&\cmark&\xmark&\cmark&\cmark&\cmark&\xmark \\
\texttt{CasinoLimit}~\cite{casinolimit2025} & 2024 & \xmark & \cmark & Zeek & \faLockOpen & \acqE & M & \xmark & 12\textsuperscript{*} & \xmark&\cmark&\xmark&\cmark&\cmark&\xmark&\cmark&\cmark&\xmark&\cmark \\
\rowcolor[HTML]{f0f0f0}
\texttt{CIC UNSW-NB15}~\cite{cicunswnb15}
\ \deriv{}[\cite{moustafa2015}] & 2024 & \xmark & \cmark & CFM & \faUser & \acqH & M & \cmark & 9 & \xmark&\xmark&\cmark&\xmark&\xmark&\cmark&\cmark&\cmark&\xmark&\cmark \\
\texttt{LycoS-Unicas-IDS2018}~\cite{lycosunicas2024}\ \deriv{}[\cite{cicids2018}] & 2024 & \xmark & \cmark & LST & \faLockOpen & \acqE & M & \cmark & 13 & \cmark&\cmark&\cmark&\xmark&\xmark&\xmark&\xmark&\xmark&\cmark&\xmark \\
\rowcolor[HTML]{f0f0f0}
\texttt{UWF-ZeekData24}~\cite{bagui2025zeek24} & 2024 & \xmark & \cmark & Zeek & \faLockOpen & \acqE & M & \cmark & 7\textsuperscript{*} & \xmark&\xmark&\xmark&\cmark&\xmark&\xmark&\cmark&\cmark&\xmark&\cmark \\
\texttt{BCCC-CSE-CIC-IDS2018}~\cite{bccccseids2018}\ \deriv{}[\cite{cicids2018}] & 2025 & \xmark & \cmark & NTL & \faEnvelope & \acqE & M & \cmark & -- & \cmark&\cmark&\cmark&\xmark&\cmark&\xmark&\cmark&\cmark&\cmark&\xmark \\
\rowcolor[HTML]{f0f0f0}
\texttt{DEDALE}~\cite{dedale2025} & 2025 & \cmark & \cmark & Zeek & \faLockOpen & \acqE & M & \cmark & 2 & \xmark&\cmark&\xmark&\cmark&\cmark&\cmark&\cmark&\cmark&\xmark&\cmark \\
\texttt{GeNIS}~\cite{genis2025} & 2025 & \cmark & \cmark & HERA, Argus & \faLockOpen & \acqH & M & \cmark & 3 & \cmark&\xmark&\cmark&\xmark&\xmark&\xmark&\cmark&\xmark&\xmark&\xmark \\
\rowcolor[HTML]{f0f0f0}
\texttt{NF-CSE-CIC-IDS2018-v3}~\cite{luay2025nfv3}\ \deriv{}[\cite{cicids2018}] & 2025 & \xmark & \cmark & nProbe & \faLockOpen & \acqE & M & \cmark & 6 & \cmark&\cmark&\cmark&\xmark&\cmark&\xmark&\xmark&\cmark&\cmark&\xmark \\
\texttt{NF-UNSW-NB15-v3}~\cite{luay2025nfv3}\ \deriv{}[\cite{moustafa2015}] & 2025 & \xmark & \cmark & nProbe & \faLockOpen & \acqH & M & \cmark & 9 & \xmark&\xmark&\cmark&\xmark&\xmark&\cmark&\cmark&\cmark&\xmark&\cmark \\
\rowcolor[HTML]{f0f0f0}
\texttt{SAFE-NID}~\cite{safenid2025}\ \deriv{}[\cite{sharafaldin2018,moustafa2015}] & 2025 & \cmark & \xmark & -- & \faLockOpen & \acqH & M & \cmark & 14 / 9 & \cmark&\cmark&\cmark&\xmark&\cmark&\cmark&\cmark&\cmark&\cmark&\cmark \\
\texttt{CAM-LDS}~\cite{camlds2026}
& 2026 & \cmark & \xmark & -- & \faLockOpen & \acqE & M & \xmark & 13\textsuperscript{*} & \cmark&\xmark&\xmark&\cmark&\cmark&\cmark&\cmark&\cmark&\xmark&\cmark \\

\end{longtable}

\smallskip
\parbox{\linewidth}{\scriptsize\raggedright
\textbf{N.B:}
\textbf{Artifact:} Raw Network Traces (Raw), Processed Records (Proc.);
\textbf{Proc. Tool}: CICFlowMeter (CFM), including its modified/fixed versions; LycoSTand (LST); NTLFlowLyzer (NTL); NFStream (NFS);
\textbf{Access:} \faLockOpen~open; \faUser~registration; \faEnvelope~ on request;
\textbf{Acquisition (Acq.):} \acqR~real devices; \acqH~real and simulated devices; \acqE~emulated/simulated;
\textbf{Labeling (Lab.):} Binary (B), Multiclass (M); Weak/heuristic (W); None (--);
\textbf{Classes:} Benign (Ben.), Attack (Att.: number of attack types; \textsuperscript{*}~number of MITRE ATT\&CK tactics or techniques used as labels, mapped to the attack families);
\textbf{Attack Families:}
Brute force (B), Botnet/C2 (C), (D)DoS (D), Exfiltration (E), Infiltration/lateral movement (L), Malware/backdoor/worm (M), Reconnaissance/scan (R), Exploitation (V), Web (W), Other, incl.\ persistence, defense evasion, collection and credential theft (O);
\deriv{}~derivative of an earlier capture, followed by the reference of its source;
\cmark~present, \xmark~not present, --~not specified or not documented;
obtained by merging \texttt{NF-UNSW-NB15-v2}, \texttt{NF-ToN-IoT-v2}, \texttt{NF-BoT-IoT-v2}, and \texttt{NF-CSE-CIC-IDS2018-v2}.
}
}

\subsection{Enterprise networks and general-purpose benchmarks}\label{sec:rev-enterprise}

Table~\ref{tab:rev-enterprise} lists the $57$ enterprise and general IP datasets in the \Obs{}, from \texttt{DARPA 1998} to \texttt{CAM-LDS}.
\emph{The group is old but still growing}: $35$ datasets were released from $2021$ onwards. Sixteen of them, however, are derivatives, mostly of \texttt{CICIDS2017}, \texttt{CSE-CIC-IDS2018} and \texttt{UNSW-NB15}, so new entries often repackage existing captures rather than add new traffic (\emph{Year} column and derivative mark).
\emph{Most datasets distribute processed records only}: $31$ of the $57$ do so, which ties every result to the extractor, most often CICFlowMeter or Zeek; the errors of the former are well documented~\citep{engelen2021} (\emph{Artifact} and \emph{Proc. Tool} columns).
\emph{Real traffic is rare}: most datasets are emulated or come from testbeds, only six were recorded in real networks, and of these only \texttt{CTU-SME-11} has multiclass attack labels (\emph{Acq.} column).
\emph{Class counts are not comparable across datasets}: labels are usually multiclass and include benign traffic, but the number of attack classes ranges from $0$ to $28$, depending on whether attacks are split into subtypes or grouped into MITRE ATT\&CK tactics, as in several recent releases (\emph{Lab.} and \emph{Att.} columns).
\emph{The attack families are unbalanced}: reconnaissance ($42$ datasets), exploitation ($38$), (D)DoS ($35$) and brute force ($30$) dominate, while malware ($11$) and exfiltration ($9$) are rare (\emph{Attack families} columns).
Overall, the group offers abundant but heavily reused data: models trained and tested on different members of the CIC lineage are not evaluated on independent data.

\begin{table}[!htbp]
\centering\scriptsize\setlength{\tabcolsep}{2pt}
\providecommand{\acqR}{\CIRCLE}
\providecommand{\acqH}{\LEFTcircle}
\providecommand{\acqE}{\Circle}
\providecommand{\deriv}{\smash{\raisebox{0.32em}{\tikz[baseline=0pt]{\draw[line width=0.4pt,line cap=round] (0,0) -- (0,0.58em) (-0.16em,0.40em) -- (0.16em,0.40em);}}}}
\providecommand{\cmark}{\textcolor{green!55!black}{\ding{51}}}
\providecommand{\xmark}{\textcolor{red!55!black}{\ding{55}}}
\caption{Encrypted traffic datasets in the \Obs{}. 
}
\label{tab:rev-encrypted}
\begin{tabular}{llccccccccccccccc}
\toprule
\multirow{2}{*}{\textbf{Dataset}} &
\multirow{2}{*}{\textbf{Year$\downarrow$}} &
\multicolumn{3}{c}{\textbf{Artifact}} &
\multirow{2}{*}{\textbf{Access}} &
\multirow{2}{*}{\textbf{Acq.}} &
\multirow{2}{*}{\textbf{Lab.}} &
\multicolumn{2}{c}{\textbf{Classes}} & \multicolumn{7}{c}{\textbf{Attack families}}
\\
\cmidrule(lr){3-5}\cmidrule(lr){9-10}\cmidrule(lr){11-17}
 & & \textbf{Raw} & \textbf{Proc.} & \textbf{Proc. Tool} & & & & \textbf{Ben.} & \textbf{Att.} & \textbf{B} & \textbf{C} & \textbf{M} & \textbf{R} & \textbf{F} & \textbf{G} & \textbf{O} \\
\midrule
\texttt{USTC-TFC2016}~\cite{wang2017ustc} & 2016 & \cmark & \xmark & -- & \faLockOpen & mix & W & \cmark & 10 & \xmark&\cmark&\cmark&\xmark&\xmark&\xmark&\xmark \\
\rowcolor[HTML]{f0f0f0}
\texttt{HIKARI-2021}~\cite{hikari2021} & 2021 & \cmark & \cmark & Zeek & \faLockOpen & \acqH & M & \cmark & 4 & \cmark&\xmark&\cmark&\cmark&\xmark&\xmark&\xmark \\
\texttt{CIC-PQC\_OAV v1}~\cite{cicpqc2025} & 2025 & \xmark & \cmark & tshark & \faUser & \acqE & M & \cmark & 1 & \xmark&\xmark&\xmark&\xmark&\cmark&\cmark&\cmark \\
\rowcolor[HTML]{f0f0f0}
\texttt{TQH-C2}~\cite{tqhc22026} & 2026 & \cmark & \cmark & Zeek, JA4 & \faLockOpen & \acqE & M & -- & -- & \xmark&\cmark&\xmark&\xmark&\xmark&\xmark&\xmark \\
\bottomrule
\end{tabular}

\smallskip
\parbox{\linewidth}{\scriptsize\raggedright
\textbf{N.B:}
\textbf{Artifact:} Raw Network Traces (Raw), Processed Records (Proc.);
\textbf{Processing Tool (Proc. Tool):} tool that produced the processed records;
\textbf{Access:} \faLockOpen~open; \faUser~registration; 
\textbf{Acquisition (Acq.):} \acqH~lab capture with real users/applications (incl.\ real background traffic); \acqE~controlled testbed (C2 frameworks, PQC TLS sessions); mix = merge of sources;
\textbf{Labeling (Lab.):} Multiclass (M); Weak, per-file labels (W);
\textbf{Classes:} Benign (Ben.), Attack (Att.);
\textbf{Attack families:}
Brute force (B), Malware C2/beaconing (C), Malware, incl.\ cryptominer (M), Probing/scan (R), Fuzzing (F), Downgrade/side-channel mimicry (G), Other, e.g.\ network impairment (O);
\cmark~present, \xmark~not present, --~not specified or not documented.
}
\end{table}
\subsection{Encrypted traffic}\label{sec:rev-encrypted}

Table~\ref{tab:rev-encrypted} lists the four encrypted-traffic datasets in the \Obs{} whose labels identify attacks.%
\footnote{Datasets of encrypted traffic labeled only by application or service, such as \texttt{ISCXVPN2016}, \texttt{ISCXTor2016} and \texttt{CESNET-TLS22}, are widely used for traffic classification but contain no attacks, and are therefore not included.}
Encrypted traffic is not limited to this section: many datasets of other domains contain TLS or QUIC sessions but are assigned to the domain of their capture environment or threat.
Encryption hides the payload but leaves timing, size, direction, endpoint behavior and handshake metadata available to different degrees, so the feature-extraction process becomes part of the scientific claim.
\emph{Three of the four datasets release packet traces}: only \texttt{CIC-PQC\_OAV} distributes derived session records alone, as an aggregate \gls{csv} from which the packet sequences used by newer architectures cannot be recreated (\emph{Artifact} column). Zeek and tshark are used where documented, and only the most recent release, \texttt{TQH-C2}, adds TLS fingerprints (JA4) (\emph{Proc. Tool} column).
\emph{Acquisition settings differ}: from lab captures with real background traffic (\texttt{HIKARI-2021}) to controlled testbeds that run C2 frameworks or post-quantum TLS sessions, while \texttt{USTC-TFC2016} merges captures from different sources (\emph{Acq.} column).
\emph{The attack families show little overlap}: malware and C2 traffic in \texttt{USTC-TFC2016}, brute force, malware and probing in \texttt{HIKARI-2021}, C2 beaconing alone in \texttt{TQH-C2}, and fuzzing or downgrade attacks on post-quantum handshakes in \texttt{CIC-PQC\_OAV}; no two datasets cover the same set of families (\emph{Attack families} columns).
Overall, encrypted-traffic security datasets are few and heterogeneous, since most encrypted corpora target application identification rather than attack detection.

\begin{table}[!t]\centering\footnotesize
\caption{Uses of the catalog for AI research. \textit{Required} is the column value a dataset must have, and \textit{Datasets} counts the records of release \release{} ($n=\NFinal{}$) that have it; \textit{Related columns} narrow these candidates further before a dataset is chosen.}\label{tab:paradigms}
\begin{tabularx}{\linewidth}{@{}c>{\raggedright\arraybackslash}p{3cm}>{\raggedright\arraybackslash}p{3cm}>{\raggedright\arraybackslash}p{2cm}>{\raggedright\arraybackslash}X r@{}}
\toprule
 & \textbf{Use} & \textbf{Required} & \textbf{Related columns} & \textbf{Caution} & \textbf{Datasets} \\
\midrule
 & Supervised classification & Lab.\ = B, M or ML & Ben., Att., Attack families & Weak labels (W), assigned per file, time interval or IDS alert, are coarse or noisy; class definitions differ across datasets, so multiclass results are not directly comparable & 184 \\
\rowcolor[HTML]{f0f0f0}
\cellcolor{white} & Anomaly detection & Ben.\ = \checkmark & Acq. & Testbed benign traffic comes from few, often scripted hosts and inflates detection rates; benign traffic from real networks may contain undetected attacks & 201 \\
\multirow{-9}{*}{\rotatebox[origin=c]{90}{\textbf{Learning setting}}}
 & Self-supervised pretraining & Raw = \checkmark & Acq., lineage (\deriv{}) & Pretraining on a capture and testing on one of its derivatives (\deriv{}) leaks test data & 141 \\
\midrule
\rowcolor[HTML]{f0f0f0}
\cellcolor{white} & Packet-level and sequence models on IP traffic & Raw = \checkmark, domain $\neq$ \{automotive, V2X\} & Proc.\ Tool, Lab. & Labels are often given per flow in the processed records and must be mapped back to packets & 113 \\
\multirow{-3}{*}{\rotatebox[origin=c]{90}{\begin{tabular}{@{}c@{}}\textbf{Input}\\ \textbf{representation}\end{tabular}}}
 & CAN and V2X models & Domain = \{automotive or V2X\} & Raw, Acq., Lab. & CAN identifiers and payloads are vehicle-specific and attacks are injected; all V2X datasets are simulated & 30 \\
\bottomrule
\end{tabularx}
\end{table}

\section{Using the Catalog for AI Research}\label{sec:suitability}\label{sec:guide}

This section shows how to use the \Obs{} to find datasets suited to a given \gls{ai} experiment.
Section~\ref{sec:procedure} relates the columns of the domain tables to common uses in AI research and turns them into a step-by-step selection procedure, and Section~\ref{sec:usecases} applies the procedure to three use cases, from choosing a benchmark to planning a new release.
Since the catalog is continuously updated, the same procedure can be applied to every later release.

\subsection{Selection procedure}\label{sec:procedure}
Table~\ref{tab:paradigms} considers five common uses of the catalog, grouped by learning setting and by input representation; for each, it gives the column value a dataset must have, the related columns to inspect, the main caution, and how many records of release \release{} meet the requirement.
The counts show that the choice is narrower than the size of the catalog suggests: only \NPcap{} of the \NFinal{} datasets distribute network traces, and 30 records are derivatives of 14 earlier captures rather than independent data (Sec.~\ref{sec:pitfalls}).

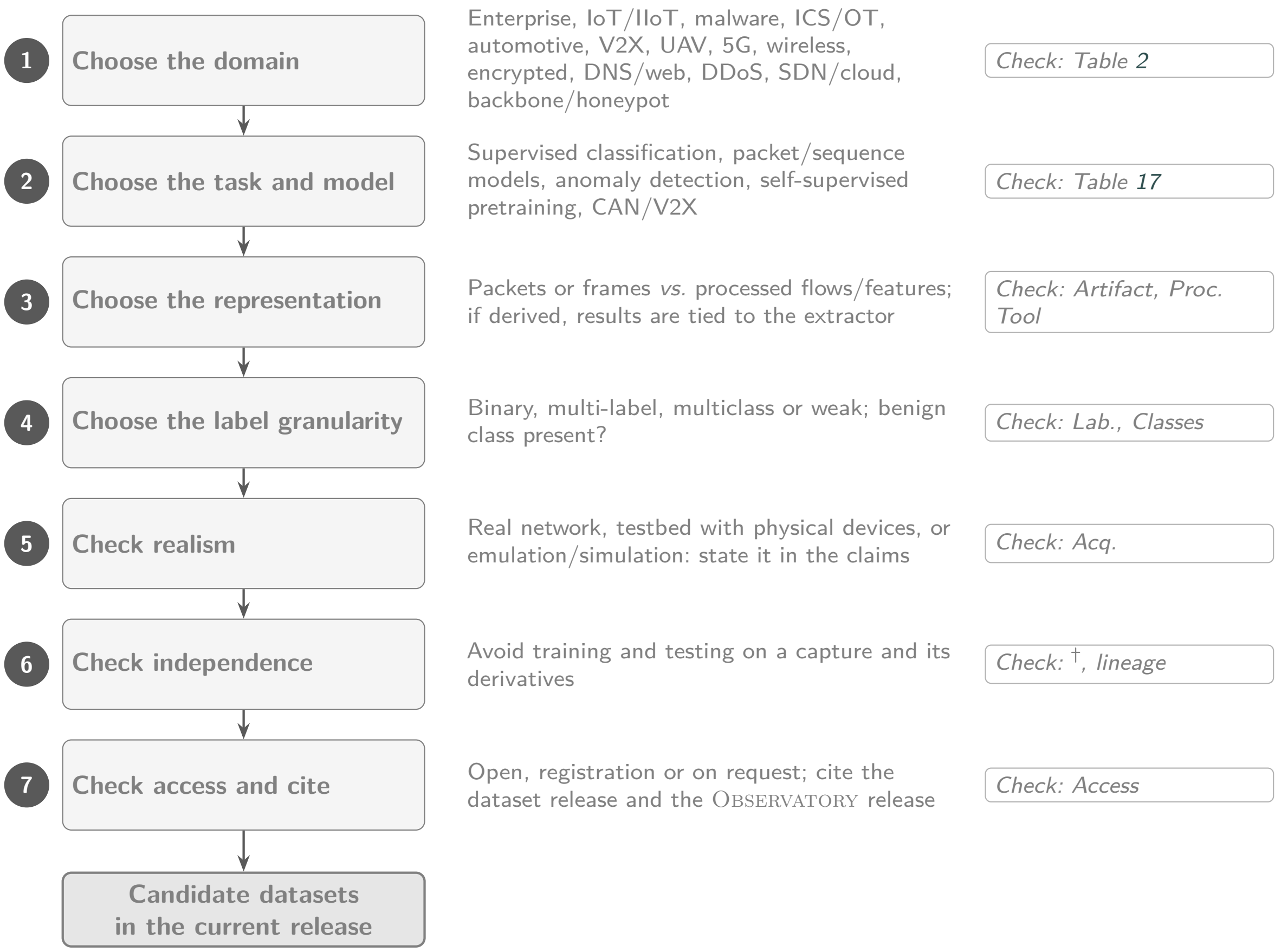
\begin{figure}[!t]
\centering
\begin{tikzpicture}[
  node distance=0.36cm,
  selstep/.style={draw, rounded corners=3pt, fill=gray!8, text width=4.2cm, minimum height=1.1cm, align=left, font=\small},
  selnum/.style={circle, fill=gray!70!black, text=white, font=\footnotesize\bfseries, inner sep=1.5pt, minimum size=0.55cm},
  selopts/.style={text width=5.9cm, align=left, font=\footnotesize},
  selchk/.style={draw=gray!60, rounded corners=2pt, text width=3.3cm, align=left, font=\footnotesize\itshape, fill=white},
  selarr/.style={-{Stealth[length=2mm]}, thick, gray!70!black}]

\node[selstep] (s1) {\textbf{Choose the domain}};
\node[selstep, below=of s1] (s2) {\textbf{Choose the task and model}};
\node[selstep, below=of s2] (s3) {\textbf{Choose the representation}};
\node[selstep, below=of s3] (s4) {\textbf{Choose the label granularity}};
\node[selstep, below=of s4] (s5) {\textbf{Check realism}};
\node[selstep, below=of s5] (s6) {\textbf{Check independence}};
\node[selstep, below=of s6] (s7) {\textbf{Check access and cite}};
\node[draw, thick, rounded corners=3pt, fill=gray!20, text width=4.2cm, minimum height=0.9cm, align=center, font=\small\bfseries, below=0.5cm of s7] (out) {Candidate datasets\\in the current release};

\foreach \i in {1,...,7} {\node[selnum, left=0.15cm of s\i] {\i};}
\foreach \a/\b in {s1/s2,s2/s3,s3/s4,s4/s5,s5/s6,s6/s7,s7/out} {\draw[selarr] (\a) -- (\b);}

\node[selopts, right=0.4cm of s1] (o1) {Enterprise, IoT/IIoT, malware, ICS/OT, automotive, V2X, UAV, 5G, wireless, encrypted, DNS/web, DDoS, SDN/cloud, backbone/honeypot};
\node[selopts, right=0.4cm of s2] (o2) {Supervised classification, packet/sequence models, anomaly detection, self-supervised pretraining, CAN/V2X};
\node[selopts, right=0.4cm of s3] (o3) {Packets or frames \emph{vs.} processed flows/features; if derived, results are tied to the extractor};
\node[selopts, right=0.4cm of s4] (o4) {Binary, multi-label, multiclass or weak; benign class present?};
\node[selopts, right=0.4cm of s5] (o5) {Real network, testbed with physical devices, or emulation/simulation: state it in the claims};
\node[selopts, right=0.4cm of s6] (o6) {Avoid training and testing on a capture and its derivatives};
\node[selopts, right=0.4cm of s7] (o7) {Open, registration or on request; cite the dataset release and the \Obs{} release};

\foreach \i/\c in {1/{Table~\ref{tab:domain}},2/{Table~\ref{tab:paradigms}},3/{Artifact, Proc. Tool},4/{Lab., Classes},5/{Acq.},6/{\deriv{}, lineage},7/{Access}} {
  \node[selchk, right=0.3cm of o\i] {Check: \c};
}
\end{tikzpicture}
\caption{Procedure for selecting a dataset in the \Obs{}. Each step narrows the candidates using one or more columns of the domain tables (``Check''), which correspond to the filters of the \Obs{} website. Because the catalog is continuously updated, the procedure should be applied to its current release.
}
\label{fig:selection}
\end{figure}

Figure~\ref{fig:selection} turns the taxonomy into a seven-step procedure. Each step narrows the candidates using one or more columns of the domain tables, which correspond to the filters of the \Obs{} website.
The first two steps fix the scope. The domain selects one of the tables of Secs.~\ref{sec:rev-iot}--\ref{sec:rev-encrypted}.
Then, the task and model determine, through Table~\ref{tab:paradigms}, which properties a dataset must have.
The next two steps determine what a model can learn. The \emph{Artifact} column separates datasets that distribute network traces from those that release only processed records, whose features are fixed by the tool in \emph{Processing Tool}; \emph{Labeling} and \emph{Classes} indicate whether the labels are fine enough for the intended task and whether a benign class is available.
The last three steps check whether a result obtained on the dataset can be trusted, compared and reproduced. \emph{Acquisition} states whether the traffic comes from a real network, a testbed or a simulation, which bounds the claims that can be made; the lineage mark (\deriv{}) reveals datasets derived from the same capture, which should not be treated as independent; and \emph{Access} indicates whether other researchers can obtain the same data.

\subsection{Use cases}\label{sec:usecases}

The following three cases show how the procedure of Fig.~\ref{fig:selection} narrows the catalog for ($i$) a researcher choosing a benchmark, ($ii$) a reviewer assessing a claim of realism, and ($iii$) a data provider looking for gaps that a new dataset release could fill and contribute through the process of Sec.~\ref{sec:observatory}.
Table~\ref{tab:funnel} reports the candidates left after each filter.

\begin{table}[!t]\centering\footnotesize
\caption{Candidates left after each step of Figure~\ref{fig:selection} in the three use cases of Section~\ref{sec:usecases} (release \release{}). The task (step~2) selects the filters of the later steps through Table~\ref{tab:paradigms} and removes no dataset by itself; -- marks a step not applied.}\label{tab:funnel}
\begin{tabular}{@{}l lr lr lr@{}}
\toprule
 & \multicolumn{2}{c}{\textbf{Case 1:} IoT intrusion detection} & \multicolumn{2}{c}{\textbf{Case 2:} 5G realism} & \multicolumn{2}{c}{\textbf{Case 3:} ICS gaps} \\
\cmidrule(lr){2-3}\cmidrule(lr){4-5}\cmidrule(l){6-7}
\textbf{Step} & Filter & \# & Filter & \# & Filter & \# \\
\midrule
1 Domain         & IoT / IIoT / IoMT & 50 & 5G / mobile core & 13 & ICS / OT & 22 \\
\rowcolor[HTML]{f0f0f0}
2 Task           & Packet-level, supervised & & Supervised & & Supervised & \\
3 Representation & Raw traces & 32 & Raw traces & 9 & Raw traces & 15 \\
\rowcolor[HTML]{f0f0f0}
4 Labels         & Multiclass & 24 & Multiclass & 5 & Multiclass & 7 \\
\rowcolor[HTML]{f0f0f0}
                 & Benign class & 19 & -- & & -- & \\
5 Realism        & -- & & Real radio equipment & 1 & -- & \\
\rowcolor[HTML]{f0f0f0}
6 Independence   & Not a derivative & 19 & -- & & -- & \\
7 Access         & Open (+ registration) & 11 (+8) & -- & & Open & 6 \\
\bottomrule
\end{tabular}
\end{table}

\paragraph{Choosing a benchmark.} A researcher who develops a packet-level intrusion detector for \gls{iot} networks needs traces, multiclass labels with a benign class, and independent captures. Four filters on the \gls{iot} table (Table~\ref{tab:rev-iot}) leave 19 candidates, none of which is a derivative, since the NetFlow re-extractions of the domain distribute processed records only. Eleven can be downloaded directly (\texttt{BoT-IoT}, \texttt{DAD}, \texttt{IoT-23}, \texttt{TON\_IoT}, \texttt{ECU-IoHT}, \texttt{MedBIoT}, \texttt{MQTTset}, \texttt{Edge-IIoTset}, \texttt{DoS/DDoS-MQTT-IoT}, \texttt{IoMT-TrafficData} and \texttt{Gotham}), and eight more after registration. 
The filters also exclude widely used datasets: \texttt{N-BaIoT}, one of the most cited \gls{iot} benchmarks, distributes only the features extracted by AfterImage and therefore cannot support a packet-level model.

\paragraph{Assessing the realism of a domain.} A reviewer who reads a claim of realistic detection in 5G core networks can check what the public evidence supports. Of the $13$ 5G datasets (Tab.~\ref{tab:rev-fiveg}), $9$ distribute traces and $5$ of these have multiclass labels, but only \texttt{5G-NIDD} was recorded with real radio equipment; the other two datasets with real radio equipment, i.e., the \texttt{5G Security \& Resource Alloc.} and \texttt{NetsLab-5GORAN-IDD}, have binary or weak labels. A claim of realistic multiclass detection in this domain therefore rests on a single public dataset, and a reviewer can ask the authors to state this limit or to test on \texttt{5G-NIDD}.

\paragraph{Planning a new release.} A team preparing an industrial dataset can use the catalog to find what is missing. Of the $22$ \gls{ics} datasets (Tab.~\ref{tab:rev-ics}), $15$ distribute traces and $7$ of these have multiclass labels; six are also open: \texttt{2017QUT\_S7comm}, \texttt{2017QUT\_DNP3}, the \texttt{DNP3} and \texttt{IEC~104 Intrusion Detection} datasets, \texttt{ICS-Flow} and \texttt{Westermo}. No \gls{ics} dataset contains malware or lateral movement, and the best-known testbeds (\texttt{SWaT}, \texttt{WADI} and \texttt{EPIC}) are available only on request, with binary labels or none. An open capture with multi-stage attacks would therefore fill a gap that the catalog makes visible.
\section{Discussion: Pitfalls and Gaps}\label{sec:pitfalls}

Some problems recur in every domain of the review. Section~\ref{sec:case} discusses the pitfalls that affect results obtained on existing datasets, and Sec.~\ref{sec:gaps} the gaps that no public dataset yet fills; each point can be checked on a column of the domain tables.

\subsection{Benchmark pitfalls}\label{sec:case}
Table~\ref{tab:pitfalls} gives, for each pitfall, the column that reveals it and the number of datasets of release \release{} concerned.
\emph{The first pitfall concerns what a model actually learns}, and has three sources: testbed artifacts, label errors and leakage between training and test data.
Laboratory testbeds can introduce differences between malicious and benign traffic that have nothing to do with the attack, such as recording windows, traffic rates, fixed IP addresses and tool fingerprints, and a model can exploit these correlates instead of the threat itself. 
Labels add a second source of error: they are often derived from attack schedules or from assumptions about infected hosts, so a label assigned to a capture window may be wrong for many of its flows, as the corrected releases of \texttt{CICIDS2017} have shown~\citep{engelen2021}. 
The review found the same weakness in the few datasets recorded in real networks, whose labels come from IDS alerts, victim addresses or per-file annotations. Data leakage has a similar effect: overlapping time windows, flows of the same connection and shared attacker infrastructure can link training and test data, and whether a random split is acceptable depends on the claim being tested~\citep{arp2022}.

\begin{table}[!t]\centering\footnotesize
\caption{Benchmark pitfalls, the column of the domain tables that reveals them, and the number of datasets of release \release{} concerned ($n=\NFinal{}$). Data leakage depends on how a dataset is split and is not revealed by any column.}\label{tab:pitfalls}
\begin{tabularx}{\linewidth}{>{\raggedright\arraybackslash}p{3.0cm}>{\raggedright\arraybackslash}p{2.8cm}r>{\raggedright\arraybackslash}X}
\toprule
\textbf{Pitfall} & \textbf{Column to check} & \textbf{Datasets} & \textbf{Examples} \\
\midrule
Testbed artifacts & Acq.\ not real & 197 & CICIDS2017, BoT-IoT, CICIoT2023 \\
\rowcolor[HTML]{f0f0f0}
Weak or missing labels & Lab.\ = W or -- & 51 & MAWILab, UGR'16, Kyoto 2006+ \\
Only processed records & Artifact: Raw = \ding{55} & 93 & NetFlow (NF-) versions, AnoShift \\
\rowcolor[HTML]{f0f0f0}
Apparent diversity & Lineage (\deriv{}) & 30 & CICIDS2017 and four derived releases \\
\bottomrule
\end{tabularx}
\end{table}

\emph{A second pitfall concerns what a dataset allows its users to choose}. 
Released feature tables are a convenience, not an obligation: users can select, discard or recompute features. 
When a dataset distributes only processed records, however, this choice is limited to what the extractor produced: flow timeouts, direction, truncation and normalization are fixed and cannot be changed or verified, and new representations cannot be tested on the same traffic. 
This is the case for $93$ of the $234$ datasets of release \release{}: $91$ distribute only processed records and $2$ only raw HTTP requests.
\begin{draft}
\end{draft}

\emph{The third pitfall is apparent diversity}. The number of datasets overstates the number of independent experiments, because many releases re-extract, relabel or augment a few earlier captures. \texttt{CICIDS2017} is the clearest example: the original release, \texttt{Improved CICIDS2017}~\citep{engelen2021}, \texttt{LYCOS-IDS2017}, \texttt{BCCC-CIC-IDS2017} and \texttt{SAFE-NID} build on the same five days of traffic but differ in extraction tool, features and labels. Two papers reporting results ``on \texttt{CICIDS2017}'' may therefore use different data, while training on one release and testing on another does not measure generalization, because the underlying traffic is the same. All four derived releases appeared years after the original, which is precisely why a catalog has to follow datasets after their publication.
\texttt{CICIDS2017} is not an isolated case: in release \release{}, \texttt{CSE-CIC-IDS2018} has seven derived releases, \texttt{UNSW-NB15} four, \texttt{CICIoT2023} three, and \texttt{BoT-IoT} and \texttt{TON\_IoT} two each.

\subsection{Gaps}\label{sec:gaps}

\begin{figure}[!t]\centering
\begin{tikzpicture}
\begin{axis}[
  xbar, width=.8\linewidth, height=5.2cm,
  bar width=7pt, xmin=0, xmax=40,
  xlabel={Datasets in release \release{}},
  symbolic y coords={itrust,uwf,uq,wustl,unsw,strat,hcrl,cic},
  ytick=data,
  yticklabels={iTrust (SUTD),UWF,UQ,WUSTL,UNSW Canberra,Stratosphere (CTU),HCRL (Korea Univ.),CIC + BCCC},
  y tick label style={font=\footnotesize},
  nodes near coords, nodes near coords align={horizontal},
  every node near coord/.append style={font=\footnotesize},
  axis x line*=bottom, axis y line*=left, enlarge y limits=0.08,
]
\addplot[fill=gray!60, draw=none] coordinates {
  (3,itrust) (3,uwf) (5,uq) (5,wustl) (7,unsw)
  (9,strat) (12,hcrl) (34,cic)};
\end{axis}
\end{tikzpicture}
\caption{Most frequent producers of the datasets in release v1.0. The eight producers shown released 78 of the 234 datasets; the remaining 156 come from many other groups, most of which released a single dataset.}\label{fig:producers}
\end{figure}
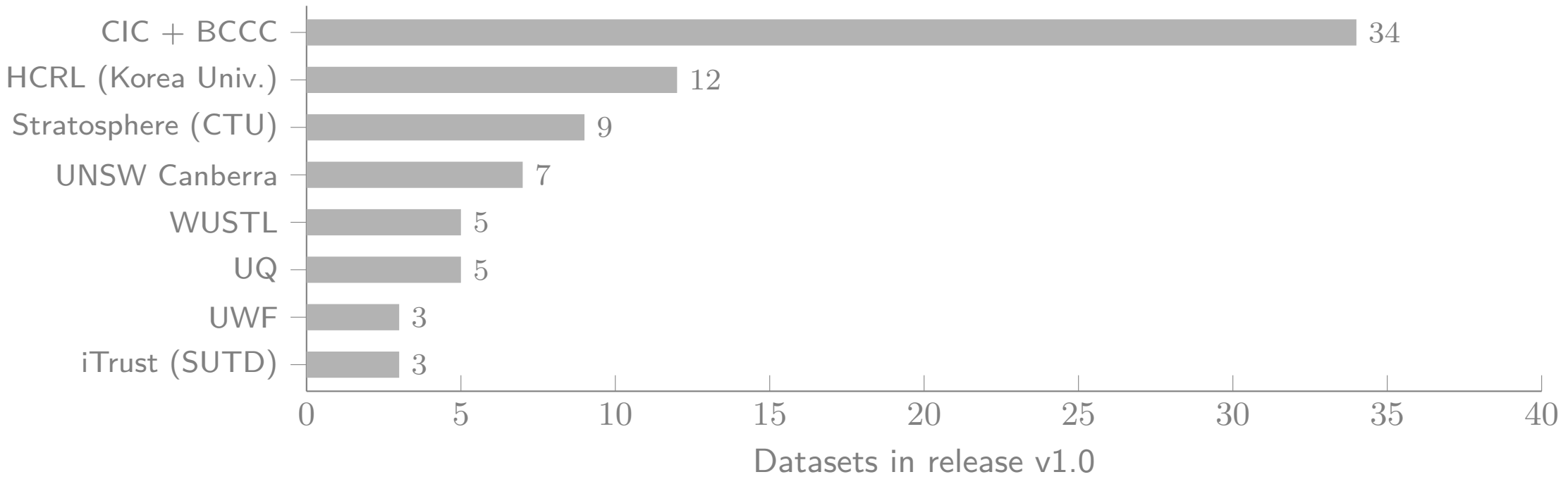

\emph{The gaps are uneven across domains}. Enterprise and \gls{iot} networks are covered by dozens of datasets, whereas UAV and wireless networks are represented by a handful each, 5G cores mostly by emulated testbeds, V2X only by simulations, and no satellite dataset was found.
Packet-level data, which recent \gls{ai} models increasingly require~\cite{holland2021nprint,lin2022etbert,zhao2023yatc}, are available for only part of the catalog and are rare in some domains, such as DNS/web, and no established practice yet exists for releasing packet traces while preserving privacy. 
\emph{Operational traffic is the most evident absence}: well-labeled datasets come almost exclusively from testbeds and simulators, while real traces exist mainly as backbone archives or honeypot collections with weak labels. 
\emph{Independent and long-running captures are also rare}. 
\emph{The catalog depends on a few producers} (Fig.~\ref{fig:producers}): the Canadian Institute for Cybersecurity (CIC) and its successor group at York University released $34$ of the $234$ datasets, and the five most frequent producers account for $67$ ($29$\%). Only a few collections, such as \textit{Stratosphere} and \textit{Malware-Traffic-Analysis.net}, span several years and support studies of drift.
Finally, \emph{the acquisition environment and the label granularity are missing or ambiguous for part of the datasets}.
Because the \Obs{} is continuously maintained, these gaps are not a static finding of this paper: each new release shows whether they are being closed, and in which domains.

\section{Conclusion}\label{sec:conclusion}
\gls{ai} has become a central tool for cybersecurity, but its models are bounded by the data they learn from.
For network-based security, those data are public datasets of very different nature, scattered across many sources and continuously released, corrected and withdrawn, so a list published in a survey starts to age at publication.
The \Obs{} replaces such a frozen list with a catalog that is verified and updated release after release.
It defines its scope by network traffic rather than by a single task, classifies every dataset along seven axes, links each record to its original source, and publishes the catalog as dated, versioned releases that can be cited.
Release \release{} covers \NFinal{} datasets across \NDomains{} domains and shows a field that grows quickly but unevenly, relies heavily on processed records and on a few laboratories, and offers little public evidence for \gls{uav}, wireless and V2X networks, and few real captures for 5G.
The same columns that describe the datasets also support their use: they turn the choice of a benchmark into a step-by-step procedure, and make each common pitfall checkable before results are reported.
We plan to automate discovery and the periodic check with link monitoring and with an AI agent that follows data repositories and new publications and proposes new datasets, releases and withdrawals; its proposals will enter the same manual validation as any other submission.
We also plan to extend the attack families with a graded correspondence between each native class and a standardized vocabulary based on CAPEC and MITRE ATT\&CK (exact, broader, narrower, partial or not mappable), proposed through the submission form and validated by the maintainers.
Future releases will show whether the gaps identified here are being closed, and we invite dataset authors and users to check, correct and extend the catalog.
The value of the \Obs{} lies less in any single release than in the fact that the next one will be more complete and more accurate than this one.



\paragraph{Limitations.}\label{sec:threats}
\emph{The catalog is not exhaustive}: it covers the datasets found through the sources and searches of Sec.~\ref{sec:method} as of \snapshot{}.
\emph{Verification confirms that a dataset and its official source exist, but does not test the files}. Because the \Obs{} links to original sources and does not redistribute data, which would conflict with the licenses and access conditions set by the authors, a dataset whose source later disappears remains in the catalog with its description and history but can no longer be obtained through it.
\emph{All seven axes were assigned manually from the authors' descriptions}, so borderline datasets, especially for labeling and acquisition, may be classified differently by another reviewer; each dataset is also assigned to a single domain, so per-domain counts understate cross-cutting topics such as DDoS.
\emph{Lineage is recorded only when the authors document it}: releases that merge earlier captures without naming their sources cannot be traced.
Finally, \emph{the release was curated by a single team}; corrections proposed by dataset authors and users through the public form provide an external check, and because every decision and source is recorded per record, each of these limitations can be reduced release after release.

\section*{Declaration of generative AI and AI-assisted technologies in the writing process}
During the preparation of this work, the author(s) used GenAI tools to provide grammatical support and minor text polishing, and to suggest candidate datasets and draft record descriptions for the catalog, as described in Section~\ref{sec:method}. After using these tools, the author(s) reviewed and edited the content as needed and take(s) full responsibility for the content of the publication.

\renewcommand{\bibpreamble}{All URLs were last accessed on \snapshot.\par\medskip}
\bibliographystyle{plainnat}
\begingroup
\footnotesize
\bibliography{references,reference_enterprise_table,reference_iot_table,reference_malware_table,reference_ics_table, reference_automotive_table,reference_v2x_table,reference_uav_table,reference_fiveg_table,reference_wireless_table, reference_encrypted_table, reference_dns_table, reference_ddos_table, reference_sdn_table, reference_honeypot_table}

@inproceedings{lee2017otids,
  author    = {Lee, Hyunsung and Jeong, Seong Hoon and Kim, Huy Kang},
  title     = {{OTIDS}: A Novel Intrusion Detection System for In-vehicle Network by Using Remote Frame},
  booktitle = {15th Annual Conference on Privacy, Security and Trust (PST)},
  pages     = {57--66},
  year      = {2017}
}

@article{han2018survival,
  author  = {Han, Mee Lan and Kwak, Byung Il and Kim, Huy Kang},
  title   = {Anomaly intrusion detection method for vehicular networks based on survival analysis},
  journal = {Vehicular Communications},
  volume  = {14},
  pages   = {52--63},
  year    = {2018}
}

@inproceedings{seo2018gids,
  author    = {Seo, Eunbi and Song, Hyun Min and Kim, Huy Kang},
  title     = {{GIDS}: {GAN} based Intrusion Detection System for In-Vehicle Network},
  booktitle = {16th Annual Conference on Privacy, Security and Trust (PST)},
  year      = {2018}
}

@misc{dupont2019tuecan,
  author       = {Dupont, Guillaume and Lekidis, Alexios and den Hartog, Jerry and Etalle, Sandro},
  title        = {Automotive Controller Area Network ({CAN}) Bus Intrusion Dataset v2},
  howpublished = {4TU.ResearchData},
  year         = {2019}
}

@article{hanselmann2020syncan,
  author  = {Hanselmann, Markus and Strauss, Thilo and Dormann, Katharina and Ulmer, Holger},
  title   = {{CANet}: An Unsupervised Intrusion Detection System for High Dimensional {CAN} Bus Data},
  journal = {IEEE Access},
  volume  = {8},
  pages   = {58194--58205},
  year    = {2020}
}

@misc{unimorebusoff,
  author       = {{WEBLab, University of Modena and Reggio Emilia}},
  title        = {{UNIMORE} Bus-Off dataset},
  howpublished = {\url{https://weblab.ing.unimore.it/people/stabili/resources/}},
  year         = {2019}
}

@misc{carhackchallenge2020,
  author       = {Kang, Hyunjae and Kwak, Byung Il and Lee, Young Hun and Lee, Haneol and Lee, Hwejae and Kim, Huy Kang},
  title        = {Car Hacking: Attack \& Defense Challenge 2020 Dataset},
  howpublished = {\url{https://ieee-dataport.org/open-access/car-hacking-attack-defense-challenge-2020-dataset}},
  year         = {2021}
}

@article{verma2024road,
  author  = {Verma, Miki E. and Bridges, Robert A. and Iannacone, Michael D. and Hollifield, Samuel C. and Moriano, Pablo and Hespeler, Steven C. and Kay, Bill and Combs, Frank L.},
  title   = {A comprehensive guide to {CAN} {IDS} data and introduction of the {ROAD} dataset},
  journal = {PLOS ONE},
  volume  = {19},
  number  = {1},
  pages   = {e0296879},
  year    = {2024}
}

@article{jeong2021avtp,
  author  = {Jeong, Seonghoon and Jeon, Boosun and Chung, Boheung and Kim, Huy Kang},
  title   = {Convolutional neural network-based intrusion detection system for {AVTP} streams in automotive {Ethernet}-based networks},
  journal = {Vehicular Communications},
  volume  = {29},
  pages   = {100338},
  year    = {2021}
}

@misc{hcrlcanfd2022,
  author       = {{Hacking and Countermeasure Research Lab (HCRL), Korea University}},
  title        = {{CAN-FD} Intrusion Dataset},
  howpublished = {IEEE Dataport, \url{https://ieee-dataport.org/documents/can-fd-intrusion-dataset}},
  year         = {2022}
}

@article{han2023towids,
  author  = {Han, Mee Lan and Kwak, Byung Il and Kim, Huy Kang},
  title   = {{TOW-IDS}: Intrusion Detection System Based on Three Overlapped Wavelets for Automotive {Ethernet}},
  journal = {IEEE Transactions on Information Forensics and Security},
  volume  = {18},
  pages   = {411--422},
  year    = {2023}
}

@misc{unimoredaga,
  author       = {{WEBLab, University of Modena and Reggio Emilia}},
  title        = {{UNIMORE} {DAGA} dataset},
  howpublished = {\url{https://weblab.ing.unimore.it/people/stabili/resources/}},
  year         = {2022}
}

@article{lampe2024cantrain,
  author  = {Lampe, Brooke and Meng, Weizhi},
  title   = {can-train-and-test: A curated {CAN} dataset for automotive intrusion detection},
  journal = {Computers \& Security},
  volume  = {140},
  pages   = {103777},
  year    = {2024}
}

@article{gazdag2023crysys,
  title={CrySyS dataset of CAN traffic logs containing fabrication and masquerade attacks},
  author={Gazdag, Andr{\'a}s and Ferenc, Rudolf and Butty{\'a}n, Levente},
  journal={Scientific Data},
  volume={10},
  number={1},
  pages={903},
  year={2023},
  publisher={Nature Publishing Group UK, London}
}

@inproceedings{rajapaksha2024canmirgu,
  author    = {Rajapaksha, Sampath and Madzudzo, Garikayi and Kalutarage, Harsha and Petrovski, Andrei and Al-Kadri, M. Omar},
  title     = {{CAN-MIRGU}: A Comprehensive {CAN} Bus Attack Dataset from Moving Vehicles for Intrusion Detection System Evaluation},
  booktitle = {Symposium on Vehicle Security and Privacy (VehicleSec), co-located with NDSS},
  year      = {2024}
}

@article{neto2024ciciov,
  author  = {Neto, Euclides Carlos Pinto and Taslimasa, Hamideh and Dadkhah, Sajjad and Iqbal, Shahrear and Xiong, Pulei and Rahman, Taufiq and Ghorbani, Ali A.},
  title   = {{CICIoV2024}: Advancing realistic {IDS} approaches against {DoS} and spoofing attack in {IoV} {CAN} bus},
  journal = {Internet of Things},
  volume  = {26},
  pages   = {101209},
  year    = {2024}
}

@misc{gemcan2026,
  title        = {{GEM-CAN} v3},
  howpublished = {Zenodo, \url{https://zenodo.org/records/19161139}},
  year         = {2026}
}

@misc{alkhatib2021someip,
  author       = {Alkhatib, Natasha and Ghauch, Hadi and Danger, Jean-Luc},
  title        = {{SOME/IP} Intrusion Detection using Deep Learning-based Sequential Models in Automotive {Ethernet} Networks},
  year         = {2021},
  howpublished = {arXiv:2108.08262},
  note         = {Dataset: \url{https://github.com/Alkhatibnatasha/SOMEIP_IDS}}
}

@misc{hcrlmcan2023,
  author       = {{Hacking and Countermeasure Research Lab (HCRL), Korea University}},
  title        = {{M-CAN} Intrusion Dataset},
  year         = {2023},
  xxxdoi       = {10.21227/mkvd-h021},
  howpublished = {IEEE DataPort, \url{https://ieee-dataport.org/documents/m-can-intrusion-dataset}}
}

@misc{hcrlbcan2023,
  author       = {{Hacking and Countermeasure Research Lab (HCRL), Korea University}},
  title        = {{B-CAN} Intrusion Dataset},
  year         = {2023},
  xxxdoi       = {10.21227/qm9j-bs98},
  howpublished = {IEEE DataPort, \url{https://sites.google.com/hksecurity.net/hcrl/Datasets/b-can-intrusion-dataset}}
}

@article{jeong2024xcanids,
  author  = {Jeong, Seonghoon and Kang, Hyunjae and Kim, Huy Kang},
  title   = {{X-CANIDS}: Signal-Aware Explainable Intrusion Detection System for Controller Area Network-Based In-Vehicle Network},
  journal = {IEEE Transactions on Vehicular Technology},
  volume  = {73},
  number  = {3},
  pages   = {3230--3246},
  year    = {2024},
  note    = {Dataset: IEEE DataPort, \url{https://ieee-dataport.org/node/12755}}
}

@inproceedings{lee2024j1939,
  author    = {Lee, Hwejae and Lee, Hyosun and Jun, Saehee and Kim, Huy Kang},
  title     = {Expanding the Attack Scenarios of {SAE J1939}: A Comprehensive Analysis of Established and Novel Vulnerabilities in Transport Protocol},
  booktitle = {escar USA},
  year      = {2024},
  note      = {Dataset: IEEE DataPort, \url{https://ieee-dataport.org/documents/sae-j1939-attack-dataset}}
}

@inproceedings{hellemans2025cards,
  author    = {Hellemans, Wouter and Hamborg, Jannis and Lauser, Timm and Rabbani, Md Masoom and Preneel, Bart and Krau{\ss}, Christoph and Mentens, Nele},
  title     = {{CarDS} -- Controller Area Network and Automotive {Ethernet} Realistic Data Set},
  booktitle = {2025 Annual Computer Security Applications Conference (ACSAC)},
  pages     = {798--814},
  year      = {2025},
  publisher = {IEEE},
  note      = {Dataset: \url{https://tudatalib.ulb.tu-darmstadt.de/handle/tudatalib/5080}}
}

@misc{song2026autohack,
  author       = {Song, Yuchan and Ahn, Sejun and Baek, Seungjin and Kim, Hyeonseong and Kim, Huy Kang and Jeon, Sanghoon},
  title        = {{AutoHack}: A Physically Verified Multi-Bus {CAN} Dataset for Intrusion Detection System Evaluation},
  year         = {2026},
  xxxdoi       = {10.5281/zenodo.19661007},
  howpublished = {Zenodo, \url{https://zenodo.org/records/19661007}}
}

@misc{darpa2000,
  author       = {{MIT Lincoln Laboratory}},
  title        = {2000 {DARPA} Intrusion Detection Scenario Specific Datasets ({LLDOS} 1.0 and 2.0.2)},
  howpublished = {\url{https://archive.ll.mit.edu/ideval/data/2000/LLS_DDOS_2.0.2.html}},
  year         = {2000}
}

@misc{caida2007ddos,
  author       = {{Center for Applied Internet Data Analysis (CAIDA)}},
  title        = {The {CAIDA} {UCSD} {DDoS} Attack 2007 Dataset},
  howpublished = {\url{https://www.caida.org/catalog/datasets/ddos-20070804_dataset/}},
  year         = {2007}
}

@article{jazi2017cicdos,
  author  = {Jazi, Hossein Hadian and Gonzalez, Hugo and Stakhanova, Natalia and Ghorbani, Ali A.},
  title   = {Detecting {HTTP}-based application layer {DoS} attacks on web servers in the presence of sampling},
  journal = {Computer Networks},
  volume  = {121},
  pages   = {25--36},
  year    = {2017}
}

@inproceedings{sharafaldin2019ddos,
  author    = {Sharafaldin, Iman and Lashkari, Arash Habibi and Hakak, Saqib and Ghorbani, Ali A.},
  title     = {Developing Realistic Distributed Denial of Service ({DDoS}) Attack Dataset and Taxonomy},
  booktitle = {International Carnahan Conference on Security Technology (ICCST)},
  year      = {2019}
}

@article{erhan2020boun,
  author  = {Erhan, Derya and Anar{\i}m, Emin},
  title   = {Bo{\u{g}}azi{\c{c}}i University distributed denial of service dataset},
  journal = {Data in Brief},
  volume  = {32},
  pages   = {106187},
  year    = {2020}
}

@inproceedings{catillo2021usbids,
  author    = {Catillo, Marta and Del Vecchio, Andrea and Ocone, Luciano and Pecchia, Antonio and Villano, Umberto},
  title     = {{USB-IDS-1}: a Public Multilayer Dataset of Labeled Network Flows for {IDS} Evaluation},
  booktitle = {51st Annual IEEE/IFIP International Conference on Dependable Systems and Networks Workshops (DSN-W)},
  pages     = {1--6},
  year      = {2021}
}

@misc{usbidstc2024,
  author       = {Catillo, Marta and others},
  title        = {{USB-IDS-TC}},
  howpublished = {\url{https://idsdata.ding.unisannio.it/usbidstc.html}},
  year         = {2024}
}

@misc{bcccudpquic2025,
  author       = {{Behaviour-Centric Cybersecurity Center (BCCC), York University}},
  title        = {{BCCC-UDP-QUIC-IDS-2025}},
  howpublished = {\url{https://www.yorku.ca/research/bccc/ucs-technical/cybersecurity-datasets-cds/}},
  year         = {2025}
}

@inproceedings{raissi2007ecml,
  author    = {Ra{\"i}ssi, Chedy and Brissaud, Johan and Dray, Gerard and Poncelet, Pascal and Roche, Mathieu and Teisseire, Maguelonne},
  title     = {Web Analyzing Traffic Challenge: Description and Results},
  booktitle = {Proceedings of the ECML/PKDD 2007 Discovery Challenge},
  pages     = {47--52},
  year      = {2007}
}

@article{gimenez2010csic,
  title={HTTP data set CSIC 2010},
  author={Gim{\'e}nez, Carmen Torrano and Villegas, Alejandro P{\'e}rez and Mara{\~n}{\'o}n, Gonzalo {\'A}lvarez},
  journal={Information Security Institute of CSIC (Spanish Research National Council)},
  volume={64},
  number={07},
  year={2010}
}

@inproceedings{diazverdejo2024biblious17,
  author    = {Diaz-Verdejo, Jesus and Estepa Alonso, Rafael and Estepa Alonso, Antonio and Munoz-Calle, Javier and Madinabeitia, German},
  title     = {{Biblio-US17}: A labeled real {URL} dataset for anomaly-based intrusion detection systems development},
  booktitle = {Proceedings of the 2024 European Interdisciplinary Cybersecurity Conference},
  pages     = {217--218},
  year      = {2024}
}

@inproceedings{montazeri2020dohbrw,
  author    = {MontazeriShatoori, Mohammadreza and Davidson, Logan and Kaur, Gurdip and Lashkari, Arash Habibi},
  title     = {Detection of {DoH} Tunnels using Time-series Classification of Encrypted Traffic},
  booktitle = {IEEE Intl Conf on Dependable, Autonomic and Secure Computing (DASC/PiCom/CBDCom/CyberSciTech)},
  pages     = {63--70},
  year      = {2020}
}

@article{riera2022srbh,
  author  = {Sureda Riera, Tomás and Bermejo Higuera, Juan-Ramón and Bermejo Higuera, Javier and Martínez Herraiz, José-Javier and Sicilia Montalvo, Juan-Antonio},
  title   = {A new multi-label dataset for Web attacks {CAPEC} classification using machine learning techniques},
  journal = {Computers \& Security},
  volume  = {120},
  pages   = {102788},
  year    = {2022},
  note    = {dataset: \url{https://doi.org/10.7910/DVN/OGOIXX}}
}

@inproceedings{mahdavifar2021dns,
  author    = {Mahdavifar, Samaneh and Maleki, Nasim and Lashkari, Arash Habibi and Broda, Matt and Razavi, Amir H.},
  title     = {Classifying Malicious Domains using {DNS} Traffic Analysis},
  booktitle = {IEEE Intl Conf on Dependable, Autonomic and Secure Computing (DASC/PiCom/CBDCom/CyberSciTech)},
  pages     = {60--67},
  year      = {2021}
}

@inproceedings{mahdavifar2021dnsexf,
  author    = {Mahdavifar, Samaneh and Salem, Amgad Hanafy and Victor, Princy and Razavi, Amir H. and Garzon, Miguel and Hellberg, Natasha and Lashkari, Arash Habibi},
  title     = {Lightweight Hybrid Detection of Data Exfiltration using {DNS} based on Machine Learning},
  booktitle = {11th International Conference on Communication and Network Security (ICCNS)},
  year      = {2021}
}

@article{chatzoglou2023h23q,
  author  = {Chatzoglou, Efstratios and Kouliaridis, Vasileios and Kambourakis, Georgios and Karopoulos, Georgios and Gritzalis, Stefanos},
  title   = {A hands-on gaze on {HTTP/3} security through the lens of {HTTP/2} and a public dataset},
  journal = {Computers \& Security},
  volume  = {125},
  pages   = {103051},
  year    = {2023}
}

@misc{bcccdohbrw2023,
  author       = {{Behaviour-Centric Cybersecurity Center (BCCC), York University}},
  title        = {{BCCC-CIRA-CIC-DoHBrw-2020}},
  howpublished = {\url{https://www.yorku.ca/research/bccc/ucs-technical/cybersecurity-datasets-cds/}},
  year         = {2023}
}

@misc{bcccbelldns2024,
  author       = {{Behaviour-Centric Cybersecurity Center (BCCC), York University}},
  title        = {{BCCC-CIC-Bell-DNS-2024}},
  howpublished = {\url{https://www.yorku.ca/research/bccc/ucs-technical/cybersecurity-datasets-cds/}},
  year         = {2024}
}

@inproceedings{wang2017ustc,
  author    = {Wang, Wei and Zhu, Ming and Zeng, Xuewen and Ye, Xiaozhou and Sheng, Yiqiang},
  title     = {Malware traffic classification using convolutional neural network for representation learning},
  booktitle = {International Conference on Information Networking (ICOIN)},
  pages     = {712--717},
  year      = {2017}
}

@misc{cicpqc2025,
  author       = {{Canadian Institute for Cybersecurity}},
  title        = {{CIC-PQC\_OAV} v1 Dataset 2025},
  howpublished = {\url{https://www.unb.ca/cic/datasets/pqc-oav-dataset-2025.html}},
  year         = {2025}
}

@misc{tqhc22026,
  title        = {{TQH-C2}: encrypted command-and-control traffic dataset},
  howpublished = {Zenodo, \url{https://zenodo.org/records/21523144}},
  year         = {2026}
}

@misc{cicids2018,
  author       = {{Communications Security Establishment (CSE) and Canadian Institute for Cybersecurity (CIC)}},
  title        = {{CSE-CIC-IDS2018} on {AWS}},
  year         = {2018},
  howpublished = {University of New Brunswick},
  url          = {https://www.unb.ca/cic/datasets/ids-2018.html}
}

@misc{abluva_cse_cic_ids2018_v2,
  author       = {Madhubalan, Akshayraj and Gautam, Amit and Tiwary, Priya},
  title        = {{CSE-CIC-IDS2018 v2}},
  howpublished = {\url{https://dataverse.harvard.edu/dataset.xhtml?persistentId=doi:10.7910/DVN/88M0DI}}
}

@misc{dhoogla2023nfcsecicids2018,
  author       = {Sarhan, Mohanad and Layeghy, Siamak and Moustafa, Nour and Portmann, Marius},
  title        = {{NF-CSE-CIC-IDS2018}: NetFlow version of {CSE-CIC-IDS2018}},
  year         = {2020},
  howpublished = {The University of Queensland},
  url          = {https://staff.itee.uq.edu.au/marius/NIDS_datasets/}
}

@inproceedings{lippmann2000darpa98,
  author    = {Lippmann, Richard P. and Fried, David J. and Graf, Isaac and Haines, Joshua W. and Kendall, Kristopher R. and McClung, David and Weber, Dan and Webster, Seth E. and Wyschogrod, Dan and Cunningham, Robert K. and Zissman, Marc A.},
  title     = {Evaluating Intrusion Detection Systems: The 1998 {DARPA} Off-line Intrusion Detection Evaluation},
  booktitle = {Proceedings of the DARPA Information Survivability Conference and Exposition (DISCEX)},
  volume    = {2},
  pages     = {12--26},
  year      = {2000}
}

@article{lippmann2000darpa99,
  author  = {Lippmann, Richard and Haines, Joshua W. and Fried, David J. and Korba, Jonathan and Das, Kumar},
  title   = {The 1999 {DARPA} off-line intrusion detection evaluation},
  journal = {Computer Networks},
  volume  = {34},
  number  = {4},
  pages   = {579--595},
  year    = {2000}
}

@misc{kddcup99,
  title        = {{KDD Cup} 1999 Data},
  howpublished = {UCI Machine Learning Repository, \url{https://kdd.ics.uci.edu/databases/kddcup99/kddcup99.html}},
  year         = {1999}
}

@inproceedings{perona2008gurekdd,
  author    = {Perona, I{\~n}igo and Gurrutxaga, Ibai and Arbelaitz, Olatz and Mart{\'i}n, Jos{\'e} I. and Muguerza, Javier and P{\'e}rez, Jes{\'u}s M.},
  title     = {Service-independent payload analysis to improve intrusion detection in network traffic},
  booktitle = {Proceedings of the 7th Australasian Data Mining Conference (AusDM)},
  pages     = {171--178},
  year      = {2008}
}

@misc{homoliak2020asnm,
  author        = {Homoliak, Ivan and Malinka, Kamil and Hanacek, Petr},
  title         = {{ASNM} Datasets: A Collection of Network Traffic Features for Testing of Adversarial Classifiers and Network Intrusion Detectors},
  year          = {2019},
  eprint        = {1910.10528},
  archivePrefix = {arXiv}
}

@misc{maccdc,
  author       = {{NETRESEC AB}},
  title        = {{MACCDC} -- Mid-Atlantic Collegiate Cyber Defense Competition captures},
  howpublished = {\url{https://www.netresec.com/?page=MACCDC}},
  year         = {2026},
}

@misc{vast2011,
  author       = {{IEEE VAST Challenge}},
  title        = {{VAST} Challenge 2011, Mini Challenge 2: Computer Networking Operations},
  howpublished = {\url{https://visualdata.wustl.edu/varepository/benchmarks.php}},
  year         = {2011}
}

@misc{ctunormal,
  author       = {{Stratosphere Laboratory}},
  title        = {Normal Captures},
  howpublished = {\url{https://www.stratosphereips.org/datasets-normal}},
  year         = {2026},
}

@incollection{kent2016lanl,
  author    = {Kent, Alexander D.},
  title     = {Cybersecurity Data Sources for Dynamic Network Research},
  booktitle = {Dynamic Networks and Cyber-Security},
  publisher = {World Scientific},
  pages     = {37--65},
  year      = {2016}
}

@inproceedings{beer2017ndsec,
  author    = {Beer, Frank and Hofer, Tim and Karimi, David and B{\"u}hler, Ulrich},
  title     = {A new Attack Composition for Network Security},
  booktitle = {10. DFN-Forum Kommunikationstechnologien},
  series    = {Lecture Notes in Informatics (LNI)},
  volume    = {P-271},
  pages     = {11--20},
  publisher = {Gesellschaft f{\"u}r Informatik},
  year      = {2017}
}

@article{haider2017ngids,
  author  = {Haider, Waqas and Hu, Jiankun and Slay, Jill and Turnbull, Benjamin P. and Xie, Yi},
  title   = {Generating realistic intrusion detection system dataset based on fuzzy qualitative modeling},
  journal = {Journal of Network and Computer Applications},
  volume  = {87},
  pages   = {185--192},
  year    = {2017}
}

@inproceedings{ring2017cidds001,
  author    = {Ring, Markus and Wunderlich, Sarah and Gr{\"u}dl, Dominik and Landes, Dieter and Hotho, Andreas},
  title     = {Flow-based benchmark data sets for intrusion detection},
  booktitle = {Proceedings of the 16th European Conference on Cyber Warfare and Security (ECCWS)},
  pages     = {361--369},
  year      = {2017}
}

@article{ring2017cidds002,
  author  = {Ring, Markus and Wunderlich, Sarah and Gr{\"u}dl, Dominik and Landes, Dieter and Hotho, Andreas},
  title   = {Creation of Flow-Based Data Sets for Intrusion Detection},
  journal = {Journal of Information Warfare},
  volume  = {16},
  number  = {4},
  pages   = {40--53},
  year    = {2017}
}

@incollection{turcotte2018uhn,
  author    = {Turcotte, Melissa J. M. and Kent, Alexander D. and Hash, Curtis},
  title     = {Unified Host and Network Data Set},
  booktitle = {Data Science for Cyber-Security},
  publisher = {World Scientific},
  pages     = {1--22},
  year      = {2018}
}

@misc{otrf,
  author       = {{Open Threat Research Forge}},
  title        = {Security Datasets},
  howpublished = {\url{https://github.com/OTRF/Security-Datasets}}
}

@inproceedings{myneni2020dapt,
  author    = {Myneni, Sowmya and Chowdhary, Ankur and Sabur, Abdulhakim and Sengupta, Sailik and Agrawal, Garima and Huang, Dijiang and Kang, Myong},
  title     = {{DAPT} 2020 -- Constructing a Benchmark Dataset for Advanced Persistent Threats},
  booktitle = {Deployable Machine Learning for Security Defense (MLHat 2020)},
  series    = {Communications in Computer and Information Science},
  volume    = {1271},
  publisher = {Springer},
  year      = {2020}
}

@article{mills2022luflow,
  author  = {Mills, Ryan and Marnerides, Angelos K. and Broadbent, Matthew and Race, Nicholas J. P.},
  title   = {Practical Intrusion Detection of Emerging Threats},
  journal = {IEEE Transactions on Network and Service Management},
  volume  = {19},
  number  = {1},
  pages   = {582--600},
  year    = {2022}
}

@misc{isecids2022,
  author       = {Serinelli, Benedetto Marco and others},
  title        = {{I-Sec-IDS} Datasets},
  howpublished = {\url{https://zenodo.org/records/6657210}},
  year         = {2022}
}

@inproceedings{rosay2022lycos,
  author    = {Rosay, Arnaud and Cheval, Eloise and Carlier, Florent and Leroux, Pascal},
  title     = {Network Intrusion Detection: A Comprehensive Analysis of {CIC-IDS2017}},
  booktitle = {Proceedings of the 8th International Conference on Information Systems Security and Privacy (ICISSP)},
  year      = {2022}
}

@article{liu2022scvic,
  author  = {Liu, Jinxin and Shen, Yu and Simsek, Murat and Kantarci, Burak and Mouftah, Hussein T. and Bagheri, Mehran and Djukic, Petar},
  title   = {A New Realistic Benchmark for Advanced Persistent Threats in Network Traffic},
  journal = {IEEE Networking Letters},
  volume  = {4},
  number  = {3},
  pages   = {162--166},
  year    = {2022}
}

@article{myneni2023unraveled,
  author  = {Myneni, Sowmya and Jangid, Kamakshi and Chowdhary, Ankur and Sabur, Abdulhakim and Huang, Dijiang and others},
  title   = {Unraveled -- A semi-synthetic dataset for Advanced Persistent Threats},
  journal = {Computer Networks},
  volume  = {227},
  pages   = {109688},
  year    = {2023}
}

@article{landauer2023ait,
  author  = {Landauer, Max and Skopik, Florian and Frank, Maximilian and Hotwagner, Wolfgang and Wurzenberger, Markus and Rauber, Andreas},
  title   = {Maintainable Log Datasets for Evaluation of Intrusion Detection Systems},
  journal = {IEEE Transactions on Dependable and Secure Computing},
  volume  = {20},
  number  = {4},
  pages   = {3466--3482},
  year    = {2023}
}

@inproceedings{dragoi2022anoshift,
  author        = {Dr{\u{a}}goi, Marius and Burceanu, Elena and Haller, Emanuela and Manolache, Andrei and Brad, Florin},
  title         = {{AnoShift}: A Distribution Shift Benchmark for Unsupervised Anomaly Detection},
  booktitle     = {Advances in Neural Information Processing Systems 35 (NeurIPS 2022), Datasets and Benchmarks Track},
  year          = {2022},
  eprint        = {2206.15476},
  archivePrefix = {arXiv}
}

@article{lawrence2022cupid,
  author  = {Lawrence, Heather and others},
  title   = {{CUPID}: A labeled dataset with Pentesting for evaluation of network intrusion detection},
  journal = {Journal of Systems Architecture},
  volume  = {129},
  pages   = {102621},
  year    = {2022}
}

@inproceedings{liu2022error,
  author    = {Liu, Lisa and Engelen, Gints and Lynar, Timothy and Essam, Daryl and Joosen, Wouter},
  title     = {Error Prevalence in {NIDS} datasets: A Case Study on {CIC-IDS-2017} and {CSE-CIC-IDS-2018}},
  booktitle = {IEEE Conference on Communications and Network Security (CNS)},
  year      = {2022}
}

@article{odids2022,
  author  = {Homaei, Mohammadhossein and Di Bartolo, Agustin and Mogoll{\'o}n-Guti{\'e}rrez, {\'O}scar and Broncano Morgado, Fernando and Garc{\'i}a Rodr{\'i}guez, Pablo},
  title   = {{OD-IDS2022}: generating a new offensive defensive intrusion detection dataset for machine learning-based attack classification},
  journal = {International Journal of Information Technology},
  volume  = {15},
  pages   = {3757--3772},
  year    = {2023}
}

@article{berady2022pwnjutsu,
  author  = {Berady, Aimad and Jaume, Mathieu and Viet Triem Tong, Val{\'e}rie and Guette, Gilles},
  title   = {{PWNJUTSU}: A Dataset and a Semantics-Driven Approach to Retrace Attack Campaigns},
  journal = {IEEE Transactions on Network and Service Management},
  volume  = {19},
  number  = {4},
  year    = {2022}
}

@article{bagui2023zeek22,
  author  = {Bagui, Sikha S. and Mink, Dustin and Bagui, Subhash C. and Ghosh, Tirthankar and Plenkers, Russel and McElroy, Tom and Dulaney, Stephan and Shabanali, Sajida},
  title   = {Introducing {UWF-ZeekData22}: A Comprehensive Network Traffic Dataset Based on the {MITRE ATT\&CK} Framework},
  journal = {Data},
  volume  = {8},
  number  = {1},
  pages   = {18},
  year    = {2023}
}

@article{bagui2023fall22,
  author  = {Bagui, Sikha S. and others},
  title   = {{UWF-ZeekDataFall22}: An Enterprise {MITRE ATT\&CK} Labeled Network Attack Traffic Dataset},
  journal = {Electronics},
  volume  = {12},
  number  = {24},
  pages   = {5039},
  year    = {2023}
}

@misc{appraise2023,
  author       = {{APPRAISE H2020 project}},
  title        = {{APPRAISE H2020} Real Labelled {NetFlow} Dataset},
  howpublished = {Kaggle, \url{https://www.kaggle.com/datasets/ittibydgoszcz/appraise-h2020-real-labelled-netflow-dataset}},
  year         = {2023}
}

@misc{cremev2,
  title        = {{CREMEv2} Datasets},
  howpublished = {Kaggle, \url{https://www.kaggle.com/datasets/masjohncook/cremev2-datasets}},
  year         = {2023}
}

@misc{bendl2023ctusme11,
  author       = {Bendl, {\v{S}}t{\v{e}}p{\'a}n and Valeros, Veronica and Garcia, Sebastian},
  title        = {{CTU-SME-11}: a labeled dataset with real benign and malicious network traffic mimicking a small medium-size enterprise environment},
  howpublished = {\url{https://zenodo.org/records/7958259}},
  year         = {2023}
}

@article{herzalla2023tii,
  author  = {Herzalla, Dania and Lunardi, Willian T. and Andreoni, Martin},
  title   = {{TII-SSRC-23} Dataset: Typological Exploration of Diverse Traffic Patterns for Intrusion Detection},
  journal = {IEEE Access},
  volume  = {11},
  year    = {2023},
  note    = {Preprint arXiv:2310.10661}
}

@misc{bccccicids2017,
  author       = {{Behaviour-Centric Cybersecurity Center (BCCC), York University}},
  title        = {Intrusion Detection Dataset {BCCC-CIC-IDS2017}},
  howpublished = {\url{https://www.yorku.ca/research/bccc/ucs-technical/cybersecurity-datasets-cds/intrusion-detection-dataset-bccc-cic-ids2017/}},
  year         = {2024}
}

@inproceedings{casinolimit2025,
  author    = {Gimenez, Pierre-Fran{\c{c}}ois and others},
  title     = {{CasinoLimit}: An Offensive Dataset Labeled with {MITRE ATT\&CK} Techniques},
  year      = {2025},
  note      = {IEEE Xplore document 11352480; HAL hal-05224264}
}

@misc{cicunswnb15,
  author       = {{Canadian Institute for Cybersecurity}},
  title        = {{UNSW-NB15} Augmented Dataset ({CIC UNSW-NB15})},
  howpublished = {\url{https://www.unb.ca/cic/datasets/cic-unsw-nb15.html}},
  year         = {2024}
}

@misc{lycosunicas2024,
  author       = {Cantone, Marco},
  title        = {{LycoS-Unicas-IDS2018}},
  howpublished = {\url{https://github.com/MarcoCantone/LycoS-Unicas-IDS2018}},
  year         = {2024}
}

@article{bagui2025zeek24,
  author  = {Bagui, Sikha S. and others},
  title   = {Introducing {UWF-ZeekData24}: An Enterprise {MITRE ATT\&CK} Labeled Network Attack Traffic Dataset for Machine Learning/{AI}},
  journal = {Data},
  volume  = {10},
  number  = {5},
  pages   = {59},
  year    = {2025}
}

@misc{bccccseids2018,
  author       = {{Behaviour-Centric Cybersecurity Center (BCCC), York University}},
  title        = {Large-Scale Intrusion Detection Dataset {BCCC-CSE-CIC-IDS2018}},
  howpublished = {\url{https://www.yorku.ca/research/bccc/ucs-technical/cybersecurity-datasets-cds/large-scale-intrusion-detection-dataset-bccc-cse-cic-ids2018/}},
  year         = {2025}
}

@inproceedings{dedale2025,
  author    = {Lanvin, Maxime and Majorczyk, Fr{\'e}d{\'e}ric},
  title     = {Get out of {DEDALE} with {RESCOUSSE}: a New Dataset and Testbed for Evaluating the Detection of {APT} Attacks among Network and System Logs},
  booktitle = {Computer Security. ESORICS 2025 International Workshops},
  series    = {Lecture Notes in Computer Science},
  volume    = {16232},
  pages     = {3--23},
  year      = {2025},
  publisher = {Springer, Cham},
  note      = {HAL hal-05329482}
}

@article{genis2025,
  author    = {Silva, Miguel and Pinto, Daniela and Vitorino, Jo{\~a}o and Gon{\c{c}}alves, Jos{\'e} and Maia, Eva and Pra{\c{c}}a, Isabel},
  title     = {{GeNIS}: A modular dataset for network intrusion detection and classification},
  journal   = {Data in Brief},
  volume    = {60},
  pages     = {111487},
  year      = {2025},
  month     = jun,
  note      = {PMID: 40226195; PMCID: PMC11986601; Dataset available at Zenodo}
}

@article{luay2025nfv3,
  title={Temporal analysis of netflow datasets for network intrusion detection systems},
  author={Luay, Majed and Layeghy, Siamak and Hosseininoorbin, Seyedehfaezeh and Sarhan, Mohanad and Moustafa, Nour and Portmann, Marius},
  journal={arXiv preprint arXiv:2503.04404},
  year={2025}
}

@misc{safenid2025,
  author       = {Matejek, Brian and Gehani, Ashish and Bastian, Nathaniel D. and Clouse, Daniel J. and Kline, Bradford J. and Jha, Susmit},
  title        = {{SAFE-NID}: Packet-level payload labels for {CIC-IDS2017} and {UNSW-NB15}},
  howpublished = {Zenodo, \url{https://zenodo.org/records/15046995}},
  year         = {2025}
}

@misc{camlds2026,
  author       = {Landauer, Max and Hotwagner, Wolfgang and Boenke, Thorina and Skopik, Florian and Wurzenberger, Markus},
  title        = {{CAM-LDS} network captures},
  howpublished = {Zenodo, \url{https://zenodo.org/records/18701095}},
  year         = {2026}
}

@inproceedings{sheikhi2023ddos,
  author    = {Sheikhi, Saeid and Kostakos, Panos},
  title     = {{DDoS} Attack Detection Using Unsupervised Federated Learning for {5G} Networks and Beyond},
  booktitle = {2023 Joint European Conference on Networks and Communications \& 6G Summit (EuCNC/6G Summit)},
  pages     = {442--447},
  year      = {2023},
  publisher = {IEEE},
  note      = {Dataset: \url{https://github.com/saeidsheikhi/5G-Intrusion-Detection-Dataset}}
}

@misc{sheikhi2026fivegcoresec,
  author       = {Sheikhi, Saeid},
  title        = {{5GCoreSec}: {5G} Core Network Intrusion Detection Dataset},
  year         = {2026},
  howpublished = {\url{https://github.com/saeidsheikhi/5GCoreSec}},
}

@misc{rosa2026cve5g,
  author       = {Rosa, Ana and Santana, Carlos and Hirata, Roberto and Batista, Daniel},
  title        = {{CVE5G}: {5G} Core Network Cyberattacks Dataset},
  year         = {2026},
  publisher    = {IEEE DataPort},
  xxxdoi          = {10.21227/9ex2-k094},
  howpublished = {\url{https://ieee-dataport.org/documents/cve5g-5g-core-network-cyberattacks-dataset}}
}

@misc{wustlhdrl2024,
  author       = {{Washington University in St. Louis}},
  title        = {{WUSTL-HDRL-2024} Dataset},
  year         = {2024},
  howpublished = {\url{https://classes.engineering.wustl.edu/~jain/hdrl/index.html}}
}

@misc{fivegcgad2025,
  author       = {Wengappuli Arachchige, Suranga Prasad and Hewa, Tharaka and Siriwardhana, Yushan},
  title        = {{5G} Core {GTP-U} Attack Dataset},
  year         = {2025},
  publisher    = {University of Oulu, Fairdata},
  xxxdoi          = {10.23729/fd-b9459ea4-88a7-35c2-992e-f178a294c12d},
  howpublished = {\url{https://etsin.fairdata.fi/dataset/5dad463d-0306-455f-b47a-bef771b819fc}}
}

@misc{pfcp2023,
  author       = {Amponis, George and Radoglou-Grammatikis, Panagiotis and Nakas, George and Zevgara, Maria and Giannakidou, Sofia and Ouzounidis, Savvas and Kakamoukas, George and Sarigiannidis, Antonios},
  title        = {{5GC} {PFCP} Intrusion Detection Dataset},
  year         = {2023},
  publisher    = {Zenodo},
  howpublished = {\url{https://zenodo.org/records/7888347}}
}

@inproceedings{fivegad2022,
  author    = {Coldwell, Cody and Conger, Denver and Goodell, Edward and Jacobson, Brendan and Petersen, Bryton and Spencer, Damon and Anderson, Matthew and Sgambati, Matthew},
  title     = {Machine Learning {5G} Attack Detection in Programmable Logic},
  booktitle = {2022 IEEE Globecom Workshops (GC Wkshps)},
  pages     = {1365--1370},
  year      = {2022},
  xxxdoi       = {10.1109/GCWkshps56602.2022.10008647}
}

@misc{samarakoon2022fivegnidd,
  author        = {Samarakoon, Sehan and Siriwardhana, Yushan and Porambage, Pawani and Liyanage, Madhusanka and Chang, Sang-Yoon and Kim, Jinoh and Kim, Jonghyun and Ylianttila, Mika},
  title         = {{5G-NIDD}: A Comprehensive Network Intrusion Detection Dataset Generated over {5G} Wireless Network},
  year          = {2022},
  eprint        = {2212.01298},
  archivePrefix = {arXiv},
  howpublished  = {arXiv:2212.01298}
}

@misc{wehbe2024fiveg,
  author       = {Wehbe, Nadine},
  title        = {{5G} Flow-Based Datasets},
  year         = {2024},
  publisher    = {IEEE DataPort},
  xxxxxxdii          = {10.21227/a3rt-eg64},
  howpublished = {\url{https://ieee-dataport.org/documents/5g-flow-based-datasets}}
}

@misc{khan2023slicing,
  author       = {Khan, Md Sajid and Farzaneh, Behnam and Shahriar, Nashid and Hasan, Md Mahibul},
  title        = {{DoS/DDoS} Attack Dataset of {5G} Network Slicing},
  year         = {2023},
  xxxdoi       = {10.21227/32k1-dr12},
  howpublished = {IEEE DataPort, \url{https://ieee-dataport.org/documents/dosddos-attack-dataset-5g-network-slicing}}
}

@article{hu2026fivegcid,
  author  = {Hu, Y. and Sun, Q. and Dang, T. and Wu, G. and Tian, L.},
  title   = {{5GCID}: Dataset of {5GC} Intrusion Detection System},
  journal = {Computer Communications},
  year    = {2026},
  xxxdoi  = {10.1016/j.comcom.2025.108375},
  note    = {Dataset: \url{https://github.com/SQ0409/5GCID}}
}

@inproceedings{civciss2025netslab,
  author    = {Civciss, A. and Ravihansa, V. and Abed Zadeh, F. and Sandeepa, C. and Liyanage, M.},
  title     = {Silent Signals, Loud Threats: Using {dApps} for Radio Signal Intelligence-based Intrusion Detection in {5G} {O-RAN}},
  booktitle = {2025 IEEE Global Communications Conference (GLOBECOM): Communication \& Information Systems Security},
  year      = {2025},
  note      = {Dataset: \url{https://www.kaggle.com/datasets/netslabdemo/netslab-5g-oran-idd}}
}

@inproceedings{fivegsra2025,
  author    = {Nugraha, Beny and Hajizadeh, Mehrdad and Niehoff, Tim and Jnanashree, Abhishek Venkatesh and Phan, Trung V. and Triantafyllopoulou, Dionysia and Krause, Oliver and Mieth, Martin and Moessner, Klaus and Bauschert, Thomas},
  title     = {A Comprehensive {5G} Network Dataset for Control and Data Plane Security and Resource Allocation Analysis},
  booktitle = {2025 IEEE International Conference on Cyber Security and Resilience (CSR)},
  address   = {Chania, Crete, Greece},
  year      = {2025},
  note      = {Dataset: \url{https://github.com/DLTeamTUC/5GDatasets}}
}

@inproceedings{song2011kyoto,
  author    = {Song, Jungsuk and Takakura, Hiroki and Okabe, Yasuo and Eto, Masashi and Inoue, Daisuke and Nakao, Koji},
  title     = {Statistical Analysis of Honeypot Data and Building of {Kyoto} 2006+ Dataset for {NIDS} Evaluation},
  booktitle = {Proceedings of the First Workshop on Building Analysis Datasets and Gathering Experience Returns for Security (BADGERS)},
  pages     = {29--36},
  year      = {2011}
}

@inproceedings{sperotto2009twente,
  author    = {Sperotto, Anna and Sadre, Ramin and van Vliet, Frank and Pras, Aiko},
  title     = {A Labeled Data Set for Flow-Based Intrusion Detection},
  booktitle = {IP Operations and Management (IPOM 2009)},
  series    = {LNCS},
  volume    = {5843},
  pages     = {39--50},
  publisher = {Springer},
  year      = {2009}
}

@article{valeros2022hornet15,
  author  = {Valeros, Veronica and Garcia, Sebastian},
  title   = {Hornet 15: Network dataset of geographically placed honeypots},
  journal = {Data in Brief},
  volume  = {40},
  pages   = {107795},
  year    = {2022}
}

@misc{hornet65niner2024,
  author       = {{Stratosphere Laboratory}},
  title        = {{CTU} Hornet 65 Niner},
  howpublished = {\url{https://www.stratosphereips.org/hornet-network-dataset-of-geographically-placed-honeypots}},
  year         = {2024}
}

@inproceedings{cho2000mawi,
  author    = {Cho, Kenjiro and Mitsuya, Koushirou and Kato, Akira},
  title     = {Traffic Data Repository at the {WIDE} Project},
  booktitle = {Proceedings of the USENIX Annual Technical Conference, FREENIX Track},
  year      = {2000}
}

@misc{caidaanon,
  author       = {{CAIDA}},
  title        = {The {CAIDA} Anonymized Internet Traces Dataset (April 2008 -- January 2019)},
  howpublished = {\url{https://www.caida.org/catalog/datasets/passive_dataset/}},
  year         = {2019}
}

@inproceedings{fontugne2010mawilab,
  author    = {Fontugne, Romain and Borgnat, Pierre and Abry, Patrice and Fukuda, Kensuke},
  title     = {{MAWILab}: Combining Diverse Anomaly Detectors for Automated Anomaly Labeling and Performance Benchmarking},
  booktitle = {Proceedings of the ACM International Conference on Emerging Networking Experiments and Technologies (CoNEXT)},
  year      = {2010}
}

@article{macia2018ugr16,
  author  = {Maci{\'a}-Fern{\'a}ndez, Gabriel and Camacho, Jos{\'e} and Mag{\'a}n-Carri{\'o}n, Roberto and Garc{\'i}a-Teodoro, Pedro and Ther{\'o}n, Roberto},
  title   = {{UGR'16}: A New Dataset for the Evaluation of Cyclostationarity-Based Network {IDSs}},
  journal = {Computers \& Security},
  volume  = {73},
  pages   = {411--424},
  year    = {2018}
}

@article{damasevicius2020litnet,
  author  = {Dama{\v{s}}evi{\v{c}}ius, Robertas and Ven{\v{c}}kauskas, Algimantas and Grigali{\={u}}nas, {\v{S}}ar{\={u}}nas and Toldinas, Jevgenijus and Morkevi{\v{c}}ius, Nerijus and Aleli{\={u}}nas, Tautvydas and Smuikys, Paulius},
  title   = {{LITNET-2020}: An Annotated Real-World Network Flow Dataset for Network Intrusion Detection},
  journal = {Electronics},
  volume  = {9},
  number  = {5},
  pages   = {800},
  year    = {2020}
}

@article{mihailescu2021simargl,
  author  = {Mihailescu, Maria-Elena and Mihai, Darius and Carabas, Mihai and Komisarek, Miko{\l}aj and Pawlicki, Marek and Ho{\l}ubowicz, Witold and Kozik, Rafa{\l}},
  title   = {The Proposition and Evaluation of the {RoEduNet-SIMARGL2021} Network Intrusion Detection Dataset},
  journal = {Sensors},
  volume  = {21},
  number  = {13},
  pages   = {4319},
  year    = {2021}
}

@inproceedings{flnet2023,
author={Kumar, Pratyay and Liu, Jiefei and Md Tayeen, Abu Saleh and Misra, Satyajayant and Cao, Huiping and Harikumar, Jayashree and Perez, Oscar},
  booktitle={MILCOM 2023 - 2023 IEEE Military Communications Conference (MILCOM)}, 
  title={FLNET2023: Realistic Network Intrusion Detection Dataset for Federated Learning}, 
  year={2023},
  volume={},
  number={},
  pages={345-350},
}

@inproceedings{morris2014ics,
  author    = {Morris, Thomas and Gao, Wei},
  title     = {Industrial Control System Traffic Data Sets for Intrusion Detection Research},
  booktitle = {Critical Infrastructure Protection VIII (ICCIP 2014)},
  series    = {IFIP Advances in Information and Communication Technology},
  volume    = {441},
  pages     = {65--78},
  publisher = {Springer},
  year      = {2014}
}

@article{pan2015power,
  author  = {Pan, Shengyi and Morris, Thomas and Adhikari, Uttam},
  title   = {Developing a Hybrid Intrusion Detection System Using Data Mining for Power Systems},
  journal = {IEEE Transactions on Smart Grid},
  volume  = {6},
  number  = {6},
  pages   = {3104--3113},
  year    = {2015}
}

@misc{fourSICS,
  author       = {{NETRESEC AB}},
  title        = {{4SICS} Geek Lounge {ICS} {PCAP} files},
  howpublished = {\url{https://www.netresec.com/?page=PCAP4SICS}},
  year         = {2015}
}

@inproceedings{lemay2016modbus,
  author    = {Lemay, Antoine and Fernandez, Jos{\'e} M.},
  title     = {Providing {SCADA} Network Data Sets for Intrusion Detection Research},
  booktitle = {9th Workshop on Cyber Security Experimentation and Test (CSET 16)},
  publisher = {USENIX Association},
  year      = {2016}
}

@inproceedings{ahmed2017wadi,
  author    = {Ahmed, Chuadhry Mujeeb and Palleti, Venkata Reddy and Mathur, Aditya P.},
  title     = {{WADI}: A Water Distribution Testbed for Research in the Design of Secure Cyber Physical Systems},
  booktitle = {Proceedings of the 3rd International Workshop on Cyber-Physical Systems for Smart Water Networks (CySWATER)},
  pages     = {25--28},
  year      = {2017}
}

@inproceedings{rodofile2017qut,
  author    = {Rodofile, Nicholas R. and Radke, Kenneth and Foo, Ernest},
  title     = {Framework for {SCADA} Cyber-Attack Dataset Creation},
  booktitle = {Proceedings of the Australasian Computer Science Week Multiconference (ACSW)},
  year      = {2017}
}

@misc{epic2018,
  author       = {{iTrust, Singapore University of Technology and Design}},
  title        = {Electric Power and Intelligent Control ({EPIC}) dataset},
  howpublished = {\url{https://www.sutd.edu.sg/itrust/itrust-labs/datasets/dataset-characteristics/epic/}},
  year         = {2018}
}

@article{perales2019electra,
  author  = {Perales G{\'o}mez, {\'A}ngel Luis and Fern{\'a}ndez Maim{\'o}, Lorenzo and Huertas Celdr{\'a}n, Alberto and Garc{\'i}a Clemente, F{\'e}lix J. and Cadenas Sarmiento, Cristian and Del Canto Masa, Carlos Javier and M{\'e}ndez Nistal, Rub{\'e}n},
  title   = {On the Generation of Anomaly Detection Datasets in Industrial Control Systems},
  journal = {IEEE Access},
  volume  = {7},
  pages   = {177460--177473},
  year    = {2019}
}

@misc{pinto2020opcua,
  author       = {Pinto, Rui and others},
  title        = {{M2M} using {OPC UA}},
  howpublished = {IEEE Dataport, \url{https://ieee-dataport.org/open-access/m2m-using-opc-ua}},
  year         = {2020}
}

@article{faramondi2021wdt,
  author  = {Faramondi, Luca and Flammini, Francesco and Guarino, Simone and Setola, Roberto},
  title   = {A Hardware-in-the-Loop Water Distribution Testbed Dataset for Cyber-Physical Security Testing},
  journal = {IEEE Access},
  volume  = {9},
  pages   = {122385--122396},
  year    = {2021}
}

@misc{dnp3ids2022,
  author       = {Radoglou-Grammatikis, Panagiotis and others},
  title        = {{DNP3} Intrusion Detection Dataset},
  howpublished = {Zenodo, \url{https://zenodo.org/records/7348493}},
  year         = {2022}
}

@misc{dtuhoneypots2022,
  author       = {Technical University of Denmark},
  title        = {A dataset of hybrid {IoT}/{OT} honeypots},
  howpublished = {Technical University of Denmark, \url{https://data.dtu.dk/articles/dataset/A_dataset_of_hybrid_IoT_OT_honeypots/21088651}},
  year         = {2022}
}

@misc{iec104ids2022,
  author       = {Radoglou-Grammatikis, Panagiotis and others},
  title        = {{IEC} 60870-5-104 Intrusion Detection Dataset},
  howpublished = {Zenodo, \url{https://zenodo.org/records/7108614}},
  year         = {2022}
}

@misc{cicmodbus2023,
  author       = {{Canadian Institute for Cybersecurity}},
  title        = {{CIC} Modbus Dataset 2023},
  howpublished = {\url{https://www.unb.ca/cic/datasets/modbus-2023.html}},
  year         = {2023}
}

@article{dehlaghi2023icsflow,
  author  = {Dehlaghi-Ghadim, Alireza and Moghadam, Mahshid Helali and Balador, Ali and Hansson, Hans},
  title   = {Anomaly Detection Dataset for Industrial Control Systems},
  journal = {IEEE Access},
  volume  = {11},
  pages   = {107982--107996},
  year    = {2023}
}

@article{spravil2023detecting,
  author  = {Spravil, Julian and Hemminghaus, Christian and von Rechenberg, M. and Padilla, E. and Bauer, Jan},
  title   = {Detecting Maritime {GPS} Spoofing Attacks Based on {NMEA} Sentence Integrity Monitoring},
  journal = {Journal of Marine Science and Engineering},
  volume  = {11},
  number  = {5},
  pages   = {928},
  year    = {2023},
  note    = {Dataset: MARitime SIMulated (MARSIM)}
}

@misc{westermo2023,
  author       = {{Westermo Network Technologies}},
  title        = {Westermo network traffic data set},
  howpublished = {\url{https://github.com/westermo/network-traffic-dataset}},
  year         = {2023}
}

@misc{cyber4ot2025,
  author       = {Cabaj, Krzysztof and Plamowski, Sebastian and Chaber, Patryk and {\L}awry{\'n}czuk, Maciej and Marusak, Piotr and Nebeluk, Robert and Wojtulewicz, Andrzej and Zarzycki, Krzysztof},
  title        = {{Cyber4OT}: {ICS} network traces containing normal activity and full attack traffic},
  howpublished = {Zenodo, \url{https://zenodo.org/records/15005190}},
  year         = {2025}
}

@misc{icscasd2025,
  author       = {Houkan, Ahmad},
  title        = {{ICSCASD-MPLC}: Industrial Control System Cyber-Attack Simulation Dataset for Mitsubishi PLCs},
  howpublished = {Zenodo, \url{https://zenodo.org/records/16351397}},
  year         = {2025}
}

@inproceedings{holland2021nprint,
  author    = {Holland, Jordan and Schmitt, Paul and Feamster, Nick and Mittal, Prateek},
  title     = {New Directions in Automated Traffic Analysis},
  booktitle = {Proceedings of the 2021 ACM SIGSAC Conference on Computer and Communications Security (CCS)},
  pages     = {3366--3383},
  year      = {2021}
}

@inproceedings{lin2022etbert,
  author    = {Lin, Xinjie and Xiong, Gang and Gou, Gaopeng and Li, Zhen and Shi, Junzheng and Yu, Jing},
  title     = {{ET-BERT}: A Contextualized Datagram Representation with Pre-training Transformers for Encrypted Traffic Classification},
  booktitle = {Proceedings of the ACM Web Conference 2022 (WWW)},
  pages     = {633--642},
  year      = {2022}
}

@inproceedings{zhao2023yatc,
  author    = {Zhao, Ruijie and Zhan, Mingwei and Deng, Xianwen and Wang, Yanhao and Wang, Yijun and Gui, Guan and Xue, Zhi},
  title     = {Yet Another Traffic Classifier: A Masked Autoencoder Based Traffic Transformer with Multi-Level Flow Representation},
  booktitle = {Proceedings of the AAAI Conference on Artificial Intelligence},
  volume    = {37},
  number    = {4},
  pages     = {5420--5427},
  year      = {2023}
}

@inproceedings{mirsky2018kitsune,
  author    = {Mirsky, Yisroel and Doitshman, Tomer and Elovici, Yuval and Shabtai, Asaf},
  title     = {Kitsune: An Ensemble of Autoencoders for Online Network Intrusion Detection},
  booktitle = {Network and Distributed System Security Symposium (NDSS)},
  year      = {2018}
}

@article{meidan2018nbaiot,
  author  = {Meidan, Yair and Bohadana, Michael and Mathov, Yael and Mirsky, Yisroel and Shabtai, Asaf and Breitenbacher, Dominik and Elovici, Yuval},
  title   = {{N-BaIoT}---Network-Based Detection of {IoT} Botnet Attacks Using Deep Autoencoders},
  journal = {IEEE Pervasive Computing},
  volume  = {17},
  number  = {3},
  pages   = {12--22},
  year    = {2018}
}

@inproceedings{hamza2019mud,
  author    = {Hamza, Ayyoob and Gharakheili, Hassan Habibi and Benson, Theophilus A. and Sivaraman, Vijay},
  title     = {Detecting Volumetric Attacks on {IoT} Devices via {SDN}-Based Monitoring of {MUD} Activity},
  booktitle = {Proceedings of the ACM Symposium on SDN Research (SOSR)},
  pages     = {36--48},
  year      = {2019}
}

@article{sivanathan2019iot,
  author  = {Sivanathan, Arunan and Gharakheili, Hassan Habibi and Loi, Franco and Radford, Adam and Wijenayake, Chamith and Vishwanath, Arun and Sivaraman, Vijay},
  title   = {Classifying {IoT} Devices in Smart Environments Using Network Traffic Characteristics},
  journal = {IEEE Transactions on Mobile Computing},
  volume  = {18},
  number  = {8},
  pages   = {1745--1759},
  year    = {2019}
}

@misc{kang2019iotnid,
  author       = {Kang, Hyunjae and Ahn, Dong Hyun and Lee, Gyung Min and Yoo, Jeong Do and Park, Kyung Ho and Kim, Huy Kang},
  title        = {{IoT} Network Intrusion Dataset},
  howpublished = {\url{https://ieee-dataport.org/open-access/iot-network-intrusion-dataset}},
  year         = {2019}
}

@misc{dad2020,
  author       = {{DAD Repository}},
  title        = {{DAD}: Dataset for Anomaly Detection in {IoT}/{MQTT} networks},
  howpublished = {\url{https://github.com/dad-repository/dad}},
  year         = {2020}
}

@article{moustafa2021toniot,
  author  = {Moustafa, Nour},
  title   = {A new distributed architecture for evaluating {AI}-based security systems at the edge: Network {TON\_IoT} datasets},
  journal = {Sustainable Cities and Society},
  volume  = {72},
  pages   = {102994},
  year    = {2021}
}

@article{ahmed2021ecuioht,
  author  = {Ahmed, Mohiuddin and Byreddy, Surender and Nutakki, Anirudh and Sikos, Leslie F. and Haskell-Dowland, Paul},
  title   = {{ECU-IoHT}: A dataset for analyzing cyberattacks in {Internet of Health Things}},
  journal = {Ad Hoc Networks},
  volume  = {122},
  pages   = {102621},
  year    = {2021}
}

@inproceedings{ullah2020iotid20,
  author    = {Ullah, Imtiaz and Mahmoud, Qusay H.},
  title     = {A Scheme for Generating a Dataset for Anomalous Activity Detection in {IoT} Networks},
  booktitle = {Advances in Artificial Intelligence (Canadian AI 2020)},
  series    = {LNCS},
  volume    = {12109},
  pages     = {508--520},
  publisher = {Springer},
  year      = {2020}
}

@inproceedings{guerra2020medbiot,
  author    = {Guerra-Manzanares, Alejandro and Medina-Galindo, Jorge and Bahsi, Hayretdin and N{\~o}mm, Sven},
  title     = {{MedBIoT}: Generation of an {IoT} Botnet Dataset in a Medium-sized {IoT} Network},
  booktitle = {Proceedings of the 6th International Conference on Information Systems Security and Privacy (ICISSP)},
  pages     = {207--218},
  year      = {2020}
}

@inproceedings{hindy2021mqttiotids,
  author    = {Hindy, Hanan and Bayne, Ethan and Bures, Miroslav and Atkinson, Robert and Tachtatzis, Christos and Bellekens, Xavier},
  title     = {Machine Learning Based {IoT} Intrusion Detection System: An {MQTT} Case Study ({MQTT-IoT-IDS2020} Dataset)},
  booktitle = {Selected Papers from the 12th International Networking Conference (INC 2020)},
  series    = {Lecture Notes in Networks and Systems},
  volume    = {180},
  pages     = {73--84},
  publisher = {Springer},
  year      = {2021}
}

@article{vaccari2020mqttset,
  author  = {Vaccari, Ivan and Chiola, Giovanni and Aiello, Maurizio and Mongelli, Maurizio and Cambiaso, Enrico},
  title   = {{MQTTset}, a New Dataset for Machine Learning Techniques on {MQTT}},
  journal = {Sensors},
  volume  = {20},
  number  = {22},
  pages   = {6578},
  year    = {2020}
}

@article{hady2020wustlehms,
  author  = {Hady, Anar A. and Ghubaish, Ali and Salman, Tara and Unal, Devrim and Jain, Raj},
  title   = {Intrusion Detection System for Healthcare Systems Using Medical and Network Data: A Comparison Study},
  journal = {IEEE Access},
  volume  = {8},
  pages   = {106576--106584},
  year    = {2020}
}

@article{alhawawreh2022xiiotid,
  author  = {Al-Hawawreh, Muna and Sitnikova, Elena and Aboutorab, Neda},
  title   = {{X-IIoTID}: A Connectivity-Agnostic and Device-Agnostic Intrusion Data Set for Industrial {Internet of Things}},
  journal = {IEEE Internet of Things Journal},
  volume  = {9},
  number  = {5},
  pages   = {3962--3977},
  year    = {2022}
}

@inproceedings{dadkhah2022ciciot2022,
  author    = {Dadkhah, Sajjad and Mahdikhani, Hassan and Danso, Priscilla Kyei and Zohourian, Alireza and Truong, Kevin Anh and Ghorbani, Ali A.},
  title     = {Towards the Development of a Realistic Multidimensional {IoT} Profiling Dataset},
  booktitle = {19th Annual International Conference on Privacy, Security \& Trust (PST)},
  year      = {2022}
}

@article{ferrag2022edgeiiotset,
  author  = {Ferrag, Mohamed Amine and Friha, Othmane and Hamouda, Djallel and Maglaras, Leandros and Janicke, Helge},
  title   = {{Edge-IIoTset}: A New Comprehensive Realistic Cyber Security Dataset of {IoT} and {IIoT} Applications for Centralized and Federated Learning},
  journal = {IEEE Access},
  volume  = {10},
  pages   = {40281--40306},
  year    = {2022}
}

@article{almaraz2022latam,
  author  = {Almaraz-Rivera, Josue Genaro and Perez-Diaz, Jesus Arturo and Cantoral-Ceballos, Jose Antonio and Botero, Juan Felipe and Trejo, Luis A.},
  title   = {Toward the Protection of {IoT} Networks: Introducing the {LATAM-DDoS-IoT} Dataset},
  journal = {IEEE Access},
  volume  = {10},
  pages   = {106909--106920},
  year    = {2022}
}

@article{sharmila2023rtiot,
  author  = {Sharmila, B. S. and Nagapadma, Rohini},
  title   = {Quantized autoencoder ({QAE}) intrusion detection system for anomaly detection in resource-constrained {IoT} devices using {RT-IoT2022} dataset},
  journal = {Cybersecurity},
  volume  = {6},
  pages   = {41},
  year    = {2023}
}

@misc{bastian2023aciiot,
  author       = {Bastian, Nathaniel and Bierbrauer, David and McKenzie, Merrill and Nack, Emily},
  title        = {{ACI IoT} Network Traffic Dataset 2023},
  howpublished = {\url{https://ieee-dataport.org/documents/aci-iot-network-traffic-dataset-2023}},
  year         = {2023}
}

@article{alatram2023dosmqtt,
  author  = {Alatram, Alaa and Sikos, Leslie F. and Johnstone, Mike and Szewczyk, Patryk and Kang, James Jin},
  title   = {{DoS/DDoS-MQTT-IoT}: A dataset for evaluating intrusions in {IoT} networks using the {MQTT} protocol},
  journal = {Computer Networks},
  volume  = {231},
  pages   = {109809},
  year    = {2023}
}

@article{areia2024iomt,
  author  = {Areia, Jos{\'e} and Bispo, Ivo A. and Santos, Leonel and Costa, Rog{\'e}rio Lu{\'i}s de C.},
  title   = {{IoMT-TrafficData}: Dataset and Tools for Benchmarking Intrusion Detection in {Internet of Medical Things}},
  journal = {IEEE Access},
  volume  = {12},
  year    = {2024}
}

@article{rosids23,
  author  = {De{\u{g}}irmenci, Elif and K{\i}rca, Yunus Sabri and {\"O}z{\c{c}}elik, {\.I}lker and Yaz{\i}c{\i}, Ahmet},
  title   = {{ROSIDS23}: Network intrusion detection dataset for robot operating system},
  journal = {Data in Brief},
  volume  = {50},
  pages   = {109739},
  year    = {2023}
}

@misc{ghiasvand2024cicapt,
  author        = {Ghiasvand, Erfan and Ray, Suprio and Iqbal, Shahrear and Dadkhah, Sajjad and Ghorbani, Ali A.},
  title         = {{CICAPT-IIOT}: A provenance-based {APT} attack dataset for {IIoT} environment},
  year          = {2024},
  eprint        = {2407.11278},
  archivePrefix = {arXiv}
}

@inproceedings{buedi2024cicevse,
  author    = {Buedi, Emmanuel Dana and Ghorbani, Ali A. and Dadkhah, Sajjad and Ferreira, Raphael Lionel},
  title     = {Enhancing {EV} Charging Station Security Using a Multi-dimensional Dataset: {CICEVSE2024}},
  booktitle = {Data and Applications Security and Privacy XXXVIII (DBSec 2024)},
  series    = {LNCS},
  publisher = {Springer},
  year      = {2024}
}

@misc{iotdh2024,
  author       = {{Saif, Syaifuddin and Ferdiana, Ridi and Widyawan, Widyawan }},
  title        = {{IoT-DH} Dataset},
  howpublished = {Mendeley Data, \url{https://data.mendeley.com/datasets/8dns3xbckv/1}},
  year         = {2024}
}

@article{dadkhah2024ciciomt,
  author  = {Dadkhah, Sajjad and Neto, Euclides Carlos Pinto and Ferreira, Raphael and Molokwu, Reginald Chukwuka and Sadeghi, Somayeh and Ghorbani, Ali A.},
  title   = {{CICIoMT2024}: A benchmark dataset for multi-protocol security assessment in {IoMT}},
  journal = {Internet of Things},
  volume  = {28},
  pages   = {101351},
  year    = {2024}
}

@article{rospace2024,
  author  = {Puccetti, Tommaso and Nardi, Simone and Cinquilli, Cosimo and Zoppi, Tommaso and Ceccarelli, Andrea},
  title   = {{ROSPaCe}: Intrusion Detection Dataset for a {ROS2}-Based Cyber-Physical System and {IoT} Networks},
  journal = {Scientific Data},
  volume  = {11},
  pages   = {503},
  year    = {2024}
}

@misc{ciciotdiad2024,
  author       = {{Canadian Institute for Cybersecurity}},
  title        = {{CIC IoT-DIAD} 2024},
  howpublished = {\url{https://www.unb.ca/cic/datasets/iot-diad-2024.html}},
  year         = {2024}
}

@misc{tabulariot2024,
  author       = {{Canadian Institute for Cybersecurity}},
  title        = {{CIC-BCCC-NRC} {TabularIoTAttack-2024}},
  howpublished = {\url{https://www.unb.ca/cic/datasets/tabular-iot-attack-2024.html}},
  year         = {2024}
}

@misc{bcccZwave2025,
  author       = {{Behaviour-Centric Cybersecurity Center (BCCC), York University}},
  title        = {{BCCC-IoT-IDS-Zwave-2025}},
  howpublished = {\url{https://www.yorku.ca/research/bccc/ucs-technical/cybersecurity-datasets-cds/}},
  year         = {2025}
}

@misc{bcccMqtt2025,
  author       = {{Behaviour-Centric Cybersecurity Center (BCCC), York University}},
  title        = {{BCCC-IoT-MQTT-IDS-2025}},
  howpublished = {\url{https://www.yorku.ca/research/bccc/ucs-technical/cybersecurity-datasets-cds/}},
  year         = {2025}
}

@misc{nims2025b,
  author       = {{NIMS Lab, Dalhousie University}},
  title        = {Dalhousie {NIMS} Lab Attack {IoT} Dataset 2025-2},
  howpublished = {IEEE Dataport, \url{https://ieee-dataport.org/documents/dalhousie-nims-lab-attack-iot-dataset-2025-2}},
  year         = {2025}
}

@misc{nims2025a,
  author       = {{NIMS Lab, Dalhousie University}},
  title        = {Dalhousie {NIMS} Lab {IoT} Attack Dataset 2025-1},
  howpublished = {IEEE Dataport, \url{https://ieee-dataport.org/documents/dalhousie-nims-lab-iot-attack-dataset-2025-1}},
  year         = {2025}
}

@misc{datasense2025,
  author       = {{Canadian Institute for Cybersecurity}},
  title        = {{DataSense}: {CIC} {IIoT} Dataset 2025},
  howpublished = {Canadian Institute for Cybersecurity, University of New Brunswick},
  year         = {2025}
}

@misc{gotham2025,
  author       = {Belarbi, Othmane and Spyridopoulos, Theodore and Anthi, Elisavet and Rana, Omer and Carnelli, Pete and Khan, Awais},
  title        = {Gotham Device-Level {IoT} Network Traffic},
  howpublished = {Zenodo, \url{https://zenodo.org/records/14502760}},
  year         = {2025}
}

@misc{mqtteebd2025,
  author       = {Aqachtoul, Abderrahmane and Karam, Khaoula and Elamrani, Abderrahmane and Najib, Mehdi and Rafalia, Najat and Bakhouya, Mohamed},
  title        = {{MQTTEEB-D}: A Real-World {IoT} Cybersecurity Dataset for {AI}-Powered Threat Detection in {MQTT} Networks},
  howpublished = {Mendeley Data, \url{https://data.mendeley.com/datasets/jfttfjn6tr/1}},
  year         = {2025}
}

@misc{mqttuad2025,
  author       = {{Héctor Alaiz-Moretón, Jose Antonio Aveleira-Mata, Saúl Díez Fernández, Ángel Luis Muñoz Castañeda, Isaías García-Rodríguez, Carmen Benavides, José Alberto Benítez-Andrades}},
  title        = {{MQTT\_UAD}: {MQTT} Under Attack Dataset},
  howpublished = {FigShare},
  year         = {2025}
}

@misc{unswiotraffic2025,
  author       = {{UNSW IoT Analytics}},
  title        = {{UNSW-IoTraffic}},
  howpublished = {\url{https://iotanalytics.unsw.edu.au/unsw-iotraffic.html}},
  year         = {2025}
}

@misc{cyberflowiot2026,
  author       = {Martinez Gonzalez, B. A. and Baseca, C. and Munoz, D.},
  title        = {{CyberFlowIoT-GICAP}: Labelled {IoT} flow-based traffic dataset},
  howpublished = {\url{https://hdl.handle.net/10259/11273}},
  year         = {2026}
}

@misc{iotwinrl2026,
  author       = {{Rodríguez-Villagrá, César and Martín-Reizábal, Sergio and Ruiz-Gonzalez, Ruben and Basurto, Nuño1 and Herrero, Alvaro}},
  title        = {Labeled {IoT} Window-Based Random Network Pattern Dataset for Reinforcement Learning},
  howpublished = {Zenodo, \url{https://zenodo.org/records/19238988}},
  year         = {2026}
}

@misc{iotsfc2026,
  author       = {{Oubaalla, Abdelwahid}},
  title        = {{IoT-SFC} v2},
  howpublished = {Mendeley Data, \url{https://data.mendeley.com/datasets/j9xffc2psb/5}},
  year         = {2026}
}

@misc{cicynu2026,
  author       = {{Canadian Institute for Cybersecurity}},
  title        = {{CIC-YNU-IoTMal} Dataset 2026},
  howpublished = {\url{https://www.unb.ca/cic/datasets/ynu-iot-2026.html}},
  year         = {2026}
}

@article{teixeira2018wustl,
  author  = {Teixeira, Marcio Andrey and Salman, Tara and Zolanvari, Maede and Jain, Raj and Meskin, Nader and Samaka, Mohammed},
  title   = {{SCADA} System Testbed for Cybersecurity Research Using Machine Learning Approach},
  journal = {Future Internet},
  volume  = {10},
  number  = {8},
  pages   = {76},
  year    = {2018}
}

@misc{wustliiot2021,
  author       = {Zolanvari, Maede and Jain, Raj},
  title        = {{WUSTL-IIoT-2021} Dataset for {IIoT} Cybersecurity Research},
  howpublished = {\url{https://www.cse.wustl.edu/~jain/iiot2/index.html}},
  year         = {2021}
}

@inproceedings{saad2011isot,
  author    = {Saad, Sherif and Traore, Issa and Ghorbani, Ali and Sayed, Bassam and Zhao, David and Lu, Wei and Felix, John and Hakimian, Payman},
  title     = {Detecting {P2P} botnets through network behavior analysis and machine learning},
  booktitle = {Ninth Annual International Conference on Privacy, Security and Trust (PST)},
  pages     = {174--180},
  year      = {2011}
}

@misc{vast2012,
  author       = {{IEEE VAST Challenge}},
  title        = {{VAST} Challenge 2012: Bank of Money network security},
  howpublished = {\url{https://visualdata.wustl.edu/varepository/benchmarks.php}},
  year         = {2012}
}

@misc{mtanet,
  author       = {Duncan, Brad},
  title        = {Malware-Traffic-Analysis.net},
  howpublished = {\url{https://www.malware-traffic-analysis.net/}}
}

@inproceedings{beigi2014iscxbot,
  author    = {Biglar Beigi, Elaheh and Hadian Jazi, Hossein and Stakhanova, Natalia and Ghorbani, Ali A.},
  title     = {Towards effective feature selection in machine learning-based botnet detection approaches},
  booktitle = {IEEE Conference on Communications and Network Security (CNS)},
  pages     = {247--255},
  year      = {2014}
}

@misc{ctumixed,
  author       = {{Stratosphere Laboratory}},
  title        = {Mixed Captures},
  howpublished = {\url{https://www.stratosphereips.org/datasets-mixed}},
  year         = {2026},
}

@inproceedings{lashkari2017aagm,
  author    = {Lashkari, Arash Habibi and Kadir, Andi Fitriah A. and Gonzalez, Hugo and Mbah, Kenneth Fon and Ghorbani, Ali A.},
  title     = {Towards a Network-Based Framework for {Android} Malware Detection and Characterization},
  booktitle = {15th Annual Conference on Privacy, Security and Trust (PST)},
  pages     = {233--242},
  year      = {2017}
}

@inproceedings{lashkari2018andmal,
  author    = {Lashkari, Arash Habibi and Kadir, Andi Fitriah A. and Taheri, Laya and Ghorbani, Ali A.},
  title     = {Toward Developing a Systematic Approach to Generate Benchmark {Android} Malware Datasets and Classification},
  booktitle = {International Carnahan Conference on Security Technology (ICCST)},
  year      = {2018}
}

@inproceedings{alenazi2017isothttp,
  author    = {Alenazi, Abdelraman and Traore, Issa and Ganame, Karim and Woungang, Isaac},
  title     = {Holistic Model for {HTTP} Botnet Detection Based on {DNS} Traffic Analysis},
  booktitle = {Intelligent, Secure, and Dependable Systems in Distributed and Cloud Environments (ISDDC 2017)},
  series    = {LNCS},
  volume    = {10618},
  pages     = {1--18},
  publisher = {Springer},
  year      = {2017}
}

@inproceedings{taheri2019invesandmal,
  author    = {Taheri, Laya and Kadir, Andi Fitriah A. and Lashkari, Arash Habibi},
  title     = {Extensible {Android} Malware Detection and Family Classification Using Network-Flows and {API}-Calls},
  booktitle = {International Carnahan Conference on Security Technology (ICCST)},
  year      = {2019}
}

@inproceedings{letteri2020mtakdd,
  author    = {Letteri, Ivan and Della Penna, Giuseppe and Di Vita, Luca and Grifa, Maria Teresa},
  title     = {{MTA-KDD'19}: A Dataset for Malware Traffic Detection},
  booktitle = {Proceedings of the Fourth Italian Conference on Cyber Security (ITASEC 2020)},
  series    = {CEUR Workshop Proceedings},
  volume    = {2597},
  pages     = {153--165},
  year      = {2020}
}

@inproceedings{mahdavifar2020maldroid,
  author    = {Mahdavifar, Samaneh and Abdul Kadir, Andi Fitriah and Fatemi, Rasool and Alhadidi, Dima and Ghorbani, Ali A.},
  title     = {Dynamic {Android} Malware Category Classification using Semi-Supervised Deep Learning},
  booktitle = {IEEE Intl Conf on Dependable, Autonomic and Secure Computing (DASC/PiCom/CBDCom/CyberSciTech)},
  pages     = {515--522},
  year      = {2020}
}

@misc{androidmischief,
  author       = {Valeros, Veronica and Garcia, Sebastian},
  title        = {Android Mischief Dataset},
  howpublished = {Stratosphere Laboratory, \url{https://www.stratosphereips.org/android-mischief-dataset}},
  year         = {2020}
}

@article{barut2020netml,
  title={NetML: A challenge for network traffic analytics},
  author={Barut, Onur and Luo, Yan and Zhang, Tong and Li, Weigang and Li, Peilong},
  journal={arXiv preprint arXiv:2004.13006},
  year={2020}
}

@inproceedings{rahali2020andmal2020,
  author    = {Rahali, Abir and Lashkari, Arash Habibi and Kaur, Gurdip and Taheri, Laya and Gagnon, Fran{\c{c}}ois and Massicotte, Fr{\'e}d{\'e}ric},
  title     = {{DIDroid}: {Android} Malware Classification and Characterization Using Deep Image Learning},
  booktitle = {10th International Conference on Communication and Network Security (ICCNS)},
  pages     = {70--82},
  year      = {2020}
}

@misc{malwaretraffic2026,
  author       = {Root0ne},
  title        = {{MalwareTraffic}: liveness-annotated malware {PCAP} dataset},
  howpublished = {\url{https://github.com/Root0ne/malware-pcap-dataset}},
  year         = {2026}
}

@inproceedings{jordaney2017transcend,
  title={Transcend: Detecting concept drift in malware classification models},
  author={Jordaney, Roberto and Sharad, Kumar and Dash, Santanu K and Wang, Zhi and Papini, Davide and Nouretdinov, Ilia and Cavallaro, Lorenzo},
  booktitle={26th USENIX security symposium (USENIX security 17)},
  pages={625--642},
  year={2017}
}

@article{aldribi2020isotcid,
  author  = {Aldribi, Abdulaziz and Traor{\'e}, Issa and Moa, Belaid and Nwamuo, Onyekachi},
  title   = {Hypervisor-based cloud intrusion detection through online multivariate statistical change tracking},
  journal = {Computers \& Security},
  volume  = {88},
  pages   = {101646},
  year    = {2020}
}

@misc{ahuja2020sdnddos,
  author       = {Ahuja, Nisha and Singal, Gaurav and Mukhopadhyay, Debajyoti},
  title        = {{DDOS} attack {SDN} Dataset},
  year         = {2020},
  howpublished = {\url{https://data.mendeley.com/datasets/b7vw628825/1}}
}

@article{elsayed2020insdn,
  author  = {Elsayed, Mahmoud Said and Le-Khac, Nhien-An and Jurcut, Anca Delia},
  title   = {{InSDN}: A Novel {SDN} Intrusion Dataset},
  journal = {IEEE Access},
  volume  = {8},
  pages   = {165263--165284},
  year    = {2020}
}

@inproceedings{das2023unridd,
  author    = {Das, Tapadhir and Hamdan, Osama Abu and Shukla, Raj Mani and Sengupta, Shamik and Arslan, Engin},
  title     = {{UNR-IDD}: Intrusion Detection Dataset using Network Port Statistics},
  booktitle = {IEEE 20th Consumer Communications \& Networking Conference (CCNC)},
  year      = {2023}
}

@misc{bccccpacket2024,
  author       = {{Behaviour-Centric Cybersecurity Center (BCCC), York University}},
  title        = {Cloud {DDoS} Attacks ({BCCC-cPacket-Cloud-DDoS-2024})},
  howpublished = {\url{https://www.yorku.ca/research/bccc/ucs-technical/cybersecurity-datasets-cds/cloud-ddos-attacks-bccc-cpacket-cloud-ddos-2024/}},
  year         = {2024}
}

@misc{uavcyberphys2023,
  title        = {Cyber-Physical Dataset for {UAVs} Under Normal Operations and Cyber-Attacks},
  howpublished = {IEEE Dataport, \url{https://ieee-dataport.org/documents/cyber-physical-dataset-uavs-under-normal-operations-and-cyber-attacks}},
  year         = {2023}
}

@misc{isotdrone2024,
  author       = {{ISOT Research Lab, University of Victoria}},
  title        = {{ISOT} Drone Anomaly Detection Dataset},
  howpublished = {\url{https://onlineacademiccommunity.uvic.ca/isot/}},
  year         = {2024}
}

@misc{uavnidd2025,
  author       = {{CyberSaR Lab, KAUST}},
  title        = {{UAV-NIDD}: {UAV} Intrusion Detection Dataset},
  howpublished = {\url{https://github.com/CyberSaR-KAUST/UAV-Intrusion-Detection-Dataset}},
  year         = {2025}
}

@misc{uavids2025,
  title        = {{UAVIDS-2025}: Benchmark Dataset for Intrusion Detection in {UAV} Networks Using Machine Learning},
  howpublished = {IEEE Dataport, \url{https://ieee-dataport.org/documents/uavids-2025-benchmark-dataset-intrusion-detection-uav-networks-using-machine-learning}},
  year         = {2025}
}

@misc{gcsnidd2026,
  author       = {{CyberSaR Lab, KAUST}},
  title        = {{GCS-NIDD}: {UAV} Ground Control Station Intrusion Detection Dataset},
  howpublished = {\url{https://github.com/CyberSaR-KAUST/UAV-GCS-Intrusion-Detection-Dataset-UAV-GCS-IDS-}},
  year         = {2026}
}

@misc{goncalves2020v2x,
  author       = {Gon{\c{c}}alves, Fabr{\'i}cio and others},
  title        = {Synthesized datasets for {VANET} security threats},
  howpublished = {Zenodo},
  year         = {2020}
}

@inproceedings{kamel2020veremiext,
  author    = {Kamel, Joseph and Wolf, Michael and van der Heijden, Rens W. and Kaiser, Arnaud and Urien, Pascal and Kargl, Frank},
  title     = {{VeReMi} Extension: A Dataset for Comparable Evaluation of Misbehavior Detection in {VANETs}},
  booktitle = {IEEE International Conference on Communications (ICC)},
  year      = {2020}
}

@misc{yang2022vanetddos,
  author       = {Yang, Fan and others},
  title        = {{DDoS} attacks dataset in {VANETs}},
  howpublished = {\url{https://github.com/YangFanAHU/DDoS-attacks-dataset-in-VANETs}},
  year         = {2022}
}

@misc{misbehaviorx2024,
  author       = {Shahriar, Md Hasan and Ansari, Mohammad Raashid and Monteuuis, Jean-Philippe and Chen, Cong and Petit, Jonathan and Hou, Y. Thomas and Lou, Wenjing},
  title        = {MisbehaviorX: Comprehensive V2X Misbehavior Detection Dataset Enabled by the V2X Application Spoofing Platform},
  howpublished = {\url{https://ieee-dataport.org/documents/misbehaviorx-comprehensive-v2x-misbehavior-detection-dataset-enabled-v2x-application}},
  year         = {2024}
}

@misc{veremings2026,
  author       = {{VeReMi NextGen Contributors}},
  title        = {{VeReMi} {NextGen}},
  howpublished = {\url{https://github.com/VeReMi-dataset/VeReMi-NextGen}},
  year         = {2026}
}

@article{almomani2016wsnds,
  author  = {Almomani, Iman and Al-Kasasbeh, Bassam and Al-Akhras, Mousa},
  title   = {{WSN-DS}: A Dataset for Intrusion Detection Systems in Wireless Sensor Networks},
  journal = {Journal of Sensors},
  volume  = {2016},
  pages   = {4731953},
  year    = {2016}
}

@article{chatzoglou2021awid3,
  author  = {Chatzoglou, Efstratios and Kambourakis, Georgios and Kolias, Constantinos},
  title   = {Empirical Evaluation of Attacks Against {IEEE} 802.11 Enterprise Networks: The {AWID3} Dataset},
  journal = {IEEE Access},
  volume  = {9},
  pages   = {34188--34205},
  year    = {2021}
}

@misc{lorawanjam2026,
  author       = {Pourghasem, Azita},
  title        = {Semi-Synthetic {LoRaWAN} Dataset for Jamming and Battery-Depletion Attack Detection},
  version      = {1.0.0},
  howpublished = {Zenodo},
  url          = {https://zenodo.org/records/22043018},
  year         = {2026}
}

@misc{wpa3attacks2026,
  author       = {Halbouni, Asmaa and Ong, Lee-Yeng and Leow, Meng-Chew},
  title        = {{WPA3} Attacks Dataset},
  howpublished = {Mendeley Data, \url{https://data.mendeley.com/datasets/cxx5t5nw7z}},
  year         = {2026}
}

@article{ring2019,
  author  = {Ring, Markus and Wunderlich, Sarah and Scheuring, Deniz and Landes, Dieter and Hotho, Andreas},
  title   = {A survey of network-based intrusion detection data sets},
  journal = {Computers \& Security},
  volume  = {86},
  pages   = {147--167},
  year    = {2019},
  xxxdoi     = {10.1016/j.cose.2019.06.005}
}

@article{goldschmidt2025,
  author  = {Goldschmidt, Patrik and Chud{\'a}, Daniela},
  title   = {Network intrusion datasets: A survey, limitations, and recommendations},
  journal = {Computers \& Security},
  volume  = {156},
  pages   = {104510},
  year    = {2025},
  publisher = {Elsevier}
}

@inproceedings{comidds2024,
  author    = {B{\"o}nninghausen, Philipp and Uetz, Rafael and Henze, Martin},
  title     = {Introducing a comprehensive, continuous, and collaborative survey of intrusion detection datasets},
  booktitle = {Proceedings of the 17th Cyber Security Experimentation and Test Workshop (CSET)},
  pages     = {34--40},
  year      = {2024},
  xxxdoi       = {10.1145/3675741.3675754}
}

@misc{comiddsrepo,
  author       = {{COMIDDS contributors}},
  title        = {{COMIDDS}: A Comprehensive, Continuous, and Collaborative Survey of Intrusion Detection Datasets},
  howpublished = {\url{https://github.com/fkie-cad/COMIDDS}}
}

@misc{xgoldyrepo,
  author       = {Goldschmidt, Patrik and Chud{\'a}, Daniela},
  title        = {Network Intrusion Datasets Survey: analysis notebooks and supplementary lists},
  howpublished = {\url{https://github.com/xGoldy/nid-datasets}}
}

@article{pinto2025,
  author  = {Pinto, Daniela and Amorim, Ivone and Maia, Eva and Praca, Isabel},
  title   = {A review on intrusion detection datasets: tools, processes, and features},
  journal = {Computer Networks},
  volume  = {262},
  pages   = {111177},
  year    = {2025},
  publisher = {Elsevier}
}

@inproceedings{arp2022,
  author    = {Arp, Daniel and Quiring, Erwin and Pendlebury, Feargus and Warnecke, Alexander and Pierazzi, Fabio and Wressnegger, Christian and Cavallaro, Lorenzo and Rieck, Konrad},
  title     = {Dos and Don'ts of Machine Learning in Computer Security},
  booktitle = {Proceedings of the 31st USENIX Security Symposium},
  pages     = {3971--3988},
  year      = {2022}
}

@inproceedings{moustafa2015,
  author    = {Moustafa, Nour and Slay, Jill},
  title     = {{UNSW-NB15}: a comprehensive data set for network intrusion detection systems},
  booktitle = {Military Communications and Information Systems Conference (MilCIS)},
  year      = {2015},
  xxxdoi       = {10.1109/MilCIS.2015.7348942}
}

@misc{garcia2020iot23,
  author       = {Garcia, Sebastian and Parmisano, Agustin and Erquiaga, Maria Jose},
  title        = {{IoT-23}: A labeled dataset with malicious and benign {IoT} network traffic},
  howpublished = {\url{https://www.stratosphereips.org/datasets-iot23}},
  year         = {2020},
}

@article{koroniotis2019,
  author  = {Koroniotis, Nickolaos and Moustafa, Nour and Sitnikova, Elena and Turnbull, Benjamin},
  title   = {Towards the development of realistic botnet dataset in the {Internet of Things} for network forensic analytics: {Bot-IoT} dataset},
  journal = {Future Generation Computer Systems},
  volume  = {100},
  pages   = {779--796},
  year    = {2019},
  xxxdoi     = {10.1016/j.future.2019.05.041}
}

@misc{stratosphereMCF,
  author       = {{Stratosphere Laboratory}},
  title        = {Malware Capture Facility},
  howpublished = {\url{https://www.stratosphereips.org/datasets-malware}},
  year         = {2026},
}

@article{kolias2016,
  author  = {Kolias, Constantinos and Kambourakis, Georgios and Stavrou, Angelos and Gritzalis, Stefanos},
  title   = {Intrusion Detection in 802.11 Networks: Empirical Evaluation of Threats and a Public Dataset},
  journal = {IEEE Communications Surveys \& Tutorials},
  volume  = {18},
  number  = {1},
  pages   = {184--208},
  year    = {2016},
  xxxdoi     = {10.1109/COMST.2015.2402161}
}

@article{wilkinson2016fair,
  author  = {Wilkinson, Mark D. and others},
  title   = {The {FAIR} Guiding Principles for scientific data management and stewardship},
  journal = {Scientific Data},
  volume  = {3},
  pages   = {160018},
  year    = {2016},
  xxxdoi     = {10.1038/sdata.2016.18}
}

@article{shiravi2012,
  author  = {Shiravi, Ali and Shiravi, Hadi and Tavallaee, Mahbod and Ghorbani, Ali A.},
  title   = {Toward developing a systematic approach to generate benchmark datasets for intrusion detection},
  journal = {Computers \& Security},
  volume  = {31},
  number  = {3},
  pages   = {357--374},
  year    = {2012},
  xxxdoi     = {10.1016/j.cose.2011.12.012}
}

@inproceedings{sharafaldin2018,
  author    = {Sharafaldin, Iman and Lashkari, Arash Habibi and Ghorbani, Ali A.},
  title     = {Toward Generating a New Intrusion Detection Dataset and Intrusion Traffic Characterization},
  booktitle = {Proceedings of the 4th International Conference on Information Systems Security and Privacy (ICISSP)},
  pages     = {108--116},
  year      = {2018},
  xxxdoi       = {10.5220/0006639801080116}
}

@inproceedings{sommer2010,
  author    = {Sommer, Robin and Paxson, Vern},
  title     = {Outside the Closed World: On Using Machine Learning for Network Intrusion Detection},
  booktitle = {IEEE Symposium on Security and Privacy},
  pages     = {305--316},
  year      = {2010},
  xxxdoi       = {10.1109/SP.2010.25}
}

@inproceedings{tavallaee2009,
  author    = {Tavallaee, Mahbod and Bagheri, Ebrahim and Lu, Wei and Ghorbani, Ali A.},
  title     = {A Detailed Analysis of the {KDD CUP} 99 Data Set},
  booktitle = {IEEE Symposium on Computational Intelligence for Security and Defense Applications (CISDA)},
  pages     = {1--6},
  year      = {2009},
  xxxdoi       = {10.1109/CISDA.2009.5356528}
}

@article{neto2023ciciot,
  author  = {Neto, Euclides Carlos Pinto and Dadkhah, Sajjad and Ferreira, Raphael and Zohourian, Alireza and Lu, Rongxing and Ghorbani, Ali A.},
  title   = {{CICIoT2023}: A Real-Time Dataset and Benchmark for Large-Scale Attacks in {IoT} Environment},
  journal = {Sensors},
  volume  = {23},
  number  = {13},
  pages   = {5941},
  year    = {2023},
  xxxdoi     = {10.3390/s23135941}
}

@inproceedings{goh2017swat,
  author    = {Goh, Jonathan and Adepu, Sridhar and Junejo, Khurum Nazir and Mathur, Aditya P.},
  title     = {A Dataset to Support Research in the Design of Secure Water Treatment Systems},
  booktitle = {Critical Information Infrastructures Security (CRITIS 2016)},
  series    = {LNCS},
  pages     = {88--99},
  publisher = {Springer},
  year      = {2017},
  xxxxxxdoi       = {10.1007/978-3-319-71368-7_8}
}

@inproceedings{vanderheijden2018veremi,
  author    = {van der Heijden, Rens W. and Lukaseder, Thomas and Kargl, Frank},
  title     = {{VeReMi}: A Dataset for Comparable Evaluation of Misbehavior Detection in {VANETs}},
  booktitle = {Security and Privacy in Communication Networks (SecureComm 2018)},
  pages     = {318--337},
  year      = {2018},
  xxxdoi       = {10.1007/978-3-030-01701-9_18}
}

@article{garcia2014ctu13,
  author  = {Garc{\'i}a, Sebastian and Grill, Martin and Stiborek, Jan and Zunino, Alejandro},
  title   = {An empirical comparison of botnet detection methods},
  journal = {Computers \& Security},
  volume  = {45},
  pages   = {100--123},
  year    = {2014},
  xxxdoi     = {10.1016/j.cose.2014.05.011}
}

@article{hikari2021,
  author  = {Ferriyan, Andrey and Thamrin, Achmad Husni and Takeda, Keiji and Murai, Jun},
  title   = {Generating Network Intrusion Detection Dataset Based on Real and Encrypted Synthetic Attack Traffic},
  journal = {Applied Sciences},
  volume  = {11},
  number  = {17},
  pages   = {7868},
  year    = {2021},
  xxxdoi     = {10.3390/app11177868}
}

@article{ashraf2026,
  author  = {Ashraf, Wasia and Masoodi, Faheem Syeed},
  title   = {A survey of intrusion detection datasets for communication networks},
  journal = {Discover Computing},
  volume  = {29},
  number  = {1},
  pages   = {577},
  year    = {2026},
  publisher = {Springer}
}

@article{dobler2025,
  author  = {Dobler, Martin and Hellwig, Michael and Lopes, Nuno and Oakley, Ken and Winterburn, Mike},
  title   = {Systematic review and characterisation of malicious industrial network traffic datasets},
  journal = {International Journal of Information Security},
  volume  = {24},
  number  = {5},
  pages   = {208},
  year    = {2025},
  publisher = {Springer}
}

@article{appsci2026selection,
  author  = {G{\"u}ler, Hakan and Boyac{\i}, Aytu{\u{g}} and Ula{\c{s}}, Mustafa},
  title   = {Security Datasets for Intrusion Detection and Prevention: A Structured Review and Dataset-Selection Framework},
  journal = {Applied Sciences},
  volume  = {16},
  number  = {15},
  pages   = {7473},
  year    = {2026},
  xxxdoi     = {10.3390/app16157473}
}

@inproceedings{engelen2021,
  author    = {Engelen, Gints and Rimmer, Vera and Joosen, Wouter},
  title     = {Troubleshooting an intrusion detection dataset: the CICIDS2017 case study},
  booktitle = {2021 IEEE Security and Privacy Workshops (SPW)},
  pages     = {7--12},
  year      = {2021},
  organization = {IEEE}
}

@article{sarhan2022standard,
  author  = {Sarhan, Mohanad and Layeghy, Siamak and Portmann, Marius},
  title   = {Towards a standard feature set for network intrusion detection system datasets},
  journal = {Mobile Networks and Applications},
  volume  = {27},
  number  = {1},
  pages   = {357--370},
  year    = {2022},
  publisher = {Springer}
}

@article{kenyon2020,
  author  = {Kenyon, Anthony and Deka, Lipika and Elizondo, David},
  title   = {Are public intrusion datasets fit for purpose characterising the state of the art in intrusion event datasets},
  journal = {Computers \& Security},
  volume  = {99},
  pages   = {102022},
  year    = {2020},
  publisher = {Elsevier}
}

@article{gebru2021datasheets,
  author  = {Gebru, Timnit and Morgenstern, Jamie and Vecchione, Briana and Vaughan, Jennifer Wortman and Wallach, Hanna and Daum{\'e}~III, Hal and Crawford, Kate},
  title   = {Datasheets for Datasets},
  journal = {Communications of the ACM},
  volume  = {64},
  number  = {12},
  pages   = {86--92},
  year    = {2021},
  xxxdoi     = {10.1145/3458723}
}

@article{klein2014rot,
  author  = {Klein, Martin and Van de Sompel, Herbert and Sanderson, Robert and Shankar, Harihar and Balakireva, Lyudmila and Zhou, Ke and Tobin, Richard},
  title   = {Scholarly Context Not Found: One in Five Articles Suffers from Reference Rot},
  journal = {PLOS ONE},
  volume  = {9},
  number  = {12},
  pages   = {e115253},
  year    = {2014},
  xxxdoi     = {10.1371/journal.pone.0115253}
}

@article{walters2023fabrication,
  author  = {Walters, William H. and Wilder, Esther Isabelle},
  title   = {Fabrication and errors in the bibliographic citations generated by {ChatGPT}},
  journal = {Scientific Reports},
  volume  = {13},
  pages   = {14045},
  year    = {2023},
  xxxdoi     = {10.1038/s41598-023-41032-5}
}

@article{ji2023hallucination,
  author  = {Ji, Ziwei and Lee, Nayeon and Frieske, Rita and Yu, Tiezheng and Su, Dan and Xu, Yan and Ishii, Etsuko and Bang, Ye Jin and Madotto, Andrea and Fung, Pascale},
  title   = {Survey of Hallucination in Natural Language Generation},
  journal = {ACM Computing Surveys},
  volume  = {55},
  number  = {12},
  pages   = {248:1--248:38},
  year    = {2023},
  xxxdoi     = {10.1145/3571730}
}

@misc{netresec,
  author       = {{NETRESEC AB}},
  title        = {Publicly available {PCAP} files},
  howpublished = {\url{https://www.netresec.com/?page=PcapFiles}},
  year         = {2026}
}

@article{hofstede2014sshcure,
  author  = {Hofstede, Rick and Hendriks, Luuk and Sperotto, Anna and Pras, Aiko},
  title   = {{SSH} Compromise Detection using {NetFlow/IPFIX}},
  journal = {ACM SIGCOMM Computer Communication Review},
  volume  = {44},
  number  = {5},
  pages   = {20--26},
  year    = {2014}
}

@article{dener2024wsnbfsf,
  author  = {Dener, Murat and Okur, Celil and Al, Samed},
  title   = {{WSN-BFSF}: A New Dataset for Attacks Detection in Wireless Sensor Networks},
  journal = {IEEE Internet of Things Journal},
  year    = {2024},
  xxxdoi  = {10.1109/JIOT.2023.3292209}
}

@inproceedings{sangster2009cdx,
  author    = {Sangster, Benjamin and O'Connor, T. J. and Cook, Thomas and Fanelli, Robert and Dean, Erik and Adams, William J. and Morrell, Chris and Conti, Gregory},
  title     = {Toward Instrumenting Network Warfare Competitions to Generate Labeled Datasets},
  booktitle = {2nd Workshop on Cyber Security Experimentation and Test (CSET '09)},
  year      = {2009},
  publisher = {USENIX}
}

@inproceedings{gogoi2012tuids,
  author    = {Gogoi, Prasanta and Bhuyan, Monowar H. and Bhattacharyya, D. K. and Kalita, J. K.},
  title     = {Packet and Flow Based Network Intrusion Dataset},
  booktitle = {Contemporary Computing (IC3 2012)},
  series    = {Communications in Computer and Information Science},
  volume    = {306},
  pages     = {322--334},
  year      = {2012},
  publisher = {Springer},
  xxxdoi    = {10.1007/978-3-642-32129-0_34}
}

@article{povalac2023lorawan,
  author  = {Povala{\v{c}}, Ale{\v{s}} and Kr{\'a}l, Jan and others},
  title   = {Exploring {LoRaWAN} Traffic: In-Depth Analysis of {IoT} Network Communications},
  journal = {Sensors},
  volume  = {23},
  number  = {17},
  pages   = {7333},
  year    = {2023},
  xxxdoi  = {10.3390/s23177333},
  note    = {Dataset: LoRaWAN Traffic Analysis Dataset, Zenodo, \url{https://zenodo.org/record/8090619}}
}
\endgroup

\appendix





\section{Catalog versus AI assistants}\label{sec:assistants}
To compare the catalog with the answers of general-purpose AI assistants, on 9 November 2026 we asked three of them (ChatGPT, Gemini and Claude, with their default models) for the public datasets of three domains: IoT datasets with packet captures and multiclass labels, 5G datasets with labeled attacks, and in-vehicle datasets with labeled attacks (Tab.~\ref{tab:assistants}). 

None of the answers contained a dataset that could be shown not to exist, and together they suggested 14 datasets that the catalog had missed, which we verified and added. Coverage, however, differed widely: against the updated catalog, the assistants retrieved between 23\% and 61\% of the 61 matching datasets, and even their union missed 12 of them. Of the 57 distinct datasets they returned, only 10 were returned by all three. A researcher who relies on one assistant therefore sees a different, and often much smaller, part of the field than one who relies on another.
The answers also contained errors that a reader cannot detect without checking each source. The assistants disagreed on basic properties of the same dataset, such as whether it distributes packet captures or provides multiclass labels, so that the same dataset was recommended by one assistant and excluded by another. Some citations were incomplete or wrong: download links pointing to unrelated datasets, datasets attributed to the paper of a different one, missing or garbled author lists, and the same dataset listed twice under different names. Release years often differed from those of the official pages, and access conditions and label granularity, which determine whether a dataset can be used for a given experiment (Sec.~\ref{sec:suitability}), were reported only occasionally.
An assistant is therefore a useful source of candidates, which is how the \Obs{} uses one (Sec.~\ref{sec:method}), but it does not replace a catalog whose records are verified, described along common fields and kept up to date.
\begin{table}[!h]\centering\small
\caption{Datasets returned by three general-purpose AI assistants for three domains, compared with the catalog. Recall: matching catalog datasets that were returned; the catalog includes 14 datasets that the assistants suggested and that were verified and added. Outside the catalog: datasets returned that were not added after verification.}\label{tab:assistants}
\begin{tabular}{lccccc}
\toprule
 & \multicolumn{4}{c}{\textbf{Recall of matching catalog datasets}} & \\
\cmidrule(lr){2-5}
\textbf{Assistant} & \textbf{IoT}$^{a}$ & \textbf{5G}$^{b}$ & \textbf{In-vehicle}$^{c}$ & \textbf{Total} & \textbf{Outside the catalog} \\
\midrule
ChatGPT & 7/24 & 7/13 & 22/24 & 36/61 (59\%) & 7 \\
Gemini  & 5/24 & 4/13 & 5/24  & 14/61 (23\%) & 0 \\
Claude  & 14/24 & 7/13 & 16/24 & 37/61 (61\%) & 1 \\
\midrule
Any of the three & 14/24 & 12/13 & 23/24 & 49/61 (80\%) & \\
Returned by all three & 3/14 & 2/15 & 5/28 & 10/57 (18\%) & \\
\bottomrule
\end{tabular}

\smallskip
\parbox{.95\linewidth}{\scriptsize\raggedright Any of the three: matching catalog datasets returned by at least one assistant. Returned by all three: datasets returned by all three assistants over all distinct datasets returned for that domain.}

\smallskip
\begin{tcolorbox}[enhanced, width=.95\linewidth, colback=gray!5, colframe=gray!60!black,
  boxrule=0.5pt, arc=2pt, left=4pt, right=4pt, top=6pt, bottom=3pt,
  fontupper=\scriptsize, attach boxed title to top left={xshift=6pt, yshift=-2pt},
  boxed title style={colback=gray!60!black, arc=1pt, boxrule=0pt, left=3pt, right=3pt, top=1pt, bottom=1pt},
  title={\scriptsize\bfseries Prompt}]
\textit{``List all publicly available datasets of network traffic for [DOMAIN] that [PROPERTY]. For each dataset give: name, year of release, link to the official download page, the associated paper, and whether it distributes packet captures (pcap) or only extracted features/flows. Only include datasets that actually exist and can be obtained.''}

\smallskip
\begin{tabular}{@{}l@{\ }p{.78\linewidth}@{}}
$^{a}$ IoT: & [DOMAIN] = \textit{IoT, IIoT or IoMT intrusion detection}; [PROPERTY] = \textit{distribute packet captures (pcap) and provide multiclass attack labels} \\
$^{b}$ 5G: & [DOMAIN] = \textit{5G networks or the 5G core}; [PROPERTY] = \textit{contain labeled attacks} \\
$^{c}$ In-vehicle: & [DOMAIN] = \textit{in-vehicle networks (CAN bus or automotive Ethernet)}; [PROPERTY] = \textit{contain labeled attacks} \\
\end{tabular}

\smallskip
One new session per assistant and domain.
\end{tcolorbox}
\end{table}

\begin{table*}[!htbp]
\centering\scriptsize\setlength{\tabcolsep}{3pt}
\providecommand{\acqR}{\CIRCLE}
\providecommand{\acqH}{\LEFTcircle}
\providecommand{\acqE}{\Circle}
\providecommand{\cmark}{\textcolor{green!55!black}{\ding{51}}}
\providecommand{\xmark}{\textcolor{red!55!black}{\ding{55}}}
\caption{Datasets within the scope of the \Obs{} that could not be obtained at the time of release \release{}, with the domain they would belong to. They are kept on the website, marked as not available, and re-enter the catalog if their distribution is restored or published.
}
\label{tab:excluded}
\resizebox{\textwidth}{!}{%
\begin{tabular}{llcccccclp{5.6cm}}
\toprule
\textbf{Dataset} & \textbf{Domain} & \textbf{Year$\downarrow$} & \textbf{Raw} & \textbf{Proc.} & \textbf{Acq.} & \textbf{Lab.} & \textbf{Att.} & \textbf{Status} & \textbf{Evidence} \\
\midrule
\textsc{CDX CTF 2009}~\cite{sangster2009cdx} & Enterprise / general IP & 2009 & \cmark & \xmark & \acqH & -- & -- & Withdrawn & The usma.edu pages are offline after the move to westpoint.edu; derived features remain available in ASNM-CDX-2009. \\
\rowcolor[HTML]{f0f0f0}
\textsc{TUIDS}~\cite{gogoi2012tuids} & DDoS & 2012 & \cmark & \cmark & \acqH & B & 8 & Withdrawn & The paper points to a Tezpur University page that is no longer online; listed as unavailable by COMIDDS. \\
\textsc{ISCX Botnet 2014}~\cite{beigi2014iscxbot} & Malware / botnet / C2 & 2014 & \cmark & \xmark & \acqH & B & 16 & Withdrawn & Confirmed as unavailable on the official UNB/CIC download platform. \\
\rowcolor[HTML]{f0f0f0}
\textsc{SSHCure}~\cite{hofstede2014sshcure} & Backbone / ISP / Honeypot & 2014 & \xmark & \cmark & \acqR & -- & -- & Withdrawn & SimpleWeb lists the flow and log files of the University of Twente campus network, but no public download link is available. \\
\textsc{CIC DoS 2017}~\cite{jazi2017cicdos} & DDoS & 2017 & \cmark & \xmark & \acqH & -- & 8 & Withdrawn & Confirmed as currently unavailable on the official UNB/CIC download page. \\
\rowcolor[HTML]{f0f0f0}
\textsc{WSN-BFSF}~\cite{dener2024wsnbfsf} & Wireless / LPWAN / WSN & 2024 & \xmark & \cmark & \acqE & -- & 3 & Data not published & The paper (NS-2 simulation, 312k records) exists; no official download location was found. \\
\textsc{WPA2-KKID} & Wireless / LPWAN / WSN & 2025 & \cmark & \cmark & -- & -- & 2 & Not yet released & Announced as open access (pcap and CSV, KRACK and Kr00k) in a 2025 presentation; no download location has been published. \\
\rowcolor[HTML]{f0f0f0}
\textsc{MalwareTraffic}~\cite{malwaretraffic2026} & Malware / botnet / C2 & 2026 & \cmark & \cmark & \acqE & M & 362 & Not yet released & Announced with the paper; data repository is not yet published. \\
\bottomrule
\end{tabular}%
}

\smallskip
\parbox{\linewidth}{\scriptsize\raggedright
\textbf{N.B.:} columns as in the per-domain tables (Tabs.~\ref{tab:rev-iot}--\ref{tab:rev-encrypted}); \textbf{Att.}: number of attack types; \textbf{Status}: \emph{Withdrawn}, distributed in the past but no longer obtainable; \emph{Not yet released}, announced by its authors but not yet published; \emph{Data not published}, described in a paper whose data were never distributed. Last checked on \snapshot{}.}
\end{table*}

\section{Excluded datasets}\label{app:excluded}
Of the \NScreened{} candidates screened for release \release{}, \NExcluded{} were excluded by the criteria of Sec.~\ref{sec:scope}: \NExclTrace{} could not be traced to any public source, \NExclNoTraffic{} did not contain network traffic and \NExclUnavail{} could not be obtained at the time of the release. Table~\ref{tab:excluded} lists only the datasets excluded for availability, because they are within the scope of the \Obs{} and would otherwise belong to one of its domains; for each of them, the table gives that domain and the same columns as the domain tables. Five have been withdrawn after being distributed, two have been announced by their authors but not yet published, and one is described in a paper whose data were never distributed.
These datasets are kept on the website, marked as not available, for two reasons. Some of them, such as \texttt{ISCX Botnet 2014} and \texttt{CIC DoS 2017}, are still used in published work, and a reader who finds them cited needs to know that the results cannot be reproduced on the original distribution; for \texttt{CDX CTF 2009}, derived features remain available through a later release. Others may become available again, or become available for the first time: the periodic check of Section~\ref{sec:maintenance} looks for a restored or newly published distribution and, when it finds one, moves the dataset into the catalog in the next update.

\begin{table*}[!htbp]
\centering
\footnotesize
\caption{Acronyms and abbreviations used in the paper, including common networking and security terms that are not expanded in the text. Dataset names are not listed.}
\label{tab:acronyms}
\begin{tabular}{@{}>{\columncolor{gray!15}}l p{0.36\textwidth} >{\columncolor{gray!15}}l p{0.36\textwidth}@{}}
\toprule
\textbf{Acronym} & \textbf{Meaning} & \textbf{Acronym} & \textbf{Meaning} \\
\midrule
5GC & 5G Core network & LoRaWAN & Long Range Wide-Area Network \\
AI & Artificial Intelligence & LPWAN & Low-Power Wide-Area Network \\
AMF & Access and Mobility Management Function (5G core) & M2M & Machine-to-Machine \\
AP & Access Point & MAC & Medium Access Control \\
APT & Advanced Persistent Threat & MAWI & Measurement and Analysis on the WIDE Internet \\
ATT\&CK & Adversarial Tactics, Techniques, and Common Knowledge (MITRE) & MitM & Man-in-the-Middle \\
AVTP & Audio Video Transport Protocol & ML & Machine Learning \\
C2 & Command and Control & MQTT & Message Queuing Telemetry Transport \\
CAIDA & Center for Applied Internet Data Analysis & NAS & Non-Access Stratum (5G signaling) \\
CAM & Cooperative Awareness Message (V2X) & NF & Network Function (5G core) \\
CAN & Controller Area Network & NGAP & NG Application Protocol (5G) \\
CAN-FD & CAN with Flexible Data-rate & NIDS & Network Intrusion Detection System \\
CAPEC & Common Attack Pattern Enumeration and Classification & NS-2/3 & Network Simulator 2 and 3 \\
CFM & CICFlowMeter (flow extractor) & OPC UA & Open Platform Communications Unified Architecture \\
CIC & Canadian Institute for Cybersecurity & OT & Operational Technology \\
CSV & Comma-Separated Values & PCAP & Packet Capture (file format) \\
DARPA & Defense Advanced Research Projects Agency & PFCP & Packet Forwarding Control Protocol \\
DBC & CAN database file (signal definitions) & PQC & Post-Quantum Cryptography \\
DDoS & Distributed Denial of Service & PTP & Precision Time Protocol \\
DNP3 & Distributed Network Protocol 3 & QUIC & QUIC transport protocol (originally Quick UDP Internet Connections) \\
DNS & Domain Name System & RAT & Remote Access Trojan \\
DoS & Denial of Service & SAE & Society of Automotive Engineers \\
ECU & Electronic Control Unit & SDN & Software-Defined Networking \\
FAIR & Findable, Accessible, Interoperable, Reusable & SMF & Session Management Function (5G core) \\
FL & Federated Learning & SMOTE & Synthetic Minority Over-sampling Technique \\
FTP & File Transfer Protocol & SOME/IP & Scalable service-Oriented MiddlewarE over IP \\
GCS & Ground Control Station & SQL & Structured Query Language \\
GenAI & Generative AI & SSH & Secure Shell \\
gNB & Next-generation NodeB (5G base station) & SUMO & Simulation of Urban MObility \\
GPS & Global Positioning System & SYN & Synchronize (TCP connection-setup flag) \\
GTP & GPRS Tunnelling Protocol & TCP & Transmission Control Protocol \\
HCRL & Hacking and Countermeasure Research Lab & TEID & Tunnel Endpoint Identifier (GTP) \\
HIL & Hardware-in-the-Loop & TLS & Transport Layer Security \\
HTTP & Hypertext Transfer Protocol & U2R & User to Root (privilege escalation) \\
HTTPS & HTTP Secure & UAV & Unmanned Aerial Vehicle \\
ICS & Industrial Control System & UDP & User Datagram Protocol \\
IDS & Intrusion Detection System & UE & User Equipment (mobile device) \\
IEC 104 & IEC 60870-5-104 telecontrol protocol & UNSW & University of New South Wales \\
IIoT & Industrial Internet of Things & V2X & Vehicle-to-Everything \\
IoMT & Internet of Medical Things & VANET & Vehicular Ad-Hoc Network \\
IoT & Internet of Things & WAF & Web Application Firewall \\
IP & Internet Protocol & WPA3 & Wi-Fi Protected Access 3 \\
ISP & Internet Service Provider & WSN & Wireless Sensor Network \\
KDD & Knowledge Discovery and Data Mining & XML & Extensible Markup Language \\
LAN & Local Area Network & XSS & Cross-Site Scripting \\
\bottomrule
\end{tabular}
\end{table*}

\section{List of Acronyms}
\label{app:acronyms}
The datasets in the \Obs{} come from different communities, from enterprise networks to vehicles, drones and mobile cores, each with its own protocols and terminology. Table~\ref{tab:acronyms} lists the acronyms used in this paper, including well-known networking and security terms that are not expanded in the text, so that readers from one community can follow the discussion of the others.
\end{document}